\documentclass[print]{aa}

\usepackage{graphicx}

\usepackage{natbib}

\bibpunct{(}{)}{;}{a}{}{,} 

\begin{document}

\title{On the {SEDs} of passively heated condensed cores}

\author{J\"org Fischera\inst{1,}\inst{2}}
\institute{
	Research School of Astronomy \& Astrophysics, Mount Stromlo Observatory, Cotter Road, 
	Weston Creek, ACT 2611, Australia\and
	Canadian Institute for Theoretical Astrophysics, University of Toronto, 60 St. George Street, 
	ON M5S3H8, Canada }


\abstract{}{
	The dust emission spectrum and the brightness profile of passively heated 
	condensed cores is analyzed in relation to their astrophysical environment. The model is used
	to study systematically the radiative transfer effects on essential parameters such as the dust emissivity, 
	dust temperature, and luminosity of the cores and to derive uncertainties in typical estimates of the IR
	flux and size.
	}{The cores are modeled as critically stable self-gravitating spheres embedded at the center of
	self-gravitating filaments that are assumed to be either spherical or cylindrical in shape. The filaments are heated 
	by an isotropic interstellar radiation field (ISRF).
	The calculations are based on a physical dust model of stochastically heated grains 
	of diffuse interstellar dust. The spectral energy distribution (SED) of the cores is calculated
	using a ray-tracing technique
	where the effects of scattered emission and re-heating by dust grains are accurately taken into account.	
	To compare with observational studies,
	the dust re-emission spectrum is approximated by a modified black-body function and
	the brightness profile with a Gaussian source.
	A simplified single-zone model for cores is presented that incorporates on the basis of the derived
	emissivities a first order approximation of their SED.	}{
	Colder dust temperatures are, independent of the core mass, 
	related to a higher pressure both inside and around the filament. 
	The pressure-temperature relation for given external pressure 
	is found to be largely independent of the true shape of the filament. 
	The calculations show that the radiative transfer leads to a lower emission coefficient at 
	$250~\mu{\rm m}$ and to 	a flatter emissivity law of typically $\beta<1.8$ in the far-infrared sub-millimeter regime.
	These effects cause the core mass to be underestimated by more than a factor of 2 based on the typical 
	assumptions used in observational programs. A larger uncertainty
	is expected for high pressure regions. 
	Fitting the core using a Gaussian source approximation overestimates the flux by 
	$\sim 10\%$. For highly embedded
	cores and in general for cores in high pressure regions,
	the surface brightness profile is flatter with respect to the profile of the column density and a Gaussian
	profile. These effects can lead to an overestimate of the core size of $10 - 30\%$ based on marginally resolved 
	$250~\mu{\rm m}$ observations.
	}{}


\maketitle

\section{Introduction}

	Condensed cores are compact dense regions of molecular clouds that have a high column density contrast to their
	surrounding medium. They 
	are regions of star formation and appear to be physically very similar to Bok-globules \citep{Bok1947}, dense,
	almost spherically shaped, highly compact molecular clouds surrounded by non-molecular thin 
	interstellar gas \citep{Curry2000}.

	The formation of the condensed cores is potentially linked to the turbulent motion of the molecular gas that 
	produces a self-similar multi-fractal density
	structure with a log-normal distribution of the local density. In this scenario, the cores are formed out of massive
	density enhancements that became self-gravitating and stable against the turbulent motion. Turbulence naturally
	produces a mass distribution of clouds similar to that observed in the ISM and is understood to be the origin
	of the initial mass function (IMF) of the stars (see references in \citet{Fischera2008}).
	
	Because the cores are highly opaque, they are most easily studied in the far infrared (FIR) or submm regime
	by observing the dust emission spectrum where the cores are still optically thin.
	Surveys of molecular clouds performed in the FIR/submm (as are possible using BLAST, HERSCHEL, and ALMA) 
	are essential to our understanding of the very early 
	evolutionary process of star formation and will help us to
	unlock the mystery of the origin of the IMF (e.g. \citet{Netterfield2009, Olmi2009}). 
	The interpretation, however, of the dust emission spectrum 
	is  complicated by both radiative transfer effects and the unknown dust properties in the molecular phase of
	the ISM. Both effects lead to uncertainties in particular in the measurement of the core masses, which is the main
	observable derived by modeling the SED. This causes a major problem
	in gaining information about the efficiency of star formation, the stellar mass, or the masses 
	produced within a core of a certain mass.
	The radiative transfer also further affects the surface brightness profile of the core as grains in the cloud are
	heated by a gradually changing radiation field \citep{Fischera2008,Nutter2009}.
	
	Observationally, cores display, if described by a single temperature model, a large variation in dust temperatures.
	This is a direct consequence not only of their gravitational state, but also of their local environment.
	Some cores are exclusively heated by an external
	radiation field while others have already formed a protostellar object in the  center and are therefore also
	heated from the inside. Cores with a warm central object therefore have on average a higher dust temperature
	\citep{Olmi2009}.
	

The SEDs of interstellar clouds and cores have been extensively studied in the past 
\citep{Bernard1992,Bernard1993,Evans2001, Stamatellos2003,Stamatellos2004,Fischera2008}, although with different emphases. \citet{Bernard1992} studied the radiative transfer using a physical dust model and derived the brightness profiles and SEDs of interstellar clouds while making simplified  assumptions about the scattered light and density profiles in clouds (a flat center and a power law further out). They concluded that the observed brightness profiles
cannot be caused by radiative transfer effects alone. \citet{Evans2001}, \citet{Stamatellos2003}, and \citet{Stamatellos2004} studied the
infrared emission from cores to obtain more robust estimates of their dust masses. Their calculations focused on
the radiative transfer aspects and used a simplified dust re-emission model based on the mean dust properties that they adopted from \citet{Ossenkopf1994}.
\citet{Fischera2008} studied the dependence of the SED derived for stable isothermal self-gravitating interstellar clouds on the radiation field and the external pressure in the interstellar medium. 
These calculations were based on a physical dust model of interstellar dust grains. The radiative transfer problem was solved by accurately considering the complications caused by the scattered light to provide (as can also be found in the work of \citet{Stamatellos2003}) profiles of the surface brightness for the whole SED.

\citet{Stamatellos2003} studied the SEDs of cores embedded in optical thick dense spherical clouds and the present work can be compared to theirs. There are, however, basic differences. In their model, the giant molecular cloud (GMC) is not modeled as a self-gravitating entity but as a homogeneous dense sphere and the pressure inside the GMC is assumed to be independent of the extinction of the cloud. 

In this paper, I relate the SEDs of passively heated condensed cores to their physical environment.
The calculations rest upon comparable studies of interstellar clouds
that are either spherical \citep{Fischera2008} or cylindrical (Fischera, in prep.) in shape and are modeled as self-gravitating entities. 
As a first approach, I focus on the radiative transfer effects and base the SEDs of the cores on dust properties in the diffuse ISM.
Complications caused by the turbulent density structure of molecular clouds are not considered and additional studies
will be needed to quantitatively understand how the results presented in this paper are affected. The expected effects are discussed.


To compare the results with observations, the dust re-emission spectrum is described by a modified black-body spectrum. This provides the corresponding observational parameters of the dust emissivity and dust temperature. On the basis 
of these effective values, I present a luminosity-temperature relation and a pressure-temperature relation of embedded cores. I compare the surface brightness with a Gaussian function that is used observationally to extract the fluxes and the size of the cores \citep{Olmi2009, Netterfield2009}. 

In Sect.~\ref{sect_model}, I present my model of passively heated embedded cores, the dust model, and the technique used to solve the radiative transfer problem. I also provide for comparison with observational data a single zone model. My
results of the radiative transfer calculations are presented in Sect.~\ref{sect_sed}. In Sect.~\ref{sect_mbb}, I compare my model with the modified black-body spectrum and in Sect.~\ref{sect_mbb_profile} I provide the surface brightness. 
In Sect.~\ref{sect_discussion}, I discuss the effect of my results on uncertainties in the radiation field and dust properties, and provide a summary of the main results of the paper in Sect.~\ref{sect_summary}.

\section{\label{sect_model}Model}

\subsection{\label{sect_cloudmodel}Model of embedded cores}

\subsubsection{Model of the cores}

The cores are modeled as isothermal self-gravitating spheres, so-called Bonnor-Ebert spheres \citep{Ebert1955,Bonnor1956},
which are embedded in extended molecular clouds and pressurized by its central pressure $p_{\rm c}$ (see references 
in \citet{Fischera2008}).
The cores are naturally in different gravitational states. However, for simplicity I consider in this work the cores that are in the critical state against collapse.

As described in \citet{Curry2000} or \citet{Fischera2008}, critical stable clouds
are characterized by a steep density profile at the outskirts close to $1/r^2$, that becomes flat at the cloud center.
They have an overdensity or overpressure $\hat p$ of $14.04$ in the cloud center where the overpressure $\hat p$ is defined as the ratio of the pressure inside the cloud to the external pressure.
Stable clouds are characterized by a flat density profile and lower overpressure, supercritical clouds by a steep density profile (also $1/r^2$) and higher overpressure. Observationally the cores are probably in a critical or super-critical state as they have a high column density contrast and are therefore easy to identify as individual sources. 


The stability of a spherical isothermal self-gravitating cloud against gravitational collapse is determined  
by the temperature, the molecular weight, and the pressure outside the cloud. 
The temperature is considered as an effective value that contains, apart from the thermal motion,
the turbulent motion of the dusty gas.  In the critical state of the core, its mass is given by
\begin{equation}
	\label{eq_coremass}
	M_{\rm core} = 0.688\left(\frac{T}{10\,{\rm K}}\right)^2\left(\frac{\mu}{2.36}\right)^{-2}\left(\frac{p_{\rm c}/k}{2.8\times10^5\,{\rm K/cm^3}}\right)^{-\frac{1}{2}} M_{\odot},
\end{equation}
where I have assumed that the pressure {$p_{\rm c}$ at the center of the} GMC is 14 times the mean pressure in the ISM of $p/k=2\times 10^4~{\rm K/cm^3}$ proposed by \citet{Curry2000} and also adopted by \citet{Fischera2008}.
The cores are also assumed to be molecular, which implies
a mean atomic weight of $\mu\approx2.36$.
The cores' radius simply scales with the temperature and is given by
\begin{equation}
	\label{eq_coreradius}
	R_{\rm core} = 0.0345 \left(\frac{T}{10\,{\rm K}}\right)\left(\frac{\mu}{2.36}\right)^{-1}
	\left(\frac{p_{\rm c}/k}{2.8\times10^5\,{\rm K/cm^3}}\right)^{-\frac{1}{2}}~{\rm pc}.
\end{equation}
In higher pressure regions, cores of the same temperature are more compact, have 
higher density, and are less massive.
Combining both of these properties provides the mass-radius relation for cores
\begin{equation}
	\label{eq_massradius}
	M_{\rm core} = 5.78 \left(\frac{R_{\rm core}}{0.1\,{\rm pc}}\right)^2 \left(\frac{p_{\rm c}/k}{2.8\times10^5~{\rm K/cm^{3}}}\right)^{\frac{1}{2}}\,M_{\odot}.
\end{equation}

Of prime importance to the radiative transfer calculations is the optical depth through the cloud and its density profile. The column density through the cloud center for a given gravitational state is reflected by the density profile or overpressure is, independent of the true mass, determined by the pressure surrounding the cloud.  For a critically stable cloud, the column density is, assuming solar abundances, given by the expression
\begin{equation}
		\label{eq_columncore}
	N_{\rm H}(0) = 5.8\times 10^{22} \sqrt{\frac{p_{\rm c}/k}{2.8\times 10^5~{\rm K/cm^3}}}~{\rm cm^{-2}}.
\end{equation}

The column density through the center is difficult observationally to measure where the mean column density
is instead derived. The corresponding mean column density of the critical stable core is then given by
\begin{equation}
	\label{eq_meancolumncore}
	\left<N_{\rm H}\right> = \frac{2}{3}\left<N_{\rm H}\right>(0) = 1.6 \times 10^{22}\sqrt{\frac{p_{\rm c}/k}{2.8\times 10^5~{\rm K/cm^3}}}~{\rm cm^{-2}},
\end{equation}
where $\left<N_{\rm H}\right>(0)$ is the column density at the center of a cloud whose mass is distributed
homogeneously inside the cloud volume (see Appendix~\ref{sect_cloud_ext}). 

In higher pressure regions, cores of the same effective temperature are less massive, smaller, but more optically thick.  In higher pressure regions, the dust inside the core is heated by a more strongly attenuated radiation field. For the same radiation field heating the core, the dust temperatures of the cores in high pressure regions are therefore colder than the cores in low pressure regions.

\subsubsection{\label{sect_pressextrel}Model of the filaments}

As the cores, the filaments are modeled as
isothermal self-gravitating clouds surrounded by a thin interstellar medium of pressure $p_{\rm ext}$. 
For comparison, both spheres and cylinders are considered as approximations of the observed structures in the ISM.
As self-gravitating entities, they display a density and pressure profile.

It is reasonable to assume that condensed cores form out of density enhancements 
at the centers of the GMC. Here, potential cores experience the highest pressures and 
are therefore the locations where clouds can most easily become self-gravitating. Only those
cases are therefore considered in the model.
For simplicity, the mass and size of the core is assumed to be negligible 
to the mass and the extension of the filaments. 

The physical properties of a critical stable spherical GMC is described by the same equations as given for the cores (Eq.~\ref{eq_coremass} -~\ref{eq_meancolumncore}). Replacing the pressure by the mean ISM pressure of $2\times 10^4~{\rm K/cm^3}$, the mass radius relation {is in good agreement with observations 
of  interstellar molecular clouds \citep{Larson1981}.

To study the whole range of structures from Bok-globules to highly embedded cores, I consider for a given external pressure
different extinction values through the filaments.
Since the central pressure of self-gravitating clouds is related to the column density 
at its center, the central pressure is also varied.
The relation for spheres and cylinders
embedded in the ISM of our Galaxy is shown in Fig.~\ref{fig_colpresrel}. For the same central extinction of a filament,
cores situated in cylinders are subject to a higher central pressure than cores in spheres. 

The range of extinction values considered in the paper leads to the situation 
indicated in Fig.~\ref{fig_colpresrel}, where the gravitational state of the spherical filaments varies
over a wide range from highly stable to, in some cases, highly super-critical clouds.
For the physical solutions of self-gravitating isothermal cylinders, a similar stability classification as
for spheres does not exist as the cloud can always produce by compression a higher gas pressure at the cloud edge. 
The cylindrical filaments can therefore for all cases studied in the paper be regarded as stable.

Cylinders have a maximum mass-line density given by $M_{\rm cyl}/L=2K/G$, where $M_{\rm cyl}$ is the cloud 
mass, $L$ is the cloud length, and $G$ is the gravitational constant. The constant
$K$, given by $k_{\rm B}T/(\mu m_{\rm u})$, is the one generally introduced in isothermal cloud models 
where $k_{\rm B}$ is
the Boltzmann constant, $m_{\rm u}$ the atomic mass unit, and $\mu$ the mean molecular weight. 
This mass-line density is only reached in the limit of infinite overpressure (Fischera, in prep.).
By construction, the mass-line density of cylinders with finite overpressure is lower than the 
asymptotic value. If we consider two clouds
with the same mass-line density, a higher overpressure corresponds to a lower (effective) gas temperature. A
change in the external pressure clearly has no consequence on the overpressure.

The situation is fundamentally 
different for spherical clouds where the maximum mass is identical to the critical stable mass 
given by $M_{\rm sph} \propto (T/\mu)^2/\sqrt{p_{\rm ext}}$, where a higher external 
pressure will change the gravitational state and can make the cloud unstable. 
While a cooling of a spherical cloud will 
consequently lead to a critical and finally unstable cloud it will lead only to a higher gravitational but still stable state
in the case of cylinders. I note that - although a supercritical spherical cloud can be constructed - this state
can not be obtained by means of simple cooling. If pressure equilibrium between cloud and the surrounding medium 
were assumed cooling of a critical stable cloud would lead to a \emph{decrease} in the pressure at the outskirts. A spherical super-critical cloud assumed to be in pressure equilibrium with the surrounding medium  also corresponds to 
a cloud temperature \emph{above} the critical temperature (Fischera, in prep.).

Cylindrical self-gravitating clouds of high overpressure are characterized by a steep density profile
at the outskirts with $\rho\propto r^{-4}$ \citep{Stodolkiewicz1963,Ostriker1964}. 
The observed density profile of interstellar clouds as pointed out by \citet{FiegePudritz2000a} appears to be less steep and more closely represented by a
profile $\rho\propto r^{-2}$. Instabilities may explain these results as e.g. discussed by \citet{Larson1985} or \citet{FiegePudritz2000b}, which prevent the clouds producing high overpressures. 
However, cases with high overpressure are still useful in the framework of the radiative transfer problem and are therefore
included in the model.

\begin{figure}[htbp]
	\includegraphics[width=\hsize]{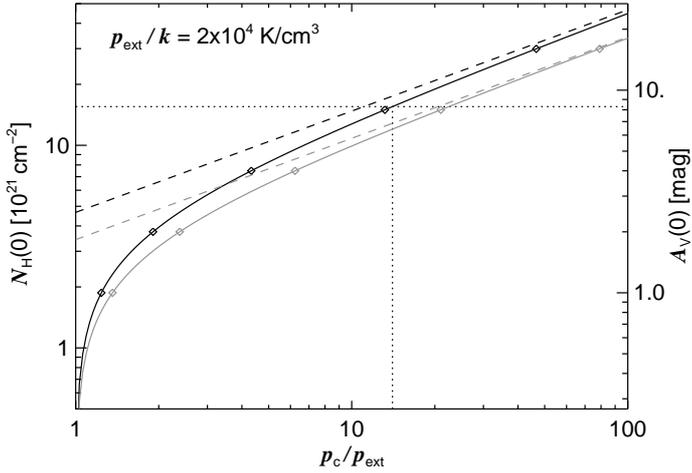}
	\caption{\label{fig_colpresrel}
	Column density at the center of a spherical (black) and cylindrical (grey) isothermal self-gravitating cloud as a
	function of the overpressure $p_c/p_{\rm ext}$, where $p_c$ is the central pressure and $p_{\rm ext}$
	is the external pressure. The dashed lines indicates the asymptotic behavior for high overpressure where $N_{\rm H}(0)\propto \sqrt{p_{\rm c}}$. The external pressure is assumed to be consistent with the mean pressure of $2\times 10^4~{\rm K/cm^3}$ in the 
	ISM of the Milky Way \citep{Boulares1990}. The grey dotted lines show the values for a critical stable sphere with an overpressure of 14.045. The diamonds mark the 
	values considered in the paper and correspond to $A_V=[1,2,4,8,16]$, where the extinction $A_V$ is
	derived from the column density using the dust-to-gas ratio $E(B-V)/N_{\rm H}=5.8\times 10^{-21} \rm mag/cm^{-2}$
	\citep{Bohlin1978}
	and the absolute to relative extinction $R_V=3.1$ \citep{Fitzpatrick1999}.
	}
\end{figure}


The column density at the center of self-gravitating cloud models is given by
\begin{equation}
	\label{eq_coltotal}
	N_{\rm H}(0) = \frac{2}{m_{\rm u}\Sigma_i \frac{n_i}{n_{\rm H}}\mu_i}\sqrt{\frac{p_c}{4\pi G}}\int\limits_0^{z_{\rm cl}}
		{\rm d}z\,e^{-\omega(z)},
\end{equation}
where $\omega$ is the potential, $G$ the gravitational constant, $\mu_i$ the weight of element $i$ in units of atomic mass units $m_{\rm u}$, and $z_{\rm cl}$ is determined by the pressure equilibrium between the cloud and the surrounding medium $p(z_{\rm cl}) = p_c\,e^{-\omega(z_{\rm cl})}=p_{\rm ext}$. The column density at the center of a  core given in Eq.~\ref{eq_columncore}
is a special case.
In the limit of high overpressure, the integral in Eq.~\ref{eq_coltotal} becomes a constant and the column density solely a function of the central pressure $p_c$. The asymptotic behavior is given by
\begin{equation}
	N_{\rm H}(0) \sim 1.5\times 10^{21}\, \xi\,\sqrt{\frac{p_c/k}{2\times 10^4~{\rm K/cm^3}}}~{\rm cm^{-2}},
\end{equation}
where
\begin{equation}
	\xi = \int_0^{\infty}{\rm d}z\,e^{-w(z)} = \left\{
		\begin{array}{ll}
			3.0279 &\quad \mathrm{(spheres)},\\
			 2.2214& \quad\mathrm{(cylinders)}.
		\end{array}
		\right.
\end{equation}
Because of this characteristic, in the case of high overpressure, the radiative transfer problem and as a consequence the dust temperature of the cores become largely independent of the external pressure.

\subsection{\label{sect_dust}Dust model}

The dust is assumed to be a specific composition of graphites, silicates, iron grains, and PAH molecules where each dust component has a given size distribution.  I use the same dust model parameters as \citet{Fischera2008}, which
were chosen to reproduce the observed dust properties of the diffuse phase of the ISM including the mean extinction curve \citep{Fitzpatrick1999}, the diffuse dust emission derived with the DIRBE experiment \citep{Arendt1998}, and the 
depletion of key elements from the gas phase. 

The dust re-emission spectrum and the PAH emission spectrum are derived by taking into account the stochastic nature of the heating process for individual grains. This heating process is of particular importance for small dust particles where the thermal energy is typically lower than the absorbed photon 
energy, leading to large temperature variations.


There are some differences between my model and that of 
\citet{DraineLi2007} as mentioned in \citet{Fischera2008}. First, the dust albedo in the optical is lower; secondly, the emission
at shorter wavelengths where the dust emission spectrum is supposedly dominated by stochastic emission is enhanced by additional iron grains; and thirdly, the graphite grains are restricted to relatively large sizes with radii larger than $0.01~\mu{\rm m}$ to avoid any contribution to the extinction bump at 2200~\AA{}, which is entirely explained by 
PAH molecules. The dust-to-gas ratio in mass is $\eta=0.59\%$. As the grains possibly accrete gas atoms or molecules, the dust-to-gas ratio might indeed be slightly higher.

\subsubsection{\label{sect_dustapprox}Mean extinction coefficient at long wavelengths}


At far infrared wavelengths, where most of the grain cooling occurs, the extinction is close to the absorption probability.
The mean cross-section for extinction per hydrogen atom in the model is given by the sum over the 
different dust components and the integrations over the individual size distributions
\begin{equation}
	\label{eq_ext}
	\left<C^{\rm ext}_\lambda\right>/H=\sum_i\zeta_i\int{\rm d}a\,f_i(a)\pi a^2\,Q^{\rm ext}_{\lambda}(i,a),
\end{equation}
where $Q^{\rm ext}_{\lambda}(i,a)$ is the extinction coefficient at wavelength $\lambda$ of a dust grain of radius $a$ and composition $i$. The size distributions $f_i(a)$ are normalized such that $\zeta_i=n_i/n_{\rm H}$ is the relative amount of a key element that is part of the dust composition $i$ (see \citet{Fischera2008}). The same expression as Eq.~\ref{eq_ext}
is used to determine the scattering cross-section and the absorption cross-section. The dust albedo is given by the ratio
$\omega_\lambda = \left<C_\lambda^{\rm sca}\right>/\left<C_{\lambda}^{\rm ext}\right>$.
\begin{figure}[htbp]
	\includegraphics[width=\hsize]{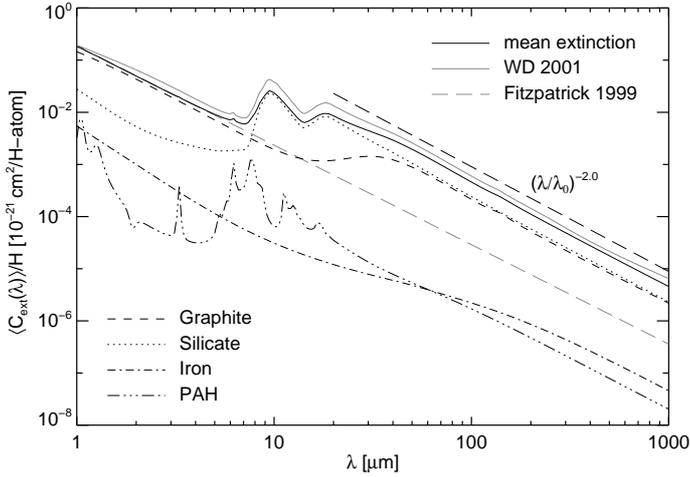}
	\caption{
	\label{fig_ext}
	Mean cross-section of the dust extinction per hydrogen atom as assumed in the radiative transfer model. 
	Its functional dependence on wavelength is compared with the observed 
	mean extinction for the diffuse ISM \citep{Fitzpatrick1999} (although extended to the FIR) and the mean extinction
	curve provided by \citet{Weingartner2001} (corrected by a factor 0.93). 
	Also shown are the contributions of the several dust
	components in the model.
		}
\end{figure}

The mean cross-section for the dust extinction assumed in the model calculations
is shown in Fig.~\ref{fig_ext}. The values in the FIR/submm regime are close, albeit slightly lower, than the ones derived
by \citet{Weingartner2001}. The behavior is predominantly determined by graphite, PAH, and silicate grains. The relative
importance of the components changes as a function of wavelength. While the extinction in the optical is caused mainly
by carbonaceous grains, silicate is the dominant absorber from $10~\mu{\rm m}$ up to $60~\mu{\rm m}$. In the FIR/submm regime, the extinction is caused almost in equal parts by silicate and graphite grains. The extinction curve at
wavelengths above $40~\mu{\rm m}$, in the spectral region where the bulk of the cold dust emission occurs, is approximated well by a power law.
The extinction coefficient per unit mass is given by
\begin{equation}
	\left< \kappa^{\rm ext}_\lambda\right> = \frac{\left<C^{\rm ext}_\lambda\right>/H}{\left<\rho_{\rm d}\right>/H},
\end{equation}
where 
\begin{equation}
	\left<\rho_{\rm d}\right>/H \approx1.384\times10^{-26}{\rm \frac{g}{H-atom}}
\end{equation}
is the mean dust density per hydrogen atom (Fischera, in prep.).  The long-wavelength behavior above $60~\mu{\rm m}$ to $1000~\mu{\rm m}$ is (within $4\%$ accuracy
to the dust model) described by
\begin{equation}
	\label{eq_irabsorption}
	\left<\kappa^{\rm ext}_\lambda\right>\approx 5.232 \left(\frac{\lambda}{250\,\mu{\rm m}}\right)^{-2.024}
		~{\rm \frac{cm^2}{g}}.
\end{equation}
The power-law behavior of the dust model also extends to longer wavelengths. The approximation up to 
$10^4~\mu{\rm m}$ is accurate to within 10\%.

The extinction coefficient per unit length is $k^{\rm ext}_\lambda = (\left<C^{\rm ext}_\lambda\right>/H)\, n_{\rm H}$.

\subsection{Radiative transfer}

The filaments are assumed to be heated by an isotropic interstellar radiation field (ISRF) adopted from \citet{Mathis1982}. The spectrum is characterized by a sharp cutoff at 13.6 eV and contains, apart from the stellar component, no re-emission from dust grains. 

The radiative transfer is solved by assuming that the mass and size of the core is negligible compared to the mass and size of the filaments, the cores are embedded in. 
Under these circumstances the radiative transfer problem can be solved in two separate steps: The radiative transfer through the filament and the radiative transfer through the core. 
The centrally located cores are heated by the mean intensity $J_{\lambda}(0)$ at the center of the filaments derived for the filaments without cores. For simplicity, the radiation field caused by 
the attenuated ISRF, the scattered emission, and the thermal dust emission from the filaments is 
assumed to be isotropic.

For fixed dust properties inside the filaments and cores, the radiative transfer problem (and therefore the SEDs of the cores) is determined by the pressure of the gas surrounding the filaments and the overpressure $p_{\rm c}/p_{\rm ext}$ in the filaments. 
The shape of the filaments has an additional effect on the radiative transfer because it determines the radial density profile,  the extinction- central pressure relation (Sect.~\ref{sect_pressextrel}) and the attenuation of the radiation inside the filaments. The influence of the shape of the filament on the SEDs of the cores is explored by considering both self-gravitating spheres and cylinders.

\subsubsection{Accurate solution of the SEDs of the cores}

\label{sect_raytrac}

For the given simplifications, the radiative transfer problem as a function of external pressure and overpressure is accurately solved using a ray-tracing technique (see \citet{Fischera2008}). This technique provides 
the angle-dependent radiation field inside the cloud caused by the attenuated radiation, the scattered radiation, and 
the thermal emission from dust grains. In the procedure,
the cloud is divided into a number of optically thin shells. The impact parameters of the individual rays, which is their nearest distance to the cloud center, is given by the mean distance to the shells. For cylindrical geometric calculations, rays at  30 additional angles around the main rotation axis are considered (Fischera, in prep.).
Effects caused by non-isotropic scattering and multiple scattering events are accurately included in the calculation as well as the effects of the re-heating of the dust  grains and subsequent emission. The polarization of the scattered emission is assumed to have a negligible effect on determining the SED and is therefore ignored.

\subsubsection{\label{sect_1zone}Approximate solution for the SEDs of embedded cores}

When observations are analyzed, the dust re-emission spectrum of condensed cores is typically modeled by assuming a single dust temperature
where the dust re-emission spectrum is approximated by a modified black-body spectrum.
The SED (in units $\rm W/m^2/\AA{}$) from the core of dust mass $M_{\rm dust}$ at distance $D_{\rm core}$ is given by
\begin{equation}
	\label{eq_modbb}
	S^{\rm dust}_{\lambda} = \frac{M_{\rm dust} \kappa^{\rm em}_{\lambda}}{D_{\rm core}^2}B_{\lambda}(T_{\rm dust}),
\end{equation}
where $T_{\rm dust}$ is the dust temperature, $B_{\lambda}(T_{\rm dust})$ the Planck-function, and $\kappa^{\rm em}_\lambda$
the emission coefficient per unit dust mass. The emissivity is commonly approximated by a power law $\kappa_\lambda^{\rm em}=\kappa_0(\lambda/\lambda_0)^{-\beta}$ by making certain assumptions about the dust emissivity $\kappa_0$ at wavelength $\lambda_0$ and the power $\beta$  \citep{Netterfield2009,Olmi2009}. 
The model is equivalent theoretically to a single-zone model for the dust temperature in the core. However, as shown in this paper, to obtain the correct shape of the dust re-emission spectrum and the correct dust mass, the intrinsic values of the dust emission coefficient need to be replaced by the appropriate effective values to take into account not only the variation in the emission spectrum with both size and composition but also the radial variation in the grain temperatures. I use
this approximation to obtain for the given dust mass and theoretical SED the appropriate effective parameters.

The single-zone model allows us not only to interpret the observations but also to test efficiently a wider range of model parameters. It is therefore a more practical model to use than time-consuming ray-tracing calculations. The model is described in detail in 
Sect.~\ref{sect_modelapprox}. In the approximate
solution of the radiative transfer problem based on energy conservation the effects of both
scattering and re-heating by dust grains inside the filaments and the cores are taken into account. 
However, in the considered cases
the contribution of self-absorption to the total heating rate inside the clouds is negligible. Of greater importance is the contribution of dust emission from the filaments to the total heating of highly embedded cores. In the model approximation this additional heating is provided  as a lower limit.

\section{Results}

The dependence of the effective parameters is analyzed for three assumptions about the external pressure. Apart
from the mean pressure in the ISM ($2\times 10^{4}~{\rm K/cm^3}$), situations where the filaments are embedded in a higher pressure gas ($10^5$ and $10^6~{\rm K/cm^3}$) are also considered. For each considered value of external pressure, the SEDs of the cores are derived for a range of extinction values $A_V$ at the center of the filaments. For both spherical and cylindrical geometries for the filaments, I considered the extinction values 0, 1, 2, 4, 8, and 16. For cores in spherical filaments, I also 
assumed higher extinction values (32 and 64~mag). In these calculations, the energy is conserved in all cases to better than 0.2\%.

In the following section \ref{sect_sed}, I describe the individual steps of the radiative transfer calculations. 
The results obtained using a modified Planck-function are described in Sect.~\ref{sect_mbb}.

%

\subsection{\label{sect_sed}The theoretical dust emission spectrum}

\subsubsection{\label{sect_sed_jmean}The mean intensity at the filaments' center}

\begin{figure*}[htbp]
%
	\includegraphics[width=0.49\hsize]{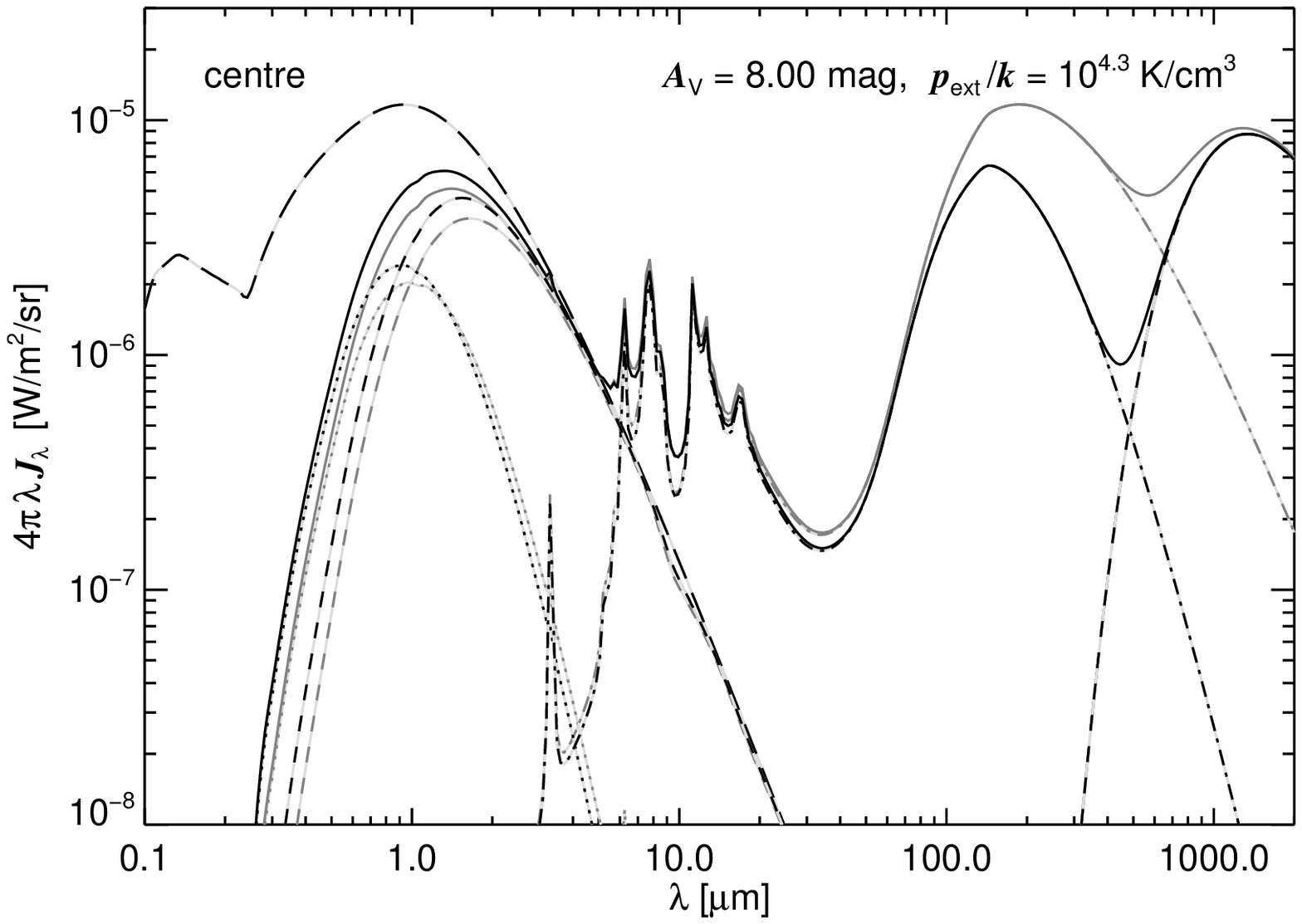}
	\hfill
	\includegraphics[width=0.49\hsize]{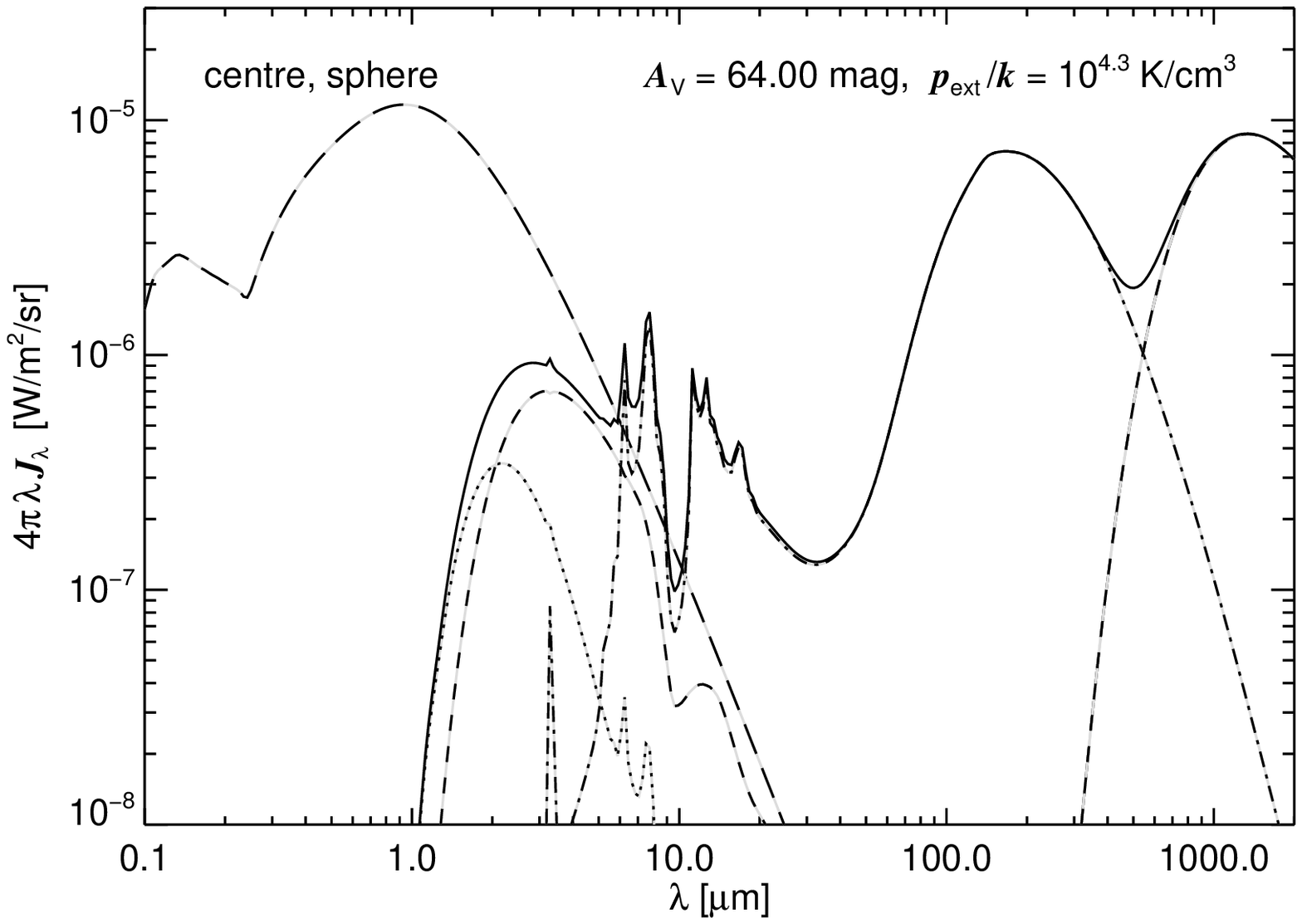}

	\includegraphics[width=0.49\hsize]{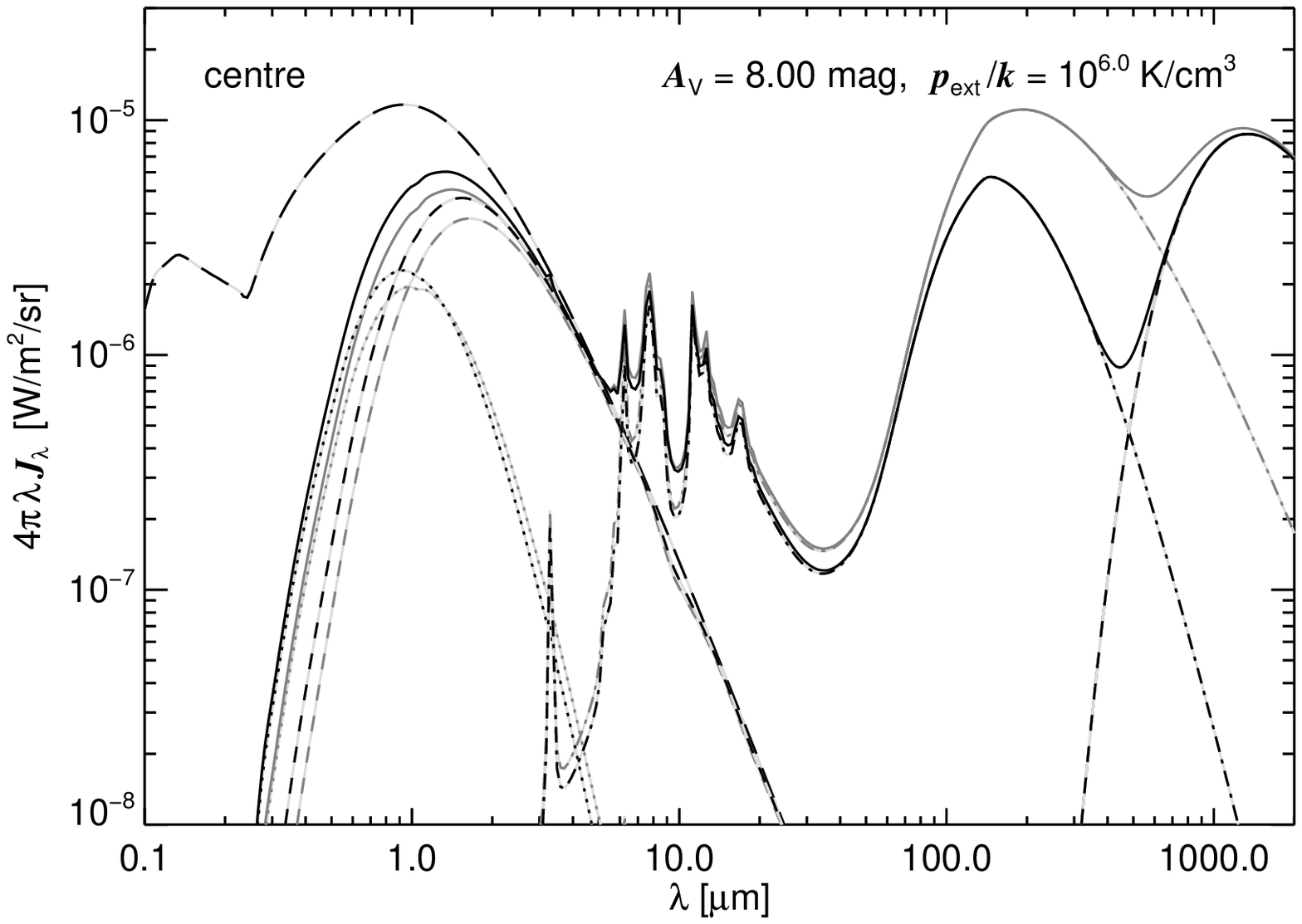}
	\hfill
	\includegraphics[width=0.49\hsize]{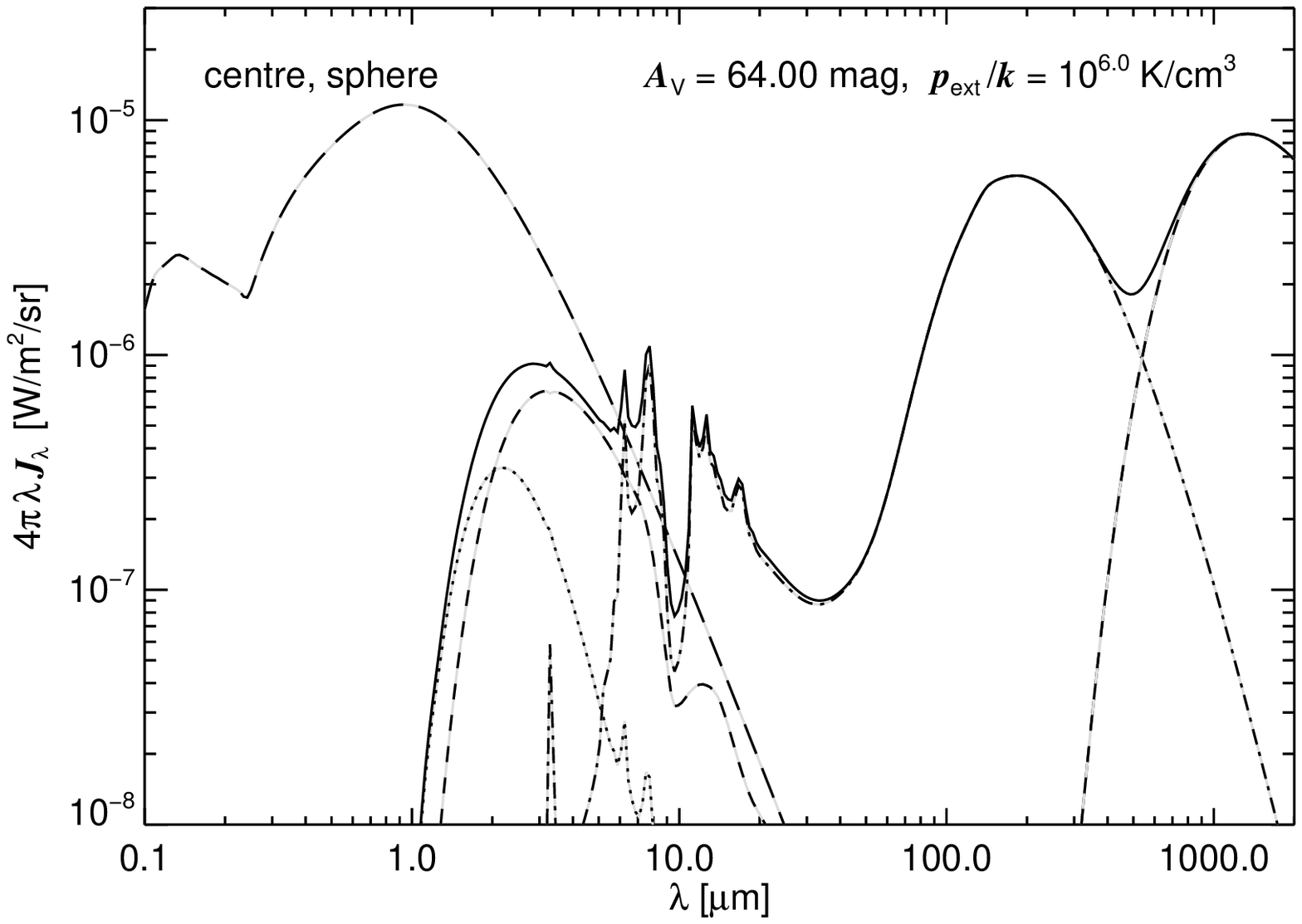}
	\caption{\label{fig_jmean_ctr}
	Mean intensity at the center of spherical (black solid lines) and cylindrical (grey solid lines) filaments
	heated by the ISRF (long dashed lines). The filaments are embedded in a medium with 
	$p_{\rm ext}/k=2\times 10^4~{\rm K/cm^3}$ (top) and $10^6~{\rm K/cm^3}$ (bottom). The extinction through the filaments' center 
	is either $A_V=8~{\rm mag}$ (left-hand figure)
	or $A_V=64~{\rm mag}$ (right-hand figure). Shown are the different components of the mean intensity:
	scattered emission (dotted lines), attenuated light (short dashed lines), and dust re-emission (dashed-dotted lines).
	{The emission peak in the millimeter regime is the CMB radiation.}
	}
\end{figure*}

The first result obtained using the ray-tracing technique is the mean intensity at the center of the filaments. The 
relative contribution of the attenuated external radiation field, the scattered emission, and the re-emission from
dust grains located in the filaments to the SED is shown in Fig.~\ref{fig_jmean_ctr} for two extinction and pressure values. 
For the ISM pressure ($p_{\rm ext}/k=2\times 10^4~{\rm K/cm^3}$), the extinction value $A_V=8~{\rm mag}$ in the spherical filament is close
to the value of a critical stable sphere with overpressure 13.2, while the higher extinction value, $A_V=64~{\rm mag}$, corresponds to a supercritical spherical cloud with high overpressure ($68.25$). In both cases, the clouds are characterized by a steep density profile in their outskirts. 
In the high pressure region ($p_{\rm ext}/k=10^6~{\rm K/cm^3}$), the high extinction value is close to the corresponding critical value  ($A_V\sim 59~{\rm mag}$), while the low extinction value corresponds to a sub-critical spherical cloud.
The clouds in the high pressure region are characterized by flat or steep density profiles with low and high overpressures (1.30 and 16.38, respectively) at the cloud center (see table~\ref{table_fitparameters}).
   
The SED of the center of the filaments is primarily determined by the radial extinction and in parts by the geometry. The effects caused by the radial density profile play only a minor role in particular for the dust re-emission spectrum. The properties
are of importance for interpreting the dust re-emission from cores and are described in the following.

Because of simple geometric effects, the radiation is, for identical values of $A_V$, more strongly attenuated by cylindrical than spherical geometry. The strength of the scattered light that dominates the SED below its maximum depends to a certain degree on the shape of the filaments. For the considered high extinction values, most of the scattered emission in the UV and optical is generated in a skin layer of the filament, so that the geometric effects
produce a lower amount of scattered emission at the center of the cylinders. The situation is different at longer wavelengths where the clouds become optically thin and the probability of scattering events is higher in cylinders than spheres. Despite its low intensity for the cases shown in Fig.~\ref{fig_jmean_ctr}, the scattered light is responsible for approximately half of the dust heating in the filaments' center \citep{Fischera2008}.

At the high extinction values considered, the amount of PAH emission produced in the filaments is rather insensitive
to the extinction value since most of the flux responsible for heating the molecules is absorbed. The emission originates mainly
in a layer close to the surface where the UV and optical light most responsible for heating the molecules is strong.
The radiation in this layer is more intense in low pressure regions causing for a given extinction a slightly stronger PAH emission 
at the filaments' center. In the case of $A_V=64~{\rm mag}$, the PAH emission itself experiences a certain extinction that is most evident
in the larger minimum at $10~{\mu\rm m}$, which is caused by silicate absorption.
It can be seen overall, that the ratio of the PAH emission relative to the dust emission peak is considerably lower than in the diffuse ISM (Fig.~\ref{fig_jmean_isrf}). 

The shape of the dust re-emission in the FIR/submm-regime depends for a given extinction and external pressure on the 
shape of the filaments. Radiation from cold dust grains located in the interior of the filament is higher in the case of a cylinder because of the infinite column density along the major axis. At relatively low extinction values ($A_V\sim 1~{\rm mag}$) where the 
UV and optical light can penetrate deeper into the cloud, this also affects the intensity of the PAH emission and the warm emission
of stochastically heated small grains. The dust emission peak at higher extinction values also appears to be broader 
because of the larger temperature variations and colder dust temperatures at the filaments' center.

\subsubsection{\label{sect_sed_heating}Grain heating}

\begin{figure*}[htbp]
	\includegraphics[width=0.49\hsize]{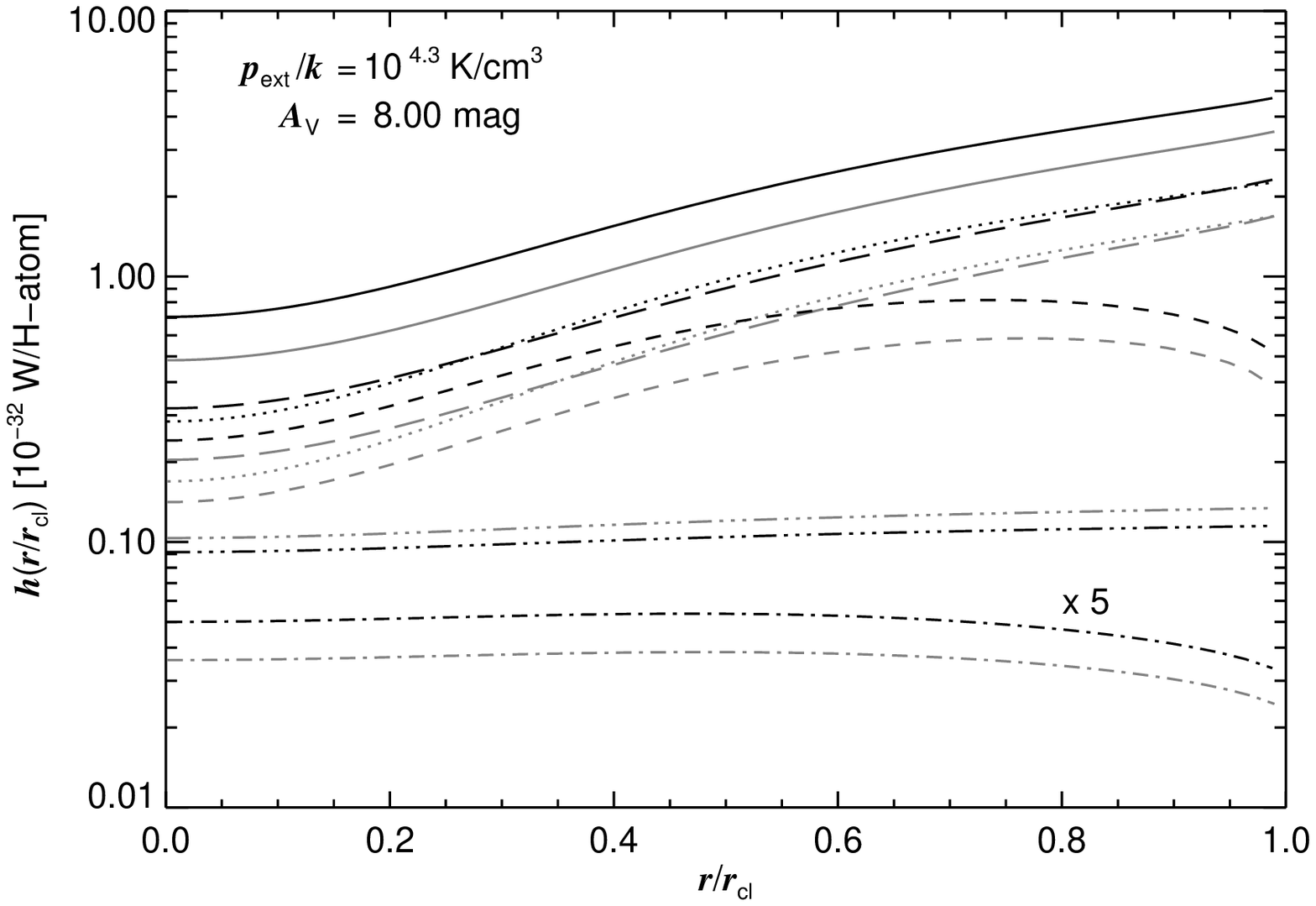}
	\hfill
	\includegraphics[width=0.49\hsize]{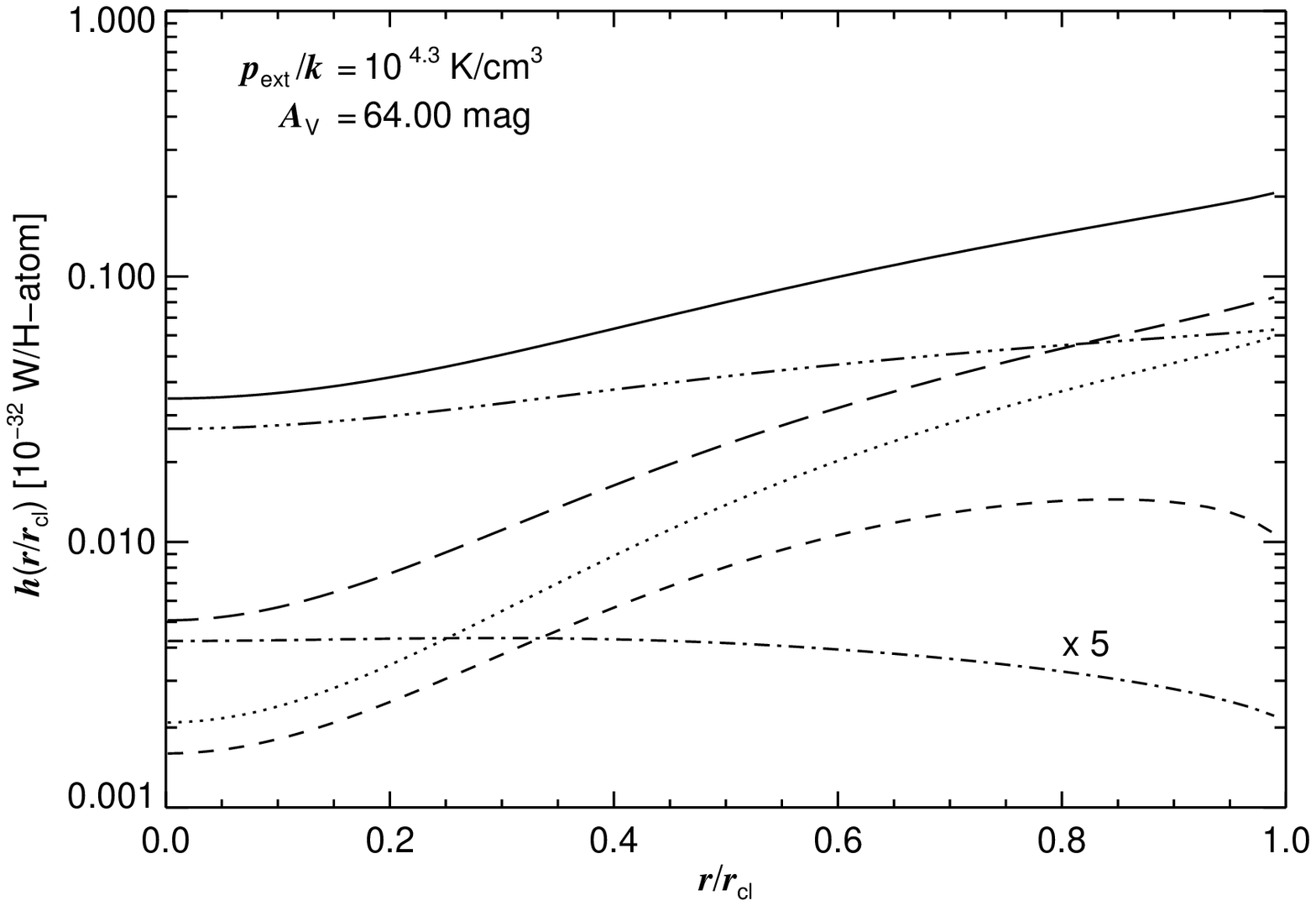}

	\includegraphics[width=0.49\hsize]{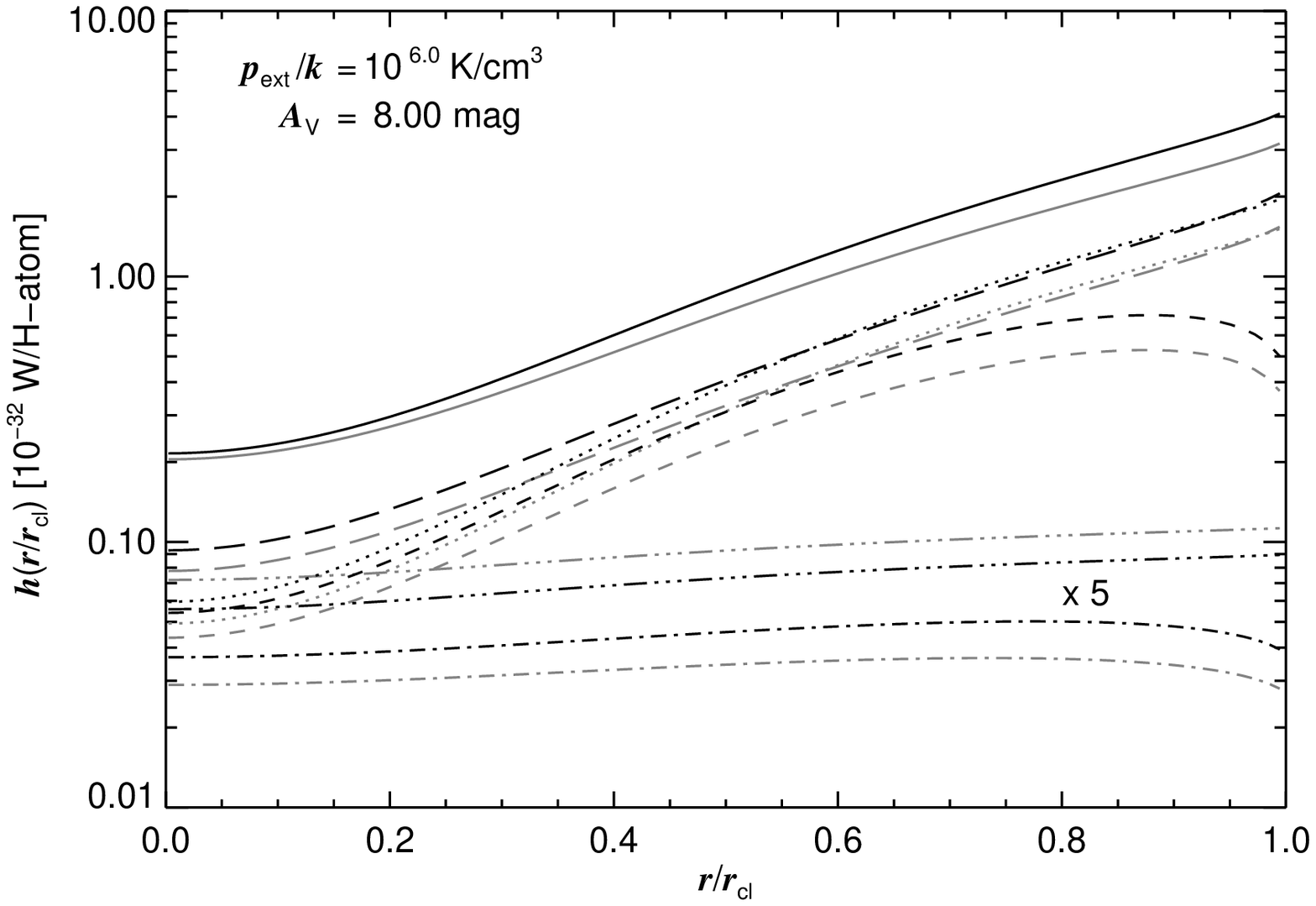}
	\hfill
	\includegraphics[width=0.49\hsize]{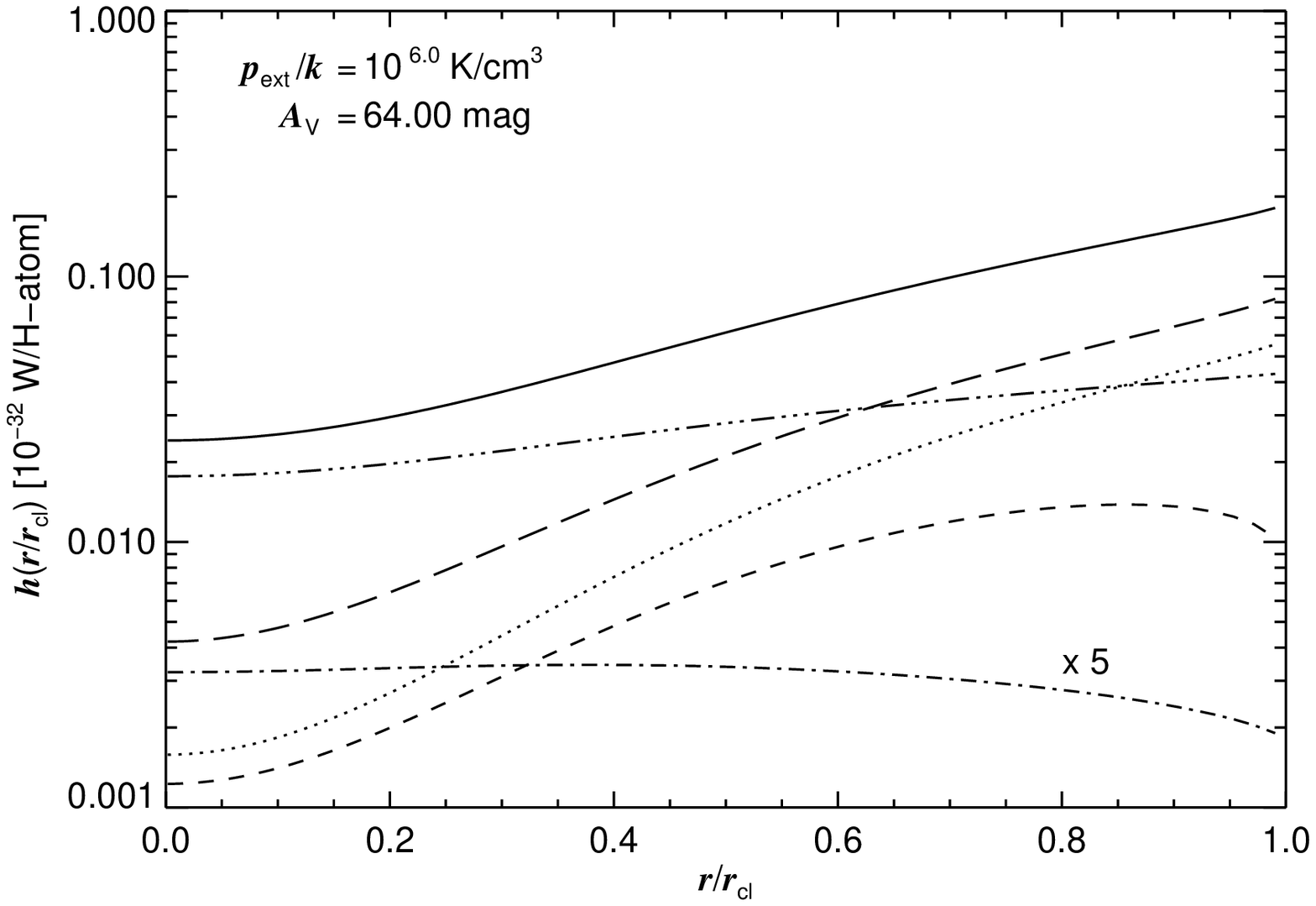}
	\caption{\label{fig_heating}
	Dust heating inside cores embedded in spherical (black curves) or cylindrical (grey curves) 
	filaments. The visible extinction $A_V$ at the center of the filaments is either $8~{\rm mag}$ (left-hand figure) 
	or $64~{\rm mag}$ (right-hand figure).  The filaments are pressurized either by the mean pressure in the ISM (top)
	or by a medium with $p_{\rm ext}/k=10^6~{\rm K/cm^3}$ (bottom). The total heating rates (solid curves) are separated
	into different components corresponding to the different radiation fields in the core: Attenuated $ISRF$ (long
	dashed curves), total scattered light (dotted curves), light scattered in the core (short dashed curves), dust emission from the
	filament (dashed-three dotted curves), and dust emission from the core (dashed-single dotted curves). The heating by 
	dust emission produced inside the core is multiplied by a factor~5.	
	}
\end{figure*}

The dust heating 
\begin{equation}
	h(r) = \int{\rm d}\lambda\,(k_\lambda^{\rm abs}/n_{\rm H})\,4\pi\,J_{\lambda}(r)
\end{equation}
at the core caused by the mean intensity in the filament varies strongly inside the core (as seen in Fig.~\ref{fig_heating}) because the light, especially in the UV and optical, is strongly attenuated by dust scattering and dust absorption towards the core center. This radial variation is even stronger in the high pressure region ($p_{\rm ext}/k=10^6~{\rm K/cm^3}$), where cores experience higher pressure and are more opaque. 

\begin{figure}[htbp]
	\includegraphics[width=0.92\hsize]{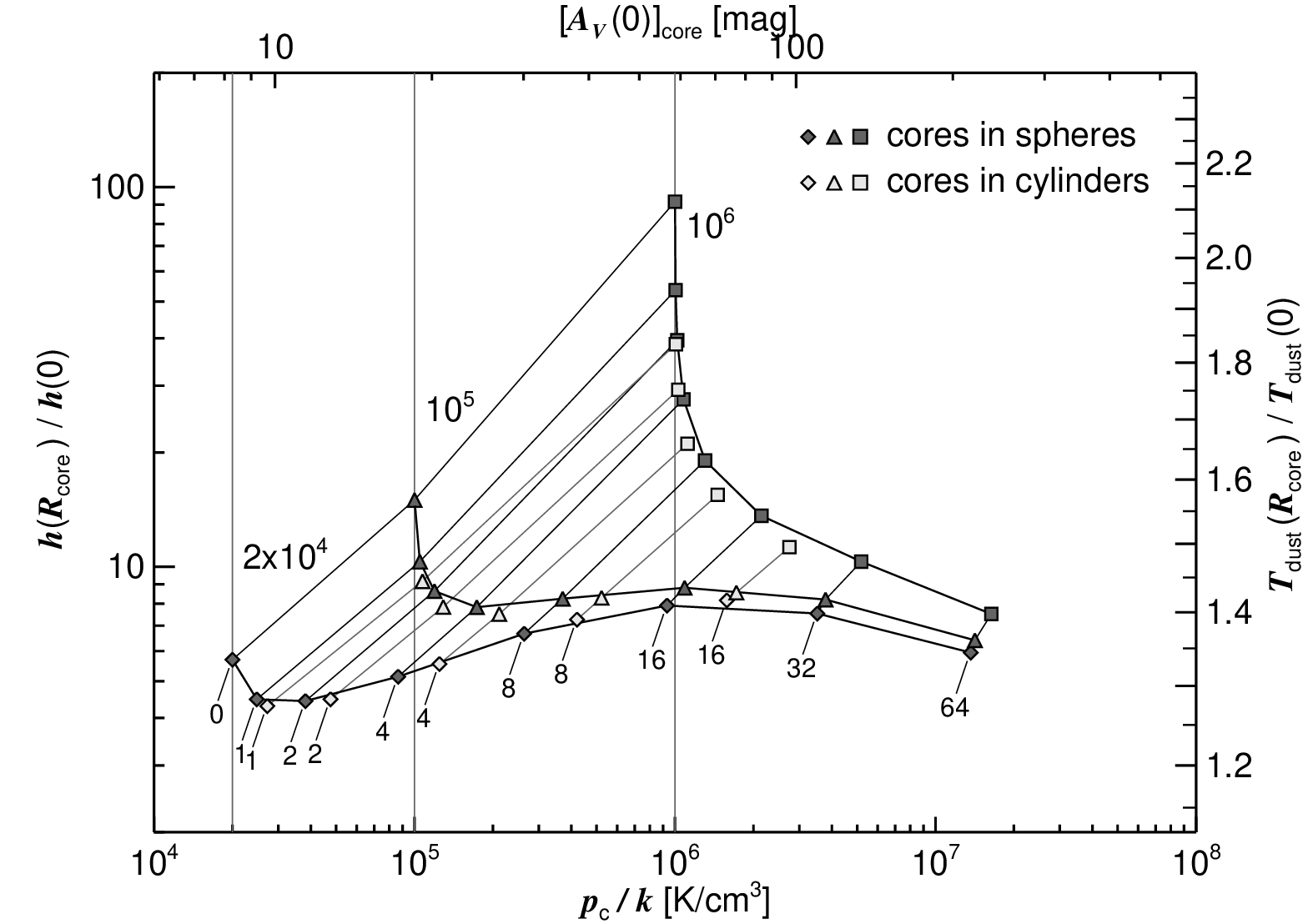}	
	\includegraphics[width=0.92\hsize]{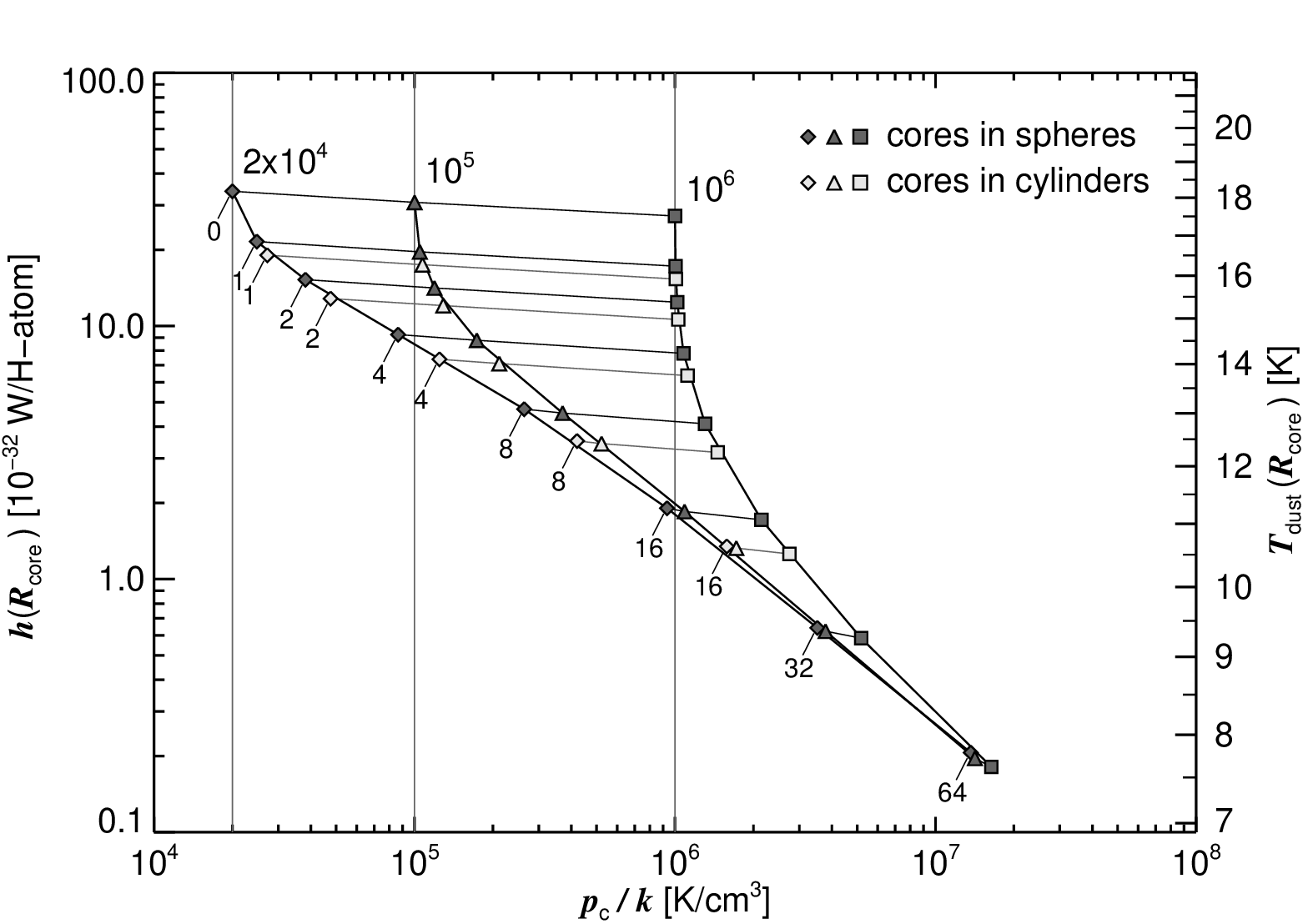}
	\includegraphics[width=0.92\hsize]{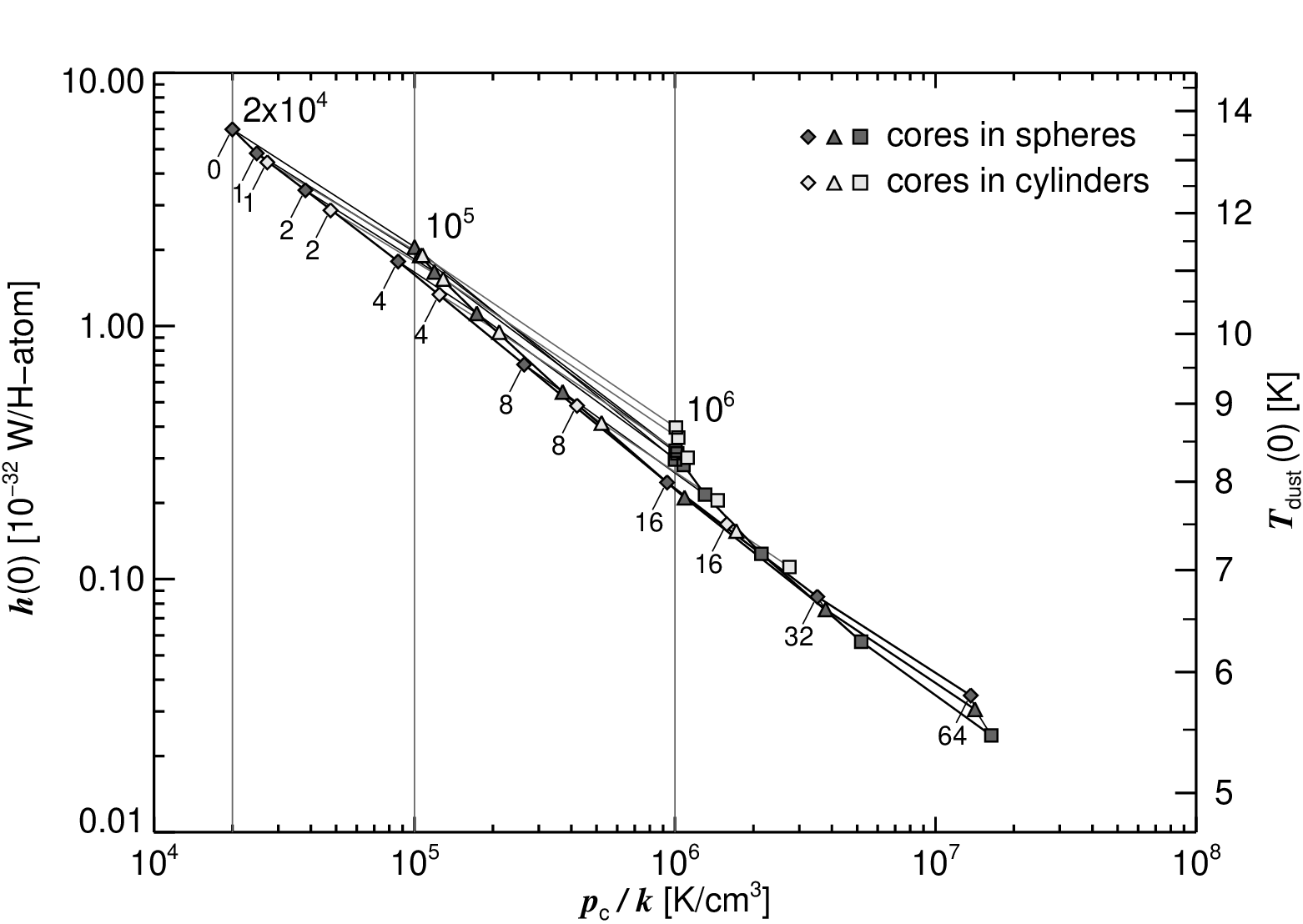}
	\caption{\label{fig_heatingratio}
	Ratio $h(R_{\rm core})/h(0)$ 
	(top figure), the heating rate $h(R_{\rm core})$ at the cores' edge (central figure), and the heating rate in the core center
	$h(0)$ (bottom figure) as a function of the central pressure in the filaments for the three different assumption of the external 
	pressure $p_{\rm ext}$. For comparison also the corresponding dust temperatures of a simplified dust model are shown.
	For sake of clarity the values obtained for cores in spherical filaments which are pressurized by the same external pressure
	$p_{\rm ext}$ are given as solid black curves. The values are labelled with the considered extinction values $A_V$ through 
	the filaments' center.}
\end{figure}

In Fig.~\ref{fig_heating}, the heating is divided into several components corresponding to the various radiation fields inside the core. For cores in filaments with $A_V=8~{\rm mag}$, the heating is 
dominated almost in equal parts by the attenuated $ISRF$ and the scattered light. As visible in the figure, inside the core the scattered radiation field
is mostly caused by photons scattered inside the core. Deep inside highly opaque cores and in the case of highly embedded
cores the scattered contribution decreases relative to the attenuated $ISRF$ as the heating shifts to longer wavelengths where the dust albedo is lower.

In case of a strongly attenuated external radiation field, the heating by dust emission from the filament and the core becomes an important component. This applies either for the central region of highly opaque cores and for highly embedded
cores in general. The size of the contribution in reality depends certainly on the dust properties in the dense phase of the ISM. The heating derived here may be an upper limit since most of the heating is caused by the PAH emission which is possibly absent in realistic clouds.
The heating by self-absorption plays a larger role in the case of non-embedded cores in high pressure
regions (see \citet{Fischera2008}). For the considered cases in Fig.~\ref{fig_heating} of embedded cores, the contribution is smaller than $\sim 10\%$. Far more important is the heating caused by the dust emission from the filaments. At $A_V=8~{\rm mag}$,
this component produces at the center of the cores embedded in spherical filaments between $13\%$ and $25\%$ of the total heating rate. The contribution is even higher if the cores are instead embedded in cylinders ($20 - 35\%$). For highly embedded cores,
the dust in the whole interior of the cores is predominantly heated by the dust emission from the filaments. This additional heating limits the variation in the heating rate inside the core. I note that the heating component is higher in the low pressure region.


The effect of the external pressure and the extinction of the filament  on the temperature variation in the heating rate 
inside the core is further visualized in Fig.~\ref{fig_heatingratio} where the heating rate in the center is compared with the heating rate at the cores' edge. For comparison with simplified radiative transfer models, the heating rate is also related to a dust temperature. The heating rates at the cores' edge
are determined in a preliminary way for a given filament shape by the extinction $A_V$ through the filaments' center. Since the
cores are more compact for a given extinction in higher pressure regions that produce a lower intensity at the cores' edge, the heating rate decreases slightly to higher pressures.  The effect reaches a maximum of only a factor of two. Clearly visible in the figure is the dependence of the overpressure on the extinction of the filaments for a given external pressure. While the overpressure changes drastically in the low pressure regions also in case of low extinction, the effect is small for filaments in high pressure regions.

The lowest dust temperature in the core is strongly dependent on the pressure surrounding the core, which determines its
extinction. However, there are some noticeable systematic effects caused by the shape of the filament and the external pressure. At a certain pressure $p_c$, the minimum temperatures are naturally lower for embedded cores in low pressure regions $p_{\rm ext}$ than for non-embedded cores in high pressure regions. 
The behavior at high pressure values $p_c$ for a given external pressure reflects the differences in the dust emission inside the filament noticeable in Fig.~\ref{fig_heating}. For given external and internal pressures, the dust at the center of cores is for example slightly warmer, if they are embedded in cylindrical filaments as the intensity of the dust emission is higher in cylinders than in spheres.

As seen in Fig.~\ref{fig_heating}, the qualitative behavior of the relative variation of the heating rate or the temperature inside the core depends on the pressure surrounding the filaments. In high pressure regions, the variation in the dust temperature inside
the cores decreases if the core becomes more embedded and shielded from the ISRF. As the cores become more optically thin to the reddened radiation, the radial variation in the heating rate 
and consequently that of the dust temperature become smaller.

The situation in the low pressure region is more complex as the increase in the extinction of the filament is
accompanied by a higher pressure at its center. The embedded cores are not only heated by a reddened radiation field but
are also more opaque than non-embedded cores. Above $A_V=1~{\rm mag}$,  this leads, assuming a mean ISM pressure,
to an increase in the dust temperature variation with central pressure. As will be shown this causes a flattening of the 
overall SED with extinction (Sect.~\ref{sect_mbb_effem}). The variation in the dust temperature inside embedded
cores is not necessarily lower than for non-embedded cores. Based on my calculation, 
the cores in molecular clouds of typical extinction $A_V=8$ display even slightly larger temperature variations than 
non-embedded cores. 

Independent of the assumed external pressure, the variation in the heating rate decreases for highly embedded cores within filaments with $A_V\gg 16~{\rm mag}$, where the interior of the cores becomes predominantly heated by the dust emission
from the filaments.


\subsubsection{\label{sect_sed_grtemp}Grain temperatures}

In the model, the SEDs of the cores are produced by the re-emission of individual grains of different sizes and compositions that have unique temperature and emission behaviors. 
The temperature of the individual grains as a function of radius
is shown in Fig.~\ref{fig_grtemp}. The grain sizes are large enough for their temperature distribution to be
close to a narrow distribution around the equilibrium temperature. The radial variation in the temperature is most visible
for the smallest grains, which are heated predominantly by the strongly diminished UV and optical light. 
The figure also shows the dependence of the dust temperature on 
composition. Small iron grains are considerably warmer than silicate or graphite grains
because of the high absorption in the UV and optical and the low emission probability in the infrared \citep{Fischera2004b}.
In cores  embedded in filaments with $A_V=8\,{\rm mag}$ that are pressurized by the mean ISM pressure, the temperatures of large grains lie between 8~K and 15~K. I note that the temperature variation for different 
grain types is smaller in both the core interior and more opaque filaments where the grains are heated by a more
reddened radiation field.
The physical explanation is the similar absorption and emission behavior of the grains at long wavelengths, which for most grains is closely described by  a power law with $Q^{\rm abs/em}_{\lambda}\propto a/\lambda^2$ where $a$ is the grain size. In this limit, the ratio of the cooling to the heating rates and therefore the grain temperatures become the same.

\begin{figure*}[htbp]
%
	\includegraphics[width=0.49\hsize]{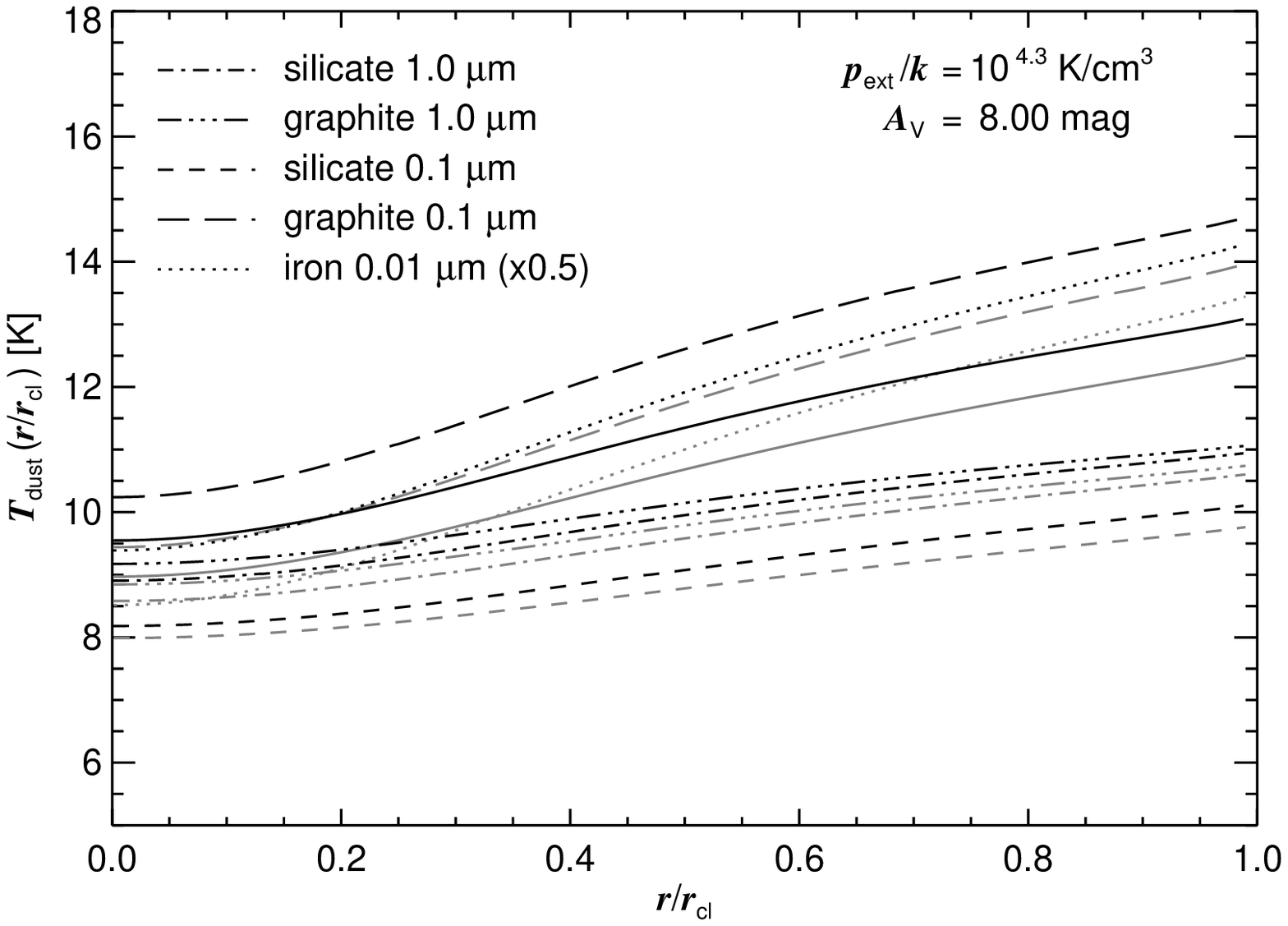}
	\hfill
	\includegraphics[width=0.49\hsize]{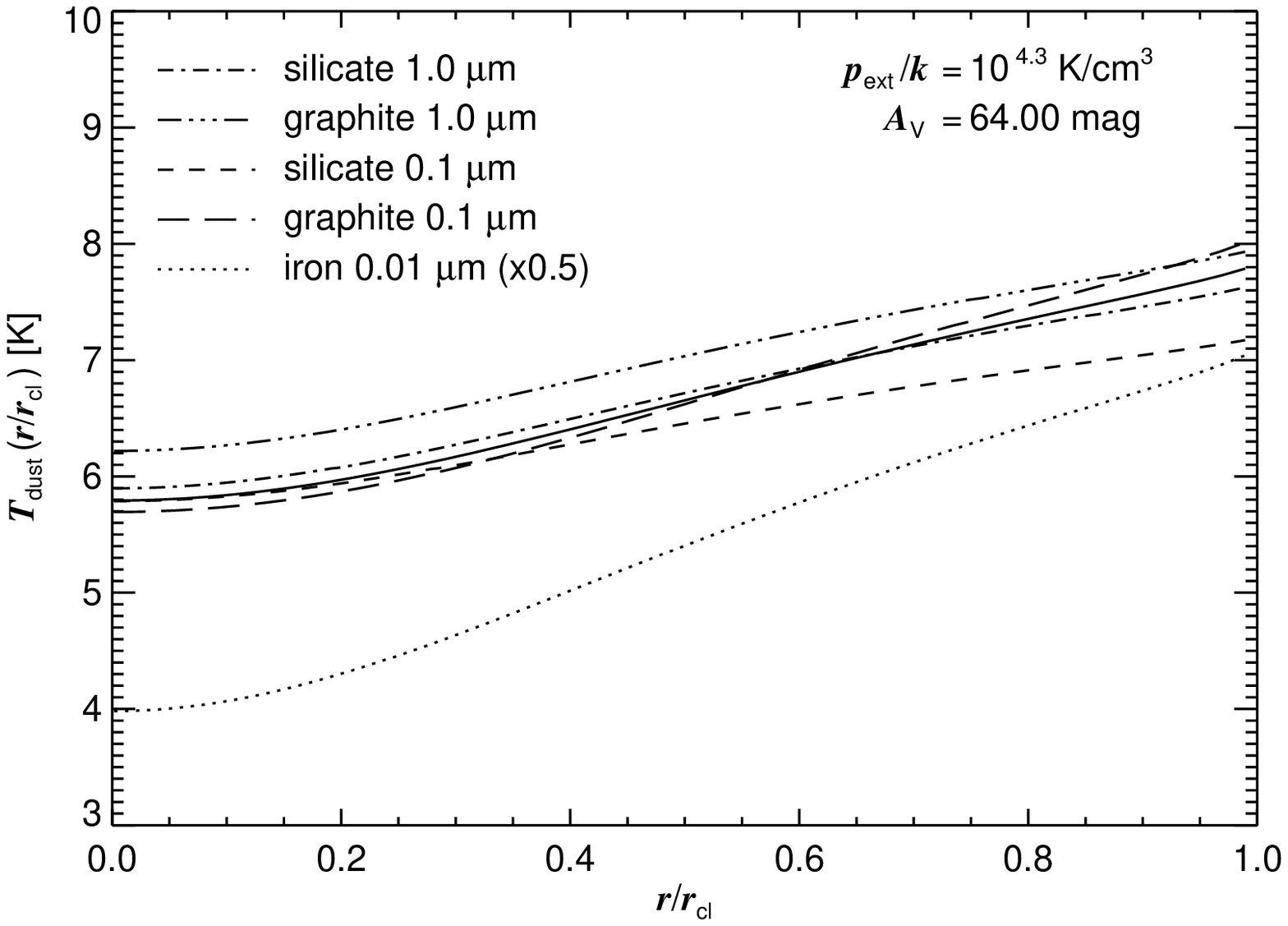}

	\includegraphics[width=0.49\hsize]{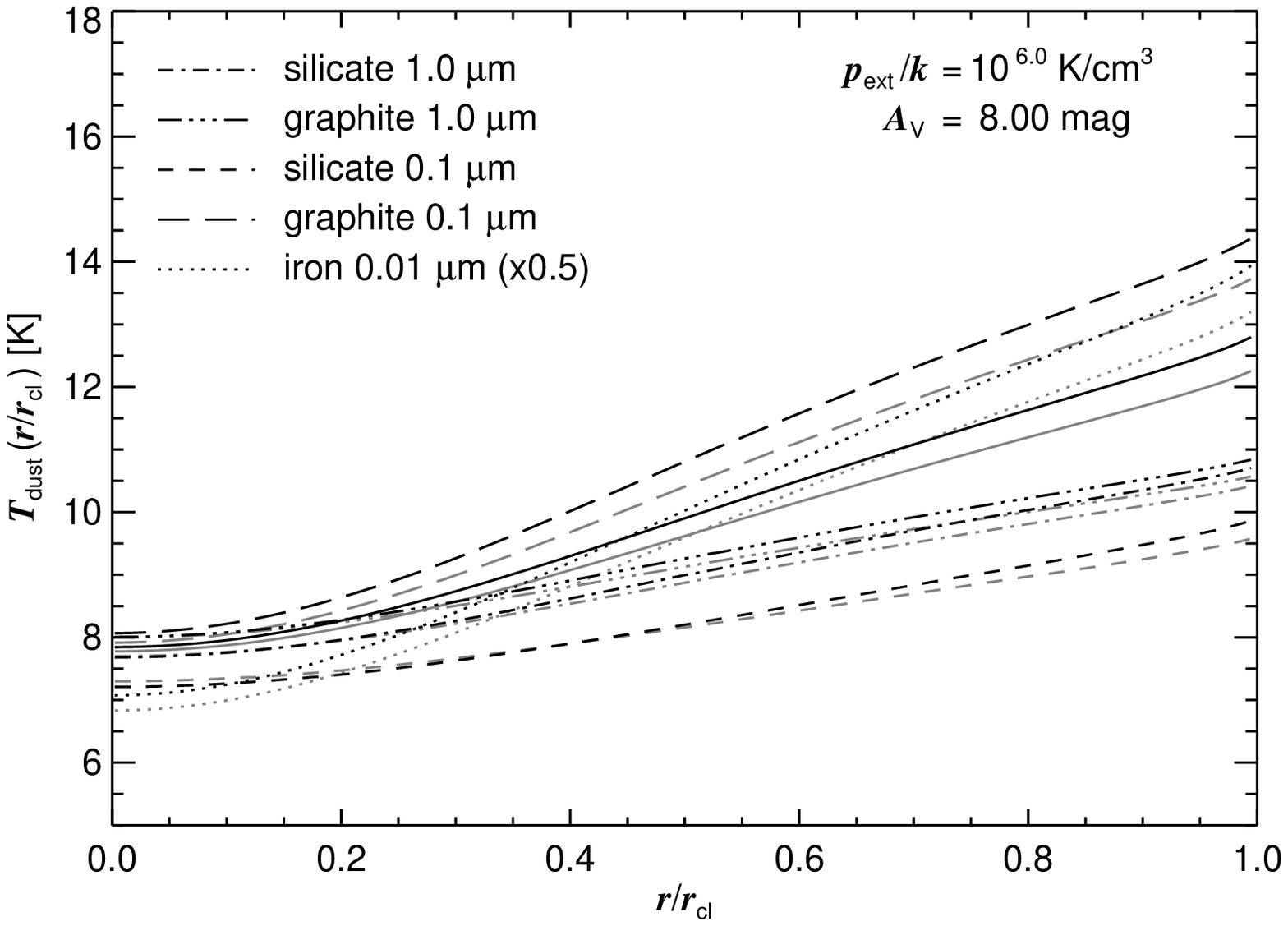}
	\hfill
	\includegraphics[width=0.49\hsize]{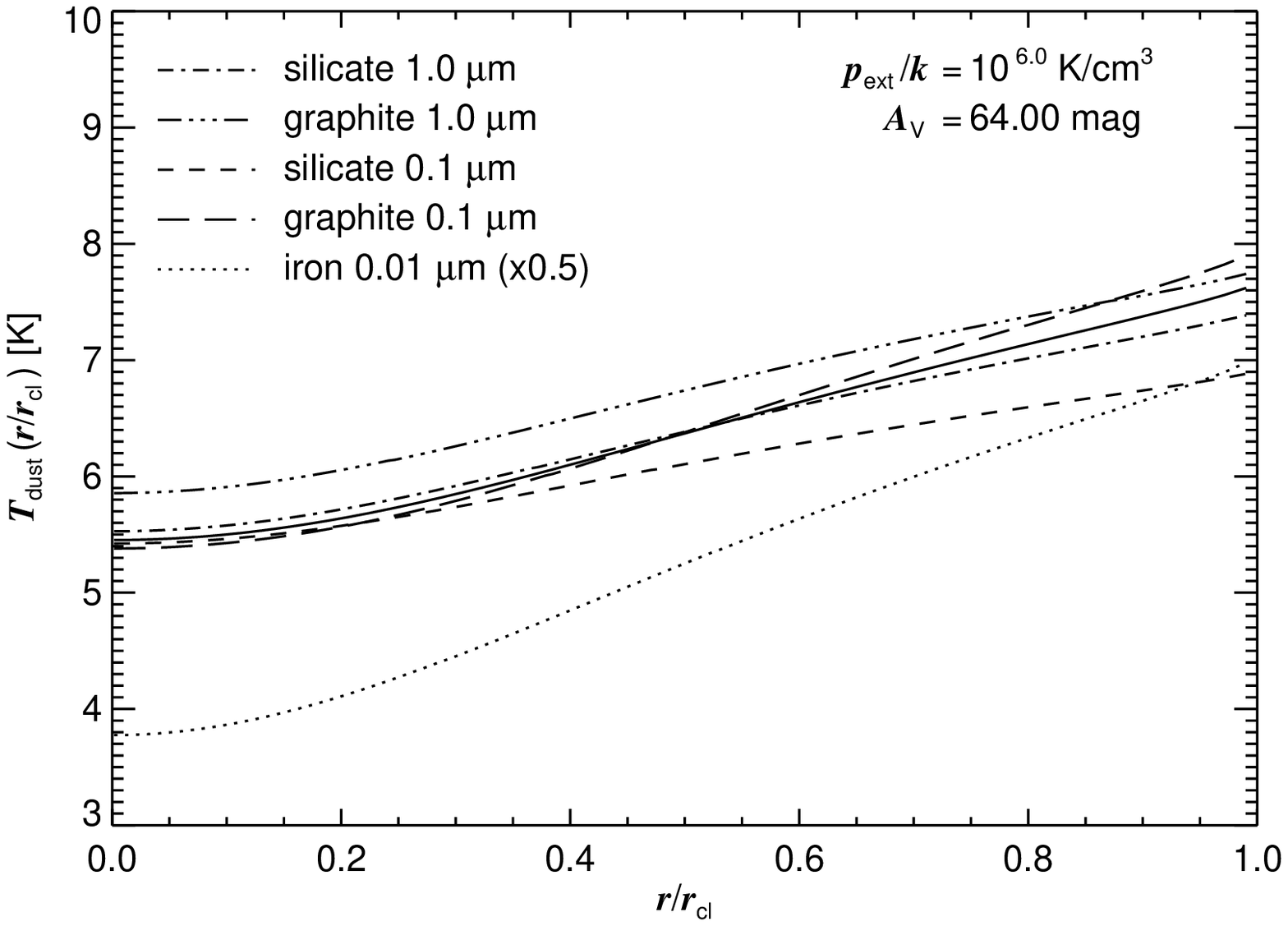}
	\caption{\label{fig_grtemp}
	Dust grain temperatures for a number of different sizes and dust compositions of cores
	embedded in spherical (black lines) or cylindrical (grey lines)
	filaments. The iron temperature is divided by a factor of 2. The effective values for a single temperature
	model of the dust grains is shown for comparison (solid curves) where the dust emissivity is approximated
	by a simple power law (Eq.~\ref{eq_irabsorption}).
	The external pressure $p_{\rm ext}/k$ 
	outside  the filaments is assumed to be either $2\times 10^4$ (top) or $10^6~{\rm K/cm^3}$ (bottom). 
	The extinction through the filaments' center is either 8~mag (left-hand figure) or 64~mag (right-hand figure).
	}
\end{figure*}

\subsubsection{\label{sect_sed_core}The SEDs from cores}

Figure~\ref{fig_sedfit} shows the obtained SEDs of cores, which are embedded in spherical filaments with $A_V=8~{\rm mag}$
and $A_V=64~{\rm mag}$. The SEDs of cores embedded in cylindrical filaments have the same properties but slightly
lower PAH emission and lower dust temperatures for the same extinction values.

The SEDs reflect the dependence of the opaqueness of the cores on both extinction and external pressure as discussed in the previous sections. For highly embedded cores, the SEDs depend to most parts on the extinction of the filament
and is largely independent of the external pressure (Sect.~\ref{sect_pressextrel}). The cores are highly opaque and are heated almost in equal part by the attenuated stellar radiation field and the PAH emission from the filaments (see Appendix~\ref{sect_modelapprox}, Fig.~\ref{fig_heatapprox}). The minimum around $10~\mu{\rm m}$ is enhanced by silicate absorption. 

At the blue side of the spectrum, the emission is dominated by light scattered within the core. For a given extinction $A_V$, the scattered radiation is only mildly dependent on the pressure since most of the scattered light originates in a layer close to the surface of the core (see Fig.~\ref{fig_heating}).  
Multiple scattered light only becomes important at wavelengths $\lambda< 3~\mu{\rm m}$.

The bulk of the dust re-emission is closely described by a modified black-body function. At shorter wavelengths, the spectral shape is dominated by the emission from stochastic dust emission, small grains, and PAH molecules. 
The effect of the dust attenuation inside the core is clearly visible in the PAH emission. Because of the low intensity of the UV/optical, the ratio of the PAH emission to the dust emission peak is lower than in the diffuse ISM or inside the filaments (Sect.~\ref{sect_sed_jmean}) and decreases systematically with extinction (Fig.~\ref{fig_sedcores}).
For highly embedded cores, the reddened radiation affects the spectral shape of the PAH emission spectrum as the photons are not sufficiently energetic to heat the molecules to high temperatures. 

For a given extinction $A_V$, the PAH emission of embedded cores is not strongly dependent on the external pressure as the heating flux for the
PAH molecules is largely absorbed in the core. The dust emission, on the other hand, appears to be colder
and stronger in higher pressure regions as the grains in the core's center are heated by a more strongly attenuated radiation field with a larger fraction of the radiation being absorbed inside the core. There is, however, a limit to the total luminosity given by the
$ISRF$ heating the filaments (see Sect.~\ref{sect_mbb_lum})

\begin{figure*}[htbp]
%
	\includegraphics[width=0.49\hsize]{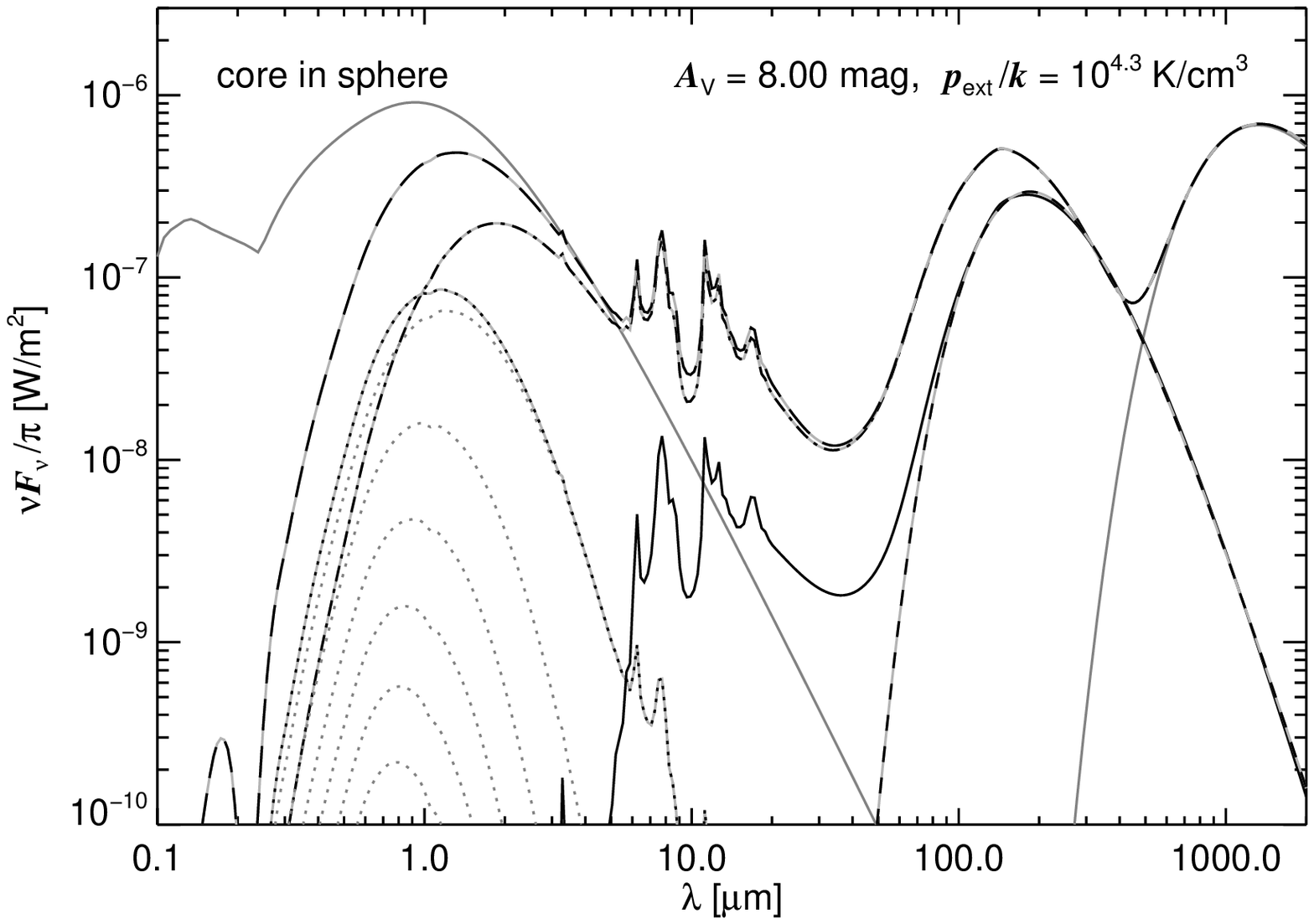}
	\hfill
	\includegraphics[width=0.49\hsize]{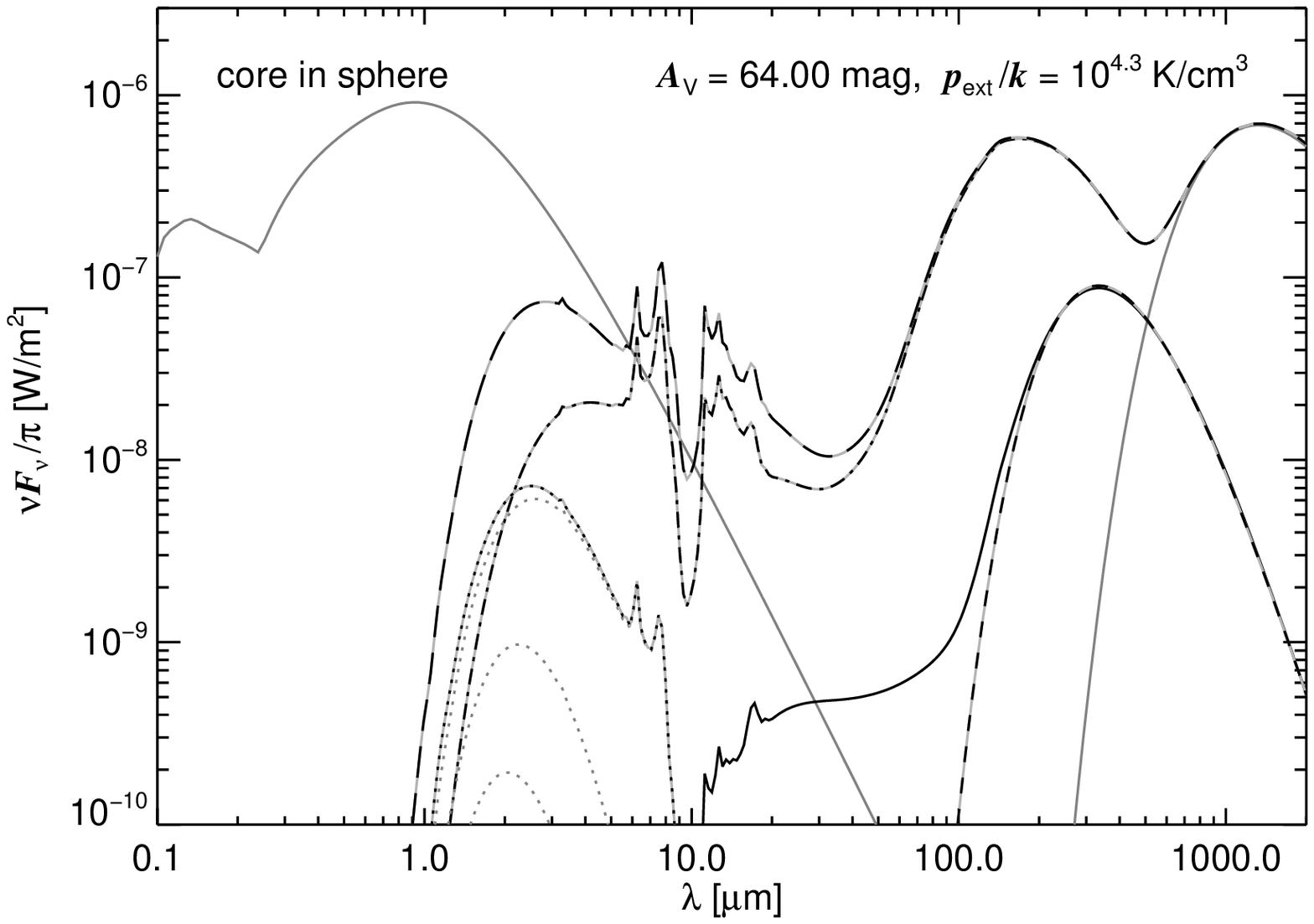}

	\includegraphics[width=0.49\hsize]{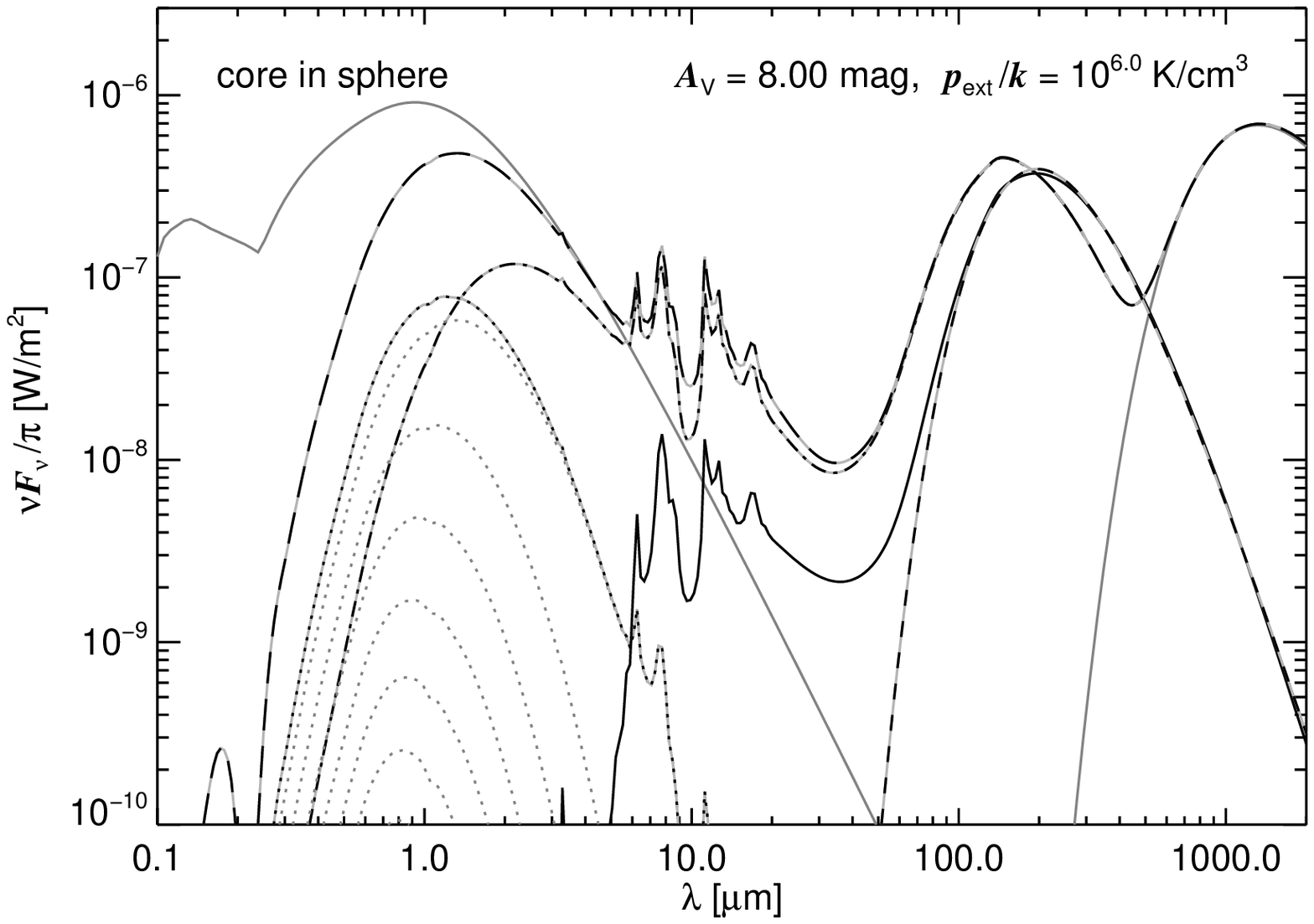}
	\hfill
	\includegraphics[width=0.49\hsize]{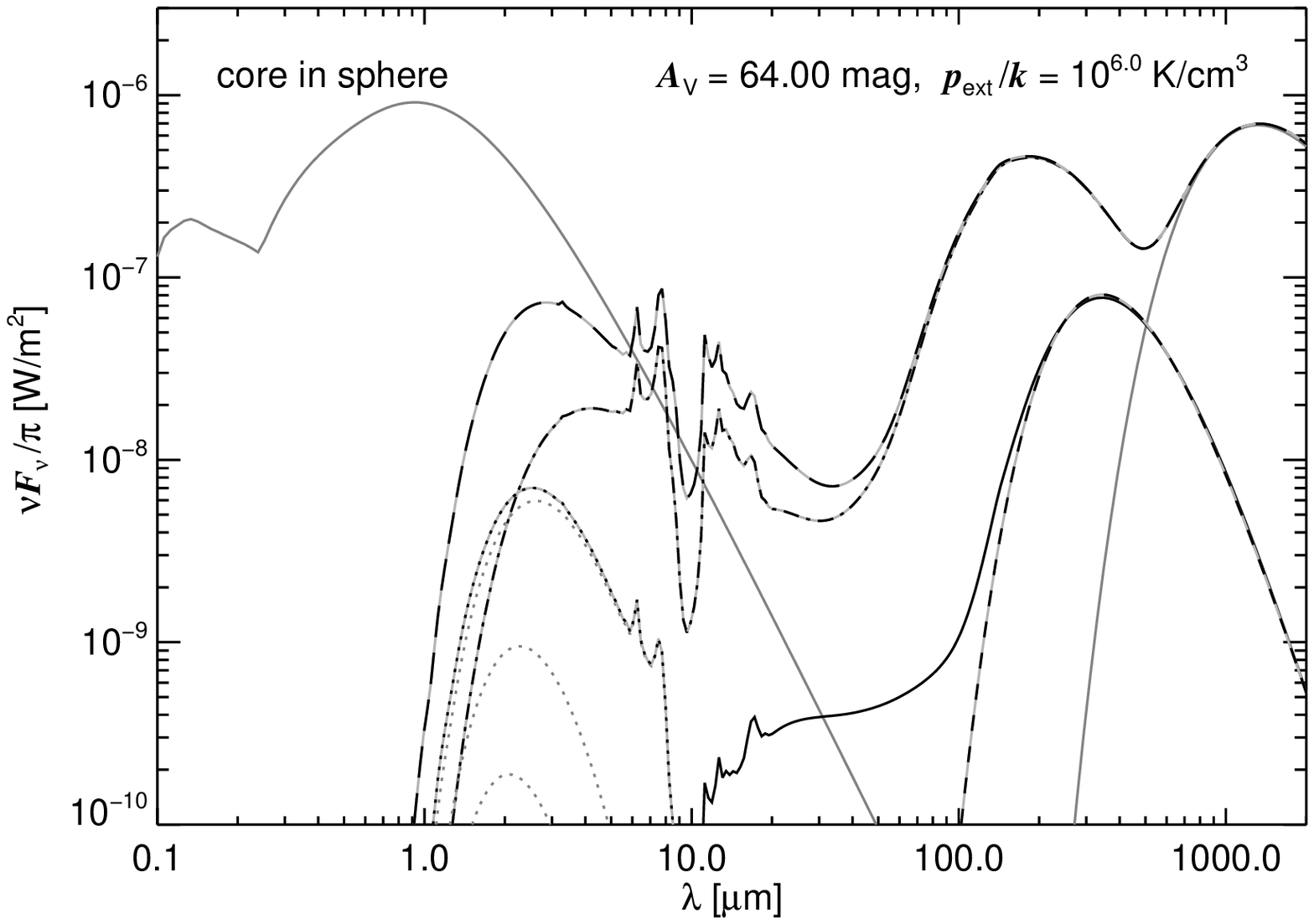}
		\caption{\label{fig_sedfit}
	SEDs of cores located at the center of a spherical filament with a total central extinction $A_V=8~\rm mag$ 
	(left-hand figure)  and $A_V=64~{\rm mag}$ (right-hand figure). The filaments are embedded in a medium with
	$p_{\rm ext}/k=2\times 10^4~{\rm K/cm^3}$ (top) or $10^6~{\rm K/cm^3}$ (bottome) and are heated by an
	unattenuated interstellar radiation field (grey solid line).
	The derived dust re-emission spectrum is shown as a solid black line, the radiation heating the core as
	a long dashed line, and the attenuated emission leaving the core as a dashed-dotted line.
	Also given is the emission produced by photons scattered inside the core (black dotted
	lines). The grey dotted lines show the individual components of multiply scattered photons. The top curve
	shows singly scattered emission, the curves below emission scattered twice and so on. The dust emission
	spectrum is fitted to a modified black-body function (short dashed line). 
	}
\end{figure*}
\begin{figure*}[htbp]
	\includegraphics[width=0.49\hsize]{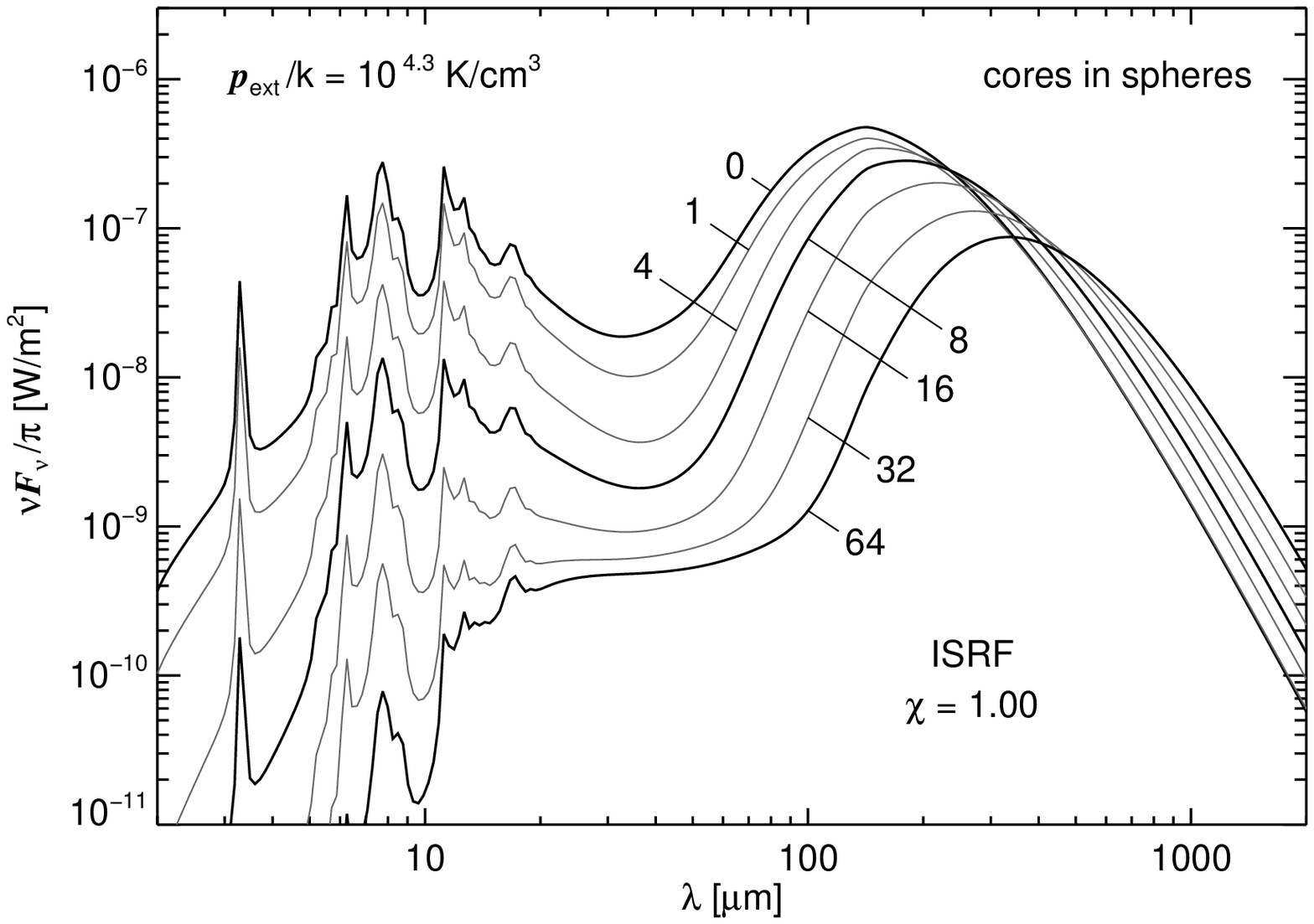}
	\hfill
	\includegraphics[width=0.49\hsize]{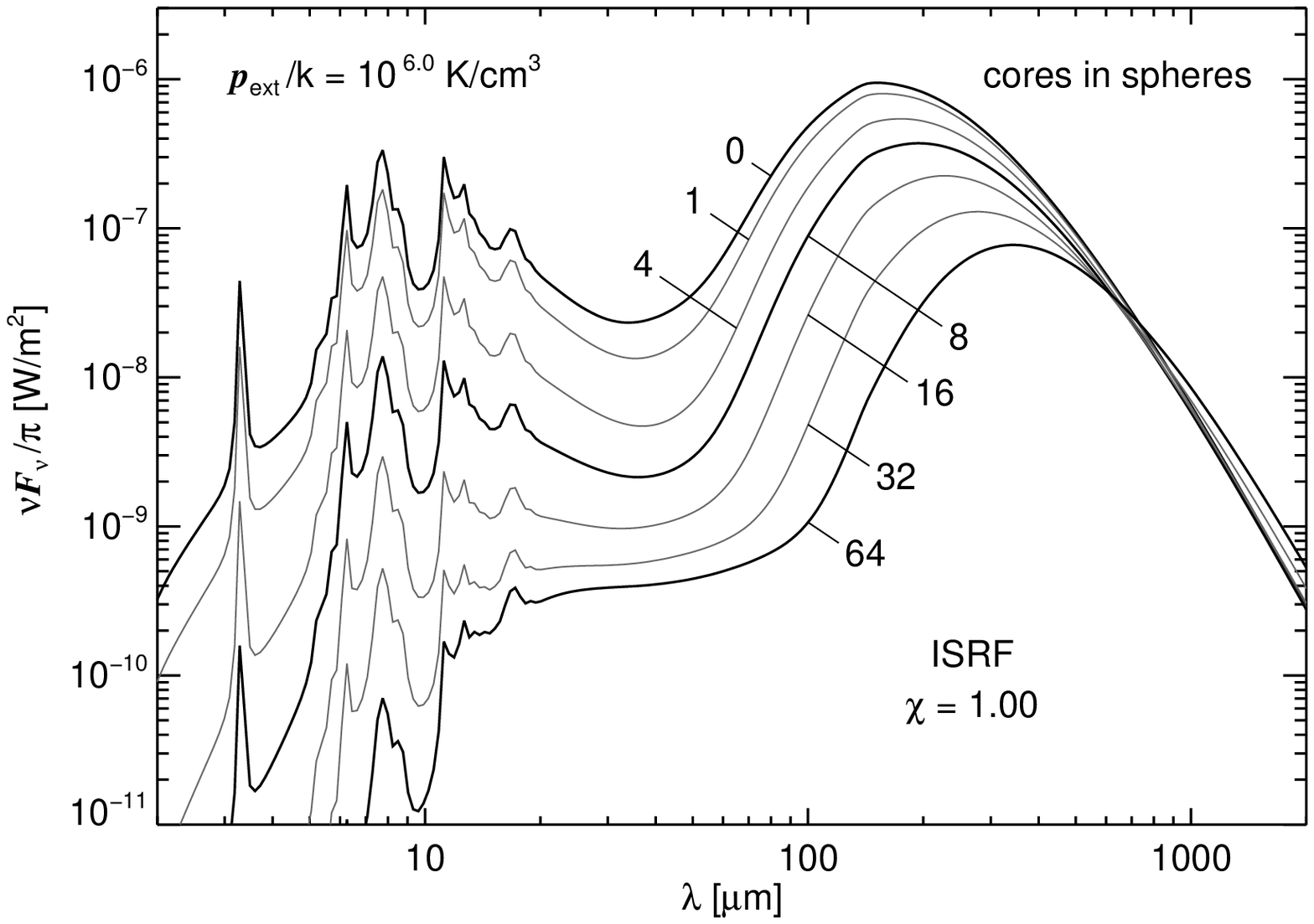}
	\caption{\label{fig_sedcores}
	SEDs of condensed cores embedded in the center of spherical self-gravitating filaments illuminated
	by the ISRF given by \citet{Mathis1982}. The external pressure $p_{\rm ext}/k$ is either $2\times 10^4$ (left-hand figure) 
	or $10^6~{\rm K/cm^3}$ (right-hand figure). The SEDs are labelled with the assumed extinction $A_V$ through the filaments'
	center. 
	}
\end{figure*}

The effect of the extinction for a given external pressure $p_{\rm ext}/k$ on the dust emission from cores is shown in
Fig.~\ref{fig_sedcores}, which covers the range from non-embedded (Bok-globules) to highly embedded cores. In low pressure
regions, a higher extinction of the filaments leads to an enhanced emission in the Rayleigh-Jeans limit of the spectrum. 
This behavior is responsible for the pressure-extinction relation of the filaments. As I have shown (Sect.~\ref{sect_sed_heating}), this relation significantly decreases the heating rate inside the core and produces above $A_V\sim 1$~mag a larger temperature variation inside the core. The effect disappears if the extinction does not lead to a strong increase in the central pressure as in the high pressure region below $A_V\sim 16~{\rm mag}$. The lower amount of radiation from embedded cores is then largely compensated by the shift to colder dust temperatures.



\subsection{\label{sect_mbb}The effective parameters for the SED}

For direct comparisons with observations, the theoretical dust emission spectrum at long wavelengths
is fitted by a modified black-body spectrum having $T_{\rm dust}$, $\beta$, and $\kappa_0$
as free parameters. The fit is achieved using a non-linear $\chi^2$-fit with
\begin{equation}
	\chi^2(\kappa_0,\beta,T_{\rm dust}) = 
	\sum_i \frac{\left(F_{\lambda_i} - S_{\lambda_i}(\kappa_0,\beta,T_{\rm dust})\right)^2}{F_{\lambda_i}^2},
\end{equation}
where $F_{\lambda_i}$ and $S_{\lambda_i}$ refer to the mean theoretical flux  and the mean fitted flux of the modified black body spectrum (Eq.~\ref{eq_modbb}) over wavelength bins $i$, respectively. 

As can be seen in Fig.~\ref{fig_sedfit}, the dust emission spectrum at shorter wavelengths from the maximum deviates considerably from the simple modified black-body spectrum. {  The fit is therefore restricted to the interval
 \begin{equation}
 [1\,400/ T_{\rm dust}[K], 10\,000]~\mu{\rm m}. 
\end{equation}

The derived parameters for the modified black-body approximation are listed in Table~\ref{table_fitparameters}, where I also provide the luminosity of the modified black-body function and the integrated luminosity. In the model, at low extinction values, a larger fraction of the light is absorbed by PAH molecules and small dust particles and subsequently re-emitted at shorter wavelengths.
This causes the modified black-body function to underestimate the total dust emission. The PAH/small dust particle contribution disappears gradually towards highly embedded cores as the UV and optical light becomes almost entirely absorbed inside the filaments heating the small dust particles.
The last two columns give the $FWHM$ and the total integrated flux of an Gaussian source approximation for the brightness profile (see Sect.~\ref{sect_mbb_profile}). 
For $\beta$, $\kappa_0$, $L/M_{\rm core}$, and the $FWHM$, I provide polynomial fits as functions of the dust temperature (Table~\ref{table_coeff}). 
I emphasize that the derived relations depend on the specific dust properties, and to an extent, on the radiation field.  The relations provided are appropriate for dust with the properties of diffuse ISM dust.

As an estimate of the accuracy of the achieved fit, I list $\sqrt{1/\chi^2_{\rm red}}$ where $\chi^2_{\rm red}=\chi^2/N_{\rm free}$ 
and $N_{\rm free}=N-3$ is the number of free parameters. The accuracy is within 2.5\% to 4\% with the largest deviation occuring
around the peak of the dust re-emission spectrum. The fit becomes worse for higher extinction values because of the 
larger temperature variations in the core. Closer agreement is achieved again at extinction values larger than $A_V=64~{\rm mag}$. Because of the larger temperature variations inside the core for a given central extinction of the filament, the agreement is also less good in high pressure regions. I note that in part the disagreement is also
caused by the optical properties used in the radiative transfer calculations, whose properties differ, as those of graphite grains,
strongly from a simple power-law behavior in the FIR.

In low pressure regions, the filaments and the condensed cores become less optically thin. In the limit of a very low
pressure region, the dust emission from the cores is identical to the one in the diffuse ISM. 
To obtain the corresponding effective values of the modified black-body approximation 
for the diffuse dust emission, I considered a spherical self-gravitating cloud with $A_V=0.01~{\rm mag}$ 
(assuming $p/k=2\times 10^4~{\rm K/cm^3}$).

The fit of the dust re-emission spectrum
is not as good as for emission from the dense cores because the contribution of stochastically heated small grains at the blue side of the emission spectrum is stronger. The derived effective temperature is also at $T_{\rm dust}=19.86~{\rm K}$ relatively high. The parameters of the emissivity are $\beta=1.83$ and $\kappa_0=3.40~{\rm cm^2/g}$. 

If the grains are heated by the non-attenuated ISRF, the ratio of the luminosity of the dust emission to the total mass of gas and dust is close to `1' (in solar units). The ratio obtained with the modified black-body fit is $0.736~L_{\odot}/M_\odot$. For the total dust emission spectrum, the ratio is higher at $L_{\rm total}/M_{\rm core}=1.149~{L_{\odot}/M_{\odot}}$.

The properties of the effective emissivity, the luminosity, and the temperature are described in the following subsections. 
It is found overall that the dependence of the effective parameters on dust temperature is not strongly affected by the shape of the filaments. For the same overpressure, the central extinction of a cylinder is indeed lower. However, when averaging over the orientations the different geometry produces in the mean an extinction close to the one of a spherical filament. The dust
temperature for a given overpressure is therefore closely the same.


As shown in Sect.~\ref{sect_pressextrel} at high overpressure, as for highly embedded cores,
the radiative transfer problem does not strongly depend on the external pressure. Therefore, the 
effect of the external
pressure on the parameters of the modified black-body spectrum weakens towards colder dust temperatures from cores.


\subsubsection{\label{sect_mbb_effem}The effective dust emissivity}

\begin{figure*}[htbp]
	\includegraphics[width=0.494\hsize]{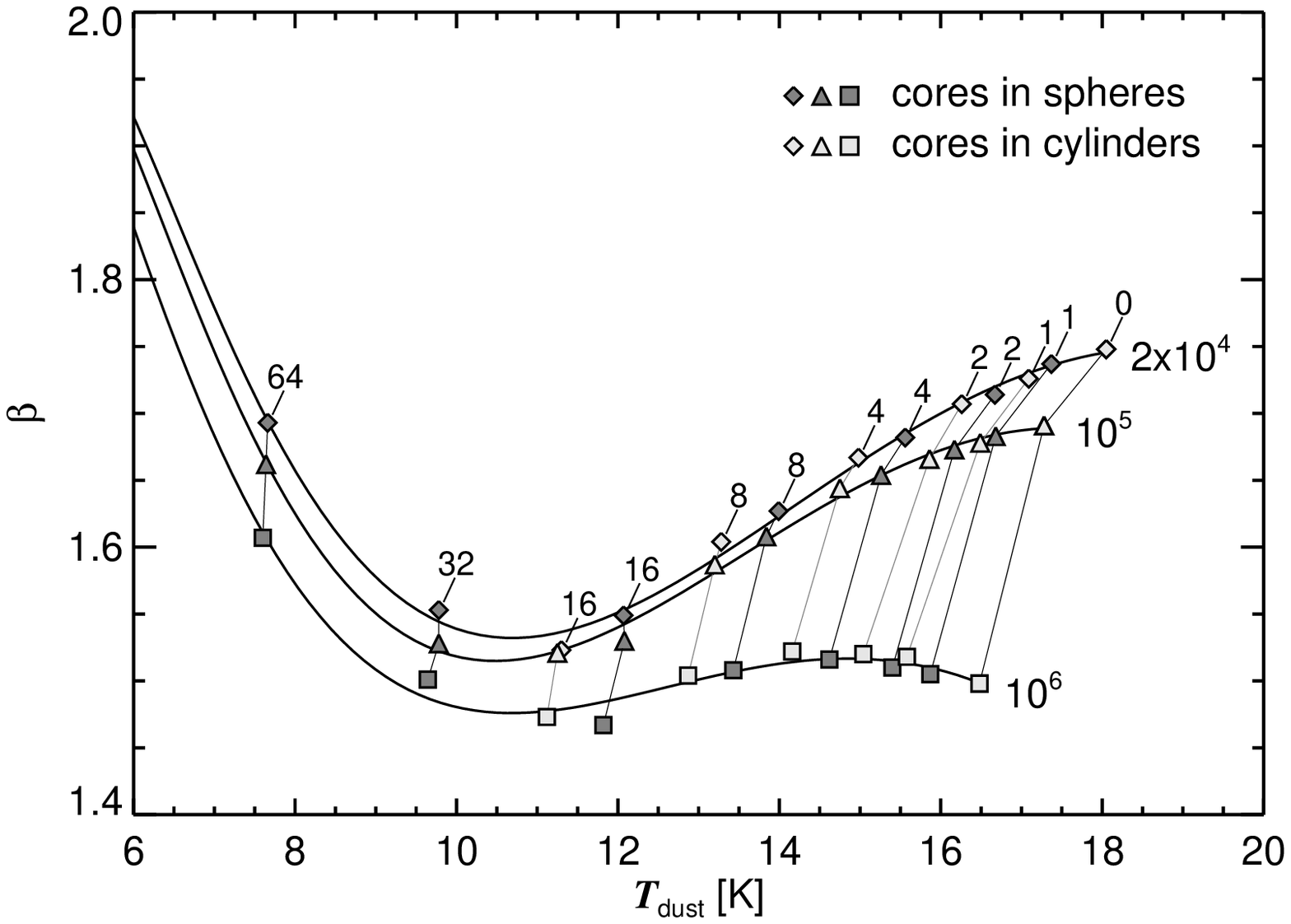}
	\hfill
	\includegraphics[width=0.49\hsize]{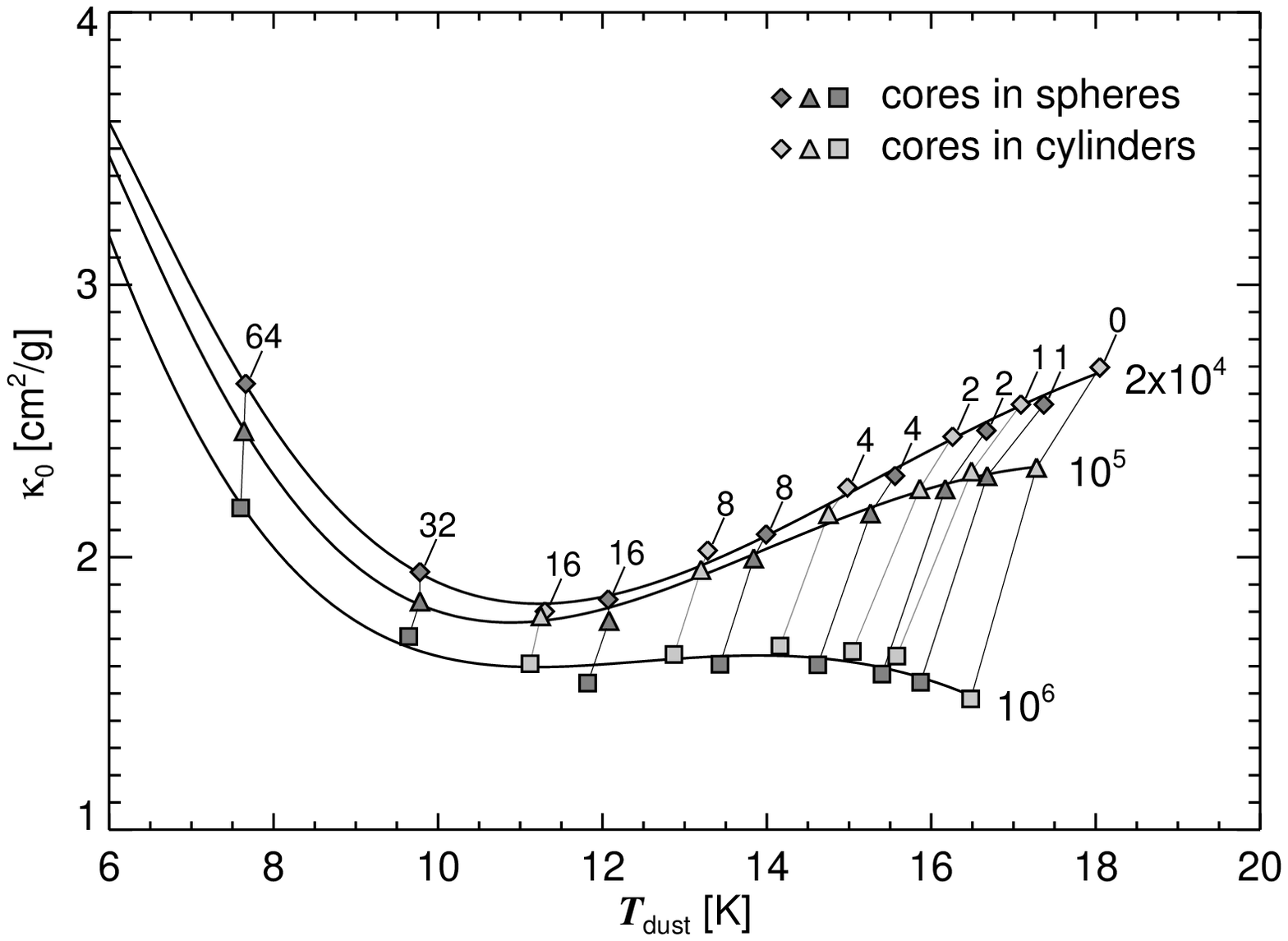}
	\caption{\label{fig_effprop}
	$\beta$ and $\kappa_0$ as functions of the effective dust temperature $T_{\rm dust}$ for the dust 
	re-emission spectrum for cores assuming diffuse dust. The cores are either embedded in spherical or
	cylindrical filaments. 
	The external pressure $p_{\rm ext}/k$ outside the filaments is assumed to be 
	$2\times 10^4$, $10^5$, or $10^5~{\rm K/cm^3}$. The corresponding theoretical values labelled by the considered
	central extinction $A_V$ of the filaments are shown either as diamonds, triangles, or squares.
	The values obtained for a given external pressure are fitted with a polynomial function shown as solid black lines. 
	}
\end{figure*}

In Fig.~\ref{fig_effprop}, I present the spectral index $\beta$ and the emissivity 
$\kappa_0$ at $\lambda_0=250~\mu{\rm m}$ as functions of the dust temperature of the cores.
In all cases, the emissivity is lower than
the corresponding mean value. 
Furthermore, the emissivity  is less steep with $\beta<1.8$ in all cases. The parameters $\beta$ and $\kappa_0$
are lower in higher pressure regions. The origin of the flatter emissivity is discussed in Sect.~\ref{sect_discussionemissivity}.


As the figure shows, the emissivity varies as a function of the dust temperature. This dependence decreases with 
the external pressure. In the high pressure region, the emissivity for $A_V<16~{\rm mag}$ shows only a mild dependence on dust temperature with $\beta\approx 1.5$. In lower pressure regions, the 
emissivity is flatter in colder cores. For the ISM pressure, $\beta$ decreases from $1.75$ to $1.55$. For 
typical GMCs with $A_V=8~{\rm mag}$, $\beta$ is close to 1.6. The flattening reflects the increasing optical depth with extinction
$A_V$ through the filaments as shown in Sect.~\ref{sect_pressextrel}. This dependence is strongest in the lowest pressure region.
For highly embedded cores with $A_V>32$, the apparent emissivity steepens again and seems to approach that
corresponding to the intrinsic dust properties. The behavior is discussed in Sect.~\ref{sect_discussionemissivity}.

\begin{figure}[htbp]
	\includegraphics[width=\hsize]{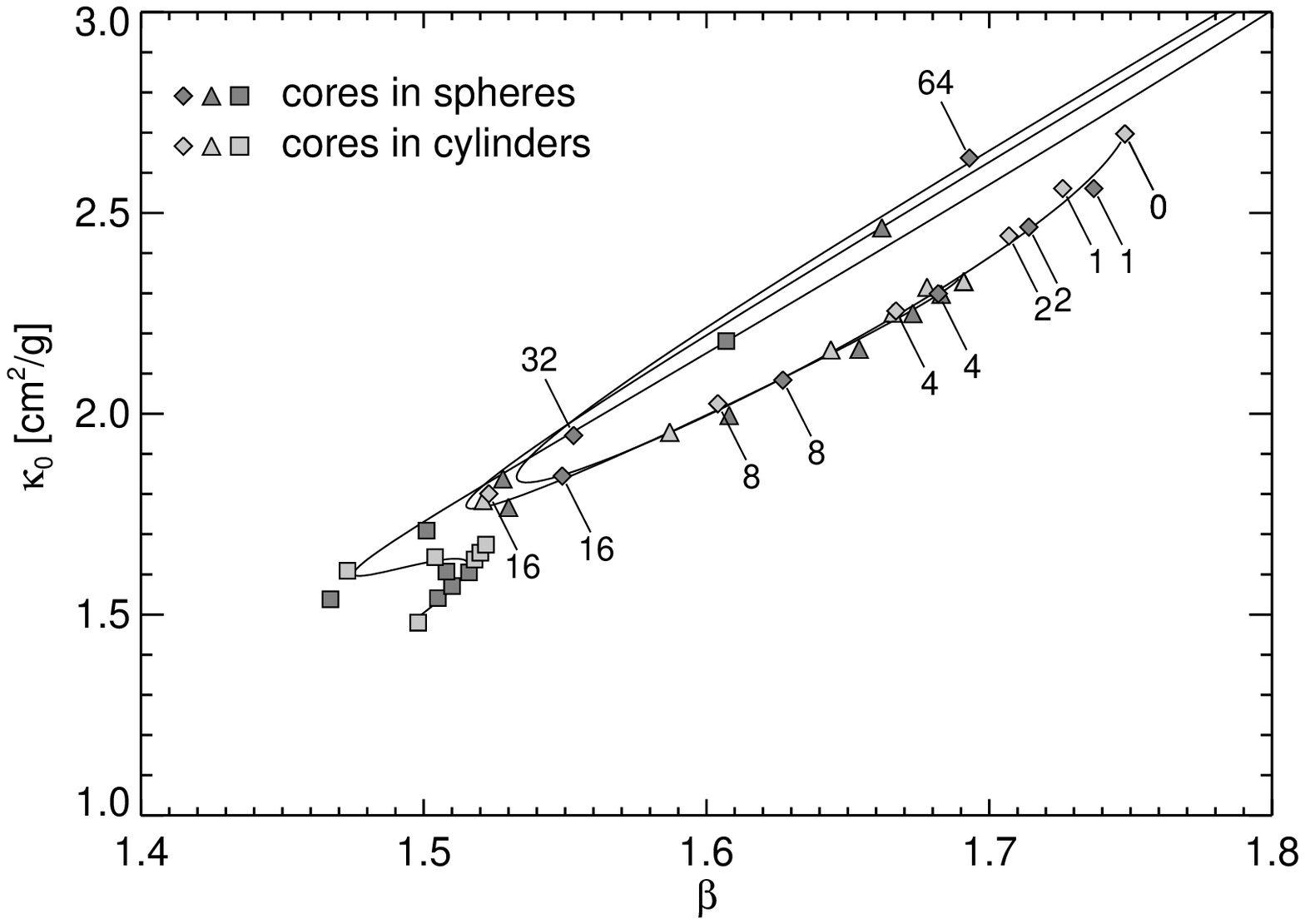}
	\caption{\label{fig_betakappa}
	Relation of the effective values $\kappa_0$ and $\beta$ derived by fitting a modified black-body function
	to the theoretical dust emission spectrum of condensed cores. The symbols represent the same external pressure
	$p_{\rm ext}/k$ of the medium surrounding the filaments as in Fig.~\ref{fig_effprop} (diamonds for 
	$2\times 10^4~{\rm K/cm^3}$, triangles for $10^5~{\rm K/cm^3}$, and squares for $10^6~{\rm K/cm^3}$.) The
	symbols corresponding to filaments pressurized by the mean ISM pressure are labeled with the assumed 
	central extinction of the filament. The curves are the derived fits for $\beta$ and $\kappa_0$ shown in Fig.~\ref{fig_effprop}.}
\end{figure}

As can be seen from their functional dependence on dust temperature, the values of $\beta$ and $\kappa_0$ are strongly correlated. A low $\beta$ is 
accompanied by a low $\kappa_0$ as shown in Fig.~\ref{fig_betakappa}. This correlation is most apparent 
for regions with $p_{\rm ext}/k<10^6~{\rm K/cm^3}$ and extinction values $A_V<16~{\rm mag}$, where the data points
exhibit only very small deviations from the fitted polynomial functions.

\begin{table*}[htbp]
	\caption{\label{table_fitparameters}
	SED Parameters of Embedded Cores ($T_{\rm eff}=10~{\rm K}$)
	}
	\begin{tabular}{ccccccccccccc}
	$A_V$	&$p_{\rm c}/k$ & $\sqrt{1/\chi^2_{\rm red}}$ & $T_{\rm dust}$\tablefootmark{a}  & $\beta$
	\tablefootmark{b} & $\kappa_0$\tablefootmark{c}  &   $L$
	\tablefootmark{d} & ${L}/{M_{\rm core}}$ & $L_{\rm total}$
	\tablefootmark{e} & $L_{\rm total}/M_{\rm core}$ & $FWHM$\tablefootmark{f} & $F_{\rm Gauss}$\tablefootmark{g}\\
		& $[10^4~\rm \frac{K}{cm^3}]$ & [\%] &  [K] &	& $[\rm \frac{cm^2}{g}]$ & $[L_{\odot}]$ & 
			$[\frac{L_{\odot}}{M_{\odot}}]$ & 
			$[L_{\odot}]$ & $[\frac{L_{\odot}}{M_{\odot}}]$ & $[R_{\rm cl}]$ & $[F_{\rm model}]$\\
\hline
\hline
	\multicolumn{12}{c}{($p_{\rm ext}=2\times 10^4~{\rm K/cm^3}$, $\chi=1$)}\\
\hline
	0	& 2.00 & 3.05 & 18.05 & 1.748 & 2.697 & 0.829 & 0.323 & 1.081 & 0.421 & 1.081 & 1.077 \\
\hline
	\multicolumn{12}{c}{Cores in the center of spherical clouds}\\
\hline
	1  &   2.48   & 2.92 & 17.37 & 1.737 & 2.561 & 0.562 & 0.244 & 0.675 & 0.293 & 1.059  & 1.062\\
	2  &   3.82 & 2.79 & 16.67 & 1.714 & 2.465 & 0.343 & 0.184 & 0.391 & 0.210 & 1.062 & 1.064\\
	4  &   8.65 & 2.83 & 15.56 & 1.682 & 2.299 & 0.142 & 0.115 & 0.152 & 0.123 & 1.086 & 1.073\\
	8  & 26.36 & 3.02 & 13.99 & 1.627 & 2.084 & 3.99(-2) & 5.61(-2) & 4.09(-2) & 5.79(-2) & 1.131 & 1.085\\
	16 & 93.22 & 3.51 & 12.07 & 1.549 & 1.845 & 8.12(-3) & 2.16(-2) & 8.13(-3) & 2.18(-2) & 1.176 & 1.090\\
	32 & 351.7 & 3.58 & 9.78 & 1.553 & 1.946 & 1.37(-3) & 7.07(-3) & 1.39(-3) &  7.19(-3) &1.223 & 1.101\\
	64 &	1365. & 2.51 & 7.66 & 1.693 & 2.637 & 2.30(-4) & 2.34(-3) & 2.35(-4)& 2.39(-3)& 1.247 & 1.105 \\	
\hline
	\multicolumn{12}{c}{Cores in the center of cylindrical clouds}\\
\hline
	1	& 2.72 & 2.84 & 17.09 & 1.726 & 2.561 & 0.489 & 0.222 & 0.577 & 0.262 & 1.056 & 1.062\\
	2	& 4.76 & 2.78 & 16.26 & 1.707 & 2.443 & 0.263 & 0.158 & 0.293 & 0.176 & 1.066 & 1.066 \\
	4 	& 12.46 & 2.88 & 14.98 & 1.667 & 2.256 & 9.30(-2) & 9.04(-2) & 9.30(-2) & 9.57(-2) & 1.099 & 1.076 \\
	8	& 42.04 & 3.18 & 13.28 & 1.604 & 2.025 & 2.27(-2) & 4.05(-2) & 2.32(-2) & 4.14(-2) & 1.151 & 1.090\\
	16	& 158.1 & 3.79 & 11.30 & 1.523 & 1.801 & 4.22(-2) & 1.46(-2)& 4.27(-3) & 1.48(-2) & 1.202 & 1.098 \\
\hline
\hline
	\multicolumn{12}{c}{($p_{\rm ext}=1\times 10^5~{\rm K/cm^3}$, $\chi=1$)}\\
\hline
	0 & 10.0 	& 3.06 & 17.28 & 1.691 & 2.329 & 0.243 & 0.211 & 0.3021 & 0.263 & 1.151 &  1.091 \\
\hline
	\multicolumn{12}{c}{Cores in the center of spherical clouds}\\
\hline
	1 &   10.48 & 2.96 & 16.68 & 1.683 & 2.298 & 0.191 & 0.170 & 0.223 & 0.199 & 1.130 & 1.083\\
	2 & 	11.92 & 2.92 & 16.17 & 1.673 & 2.250 & 0.147 & 0.139 & 0.164 & 0.156 & 1.123 & 1.082\\
	4 &	17.33 & 2.92 & 15.26 & 1.654 & 2.161 & 8.34(-2) & 9.56(-2) & 8.90(-2) & 0.102 & 1.128 & 1.084\\
	8 &	37.01 & 3.10 & 13.84 & 1.608 & 1.996 & 3.02(-2) & 5.05(-2) & 3.11(-2) & 5.20(-2) & 1.154 & 1.090\\
	16 & 108.7 & 3.56 & 12.08 & 1.530 & 1.767 & 7.06(-3) & 2.03(-3) & 7.14(-3) & 2.05(-2) & 1.202 & 1.104\\
	32 & 378.2 & 3.64 & 9.78 & 1.528 & 1.838 & 1.25(-3) & 6.70(-3) & 1.27(-3) & 6.81(-3) & 1.233 & 1.103\\
	64 & 1417. & 2.57 & 7.64 & 1.662 & 2.463 & 2.10(-4) & 2.18(-3) & 2.14(-4) & 2.22(-3) & 1.262 & 1.110\\
\hline
	\multicolumn{12}{c}{Cores in the center of cylindrical clouds}\\
\hline
	1 & 	10.73 & 3.00 & 16.49 & 1.678 & 2.315 & 0.178 & 0.160 & 0.203 & 0.184 & 1.122 & 1.081 \\
	2 & 	12.89 & 2.96 & 15.86 & 1.666 & 2.251 & 0.126 & 0.124 & 0.138 & 0.137 & 1.118 & 1.081\\
	4 & 	21.14 & 2.97 & 14.75 & 1.644 & 2.159 & 6.22(-2) & 7.87(-2) & 6.55(-2) & 8.30(-2) & 1.131 & 1.085\\
	8 &	52.21 & 3.24 & 13.20 & 1.587 & 1.954 & 1.90(-2) & 3.78(-2) & 1.94(-2) & 3.86(-2) & 1.161 & 1.088\\
	16 & 	171.9 & 3.82 & 11.25 & 1.521 & 1.784 & 3.92(-3) & 1.41(-3) & 3.97(-3) & 1.42(-2) & 1.208 & 1.099\\
\hline
\hline
	\multicolumn{12}{c}{($p_{\rm ext}=1\times 10^6~{\rm K/cm^3}$, $\chi=1$)}\\
\hline
	0 & 100.0 & 3.79 & 16.48 & 1.498 & 1.480 & 3.47(-2) & 9.57(-2) & 0.0409 & 0.113 & 1.311 & 1.125 \\
\hline
	\multicolumn{12}{c}{Cores in the center of spherical clouds}\\
\hline
	1 	& 100.5 & 3.70 & 15.87 & 1.505 & 1.541 & 2.94(-2) & 8.13(-2) & 3.29(-2) & 9.09(-2) & 1.283 & 1.111\\
	2 	& 102.0 & 3.59 & 15.40 & 1.510 & 1.571 & 2.53(-2) & 7.02(-2) & 2.74(-2) & 7.62(-2) & 1.280 & 1.119\\
	4 	& 107.8 & 3.47 & 14.62 & 1.516 & 1.605 & 1.89(-2) & 5.39(-2) & 1.98(-2) & 5.67(-2) & 1.256 & 1.108\\
	8  	& 130.4 & 3.50 & 13.43 & 1.508 & 1.607 & 1.08(-2) & 3.38(-2) & 1.10(-2) & 3.47(-2) & 1.245 & 1.105\\
	16 	& 214.4 & 3.83 & 11.82 & 1.467 & 1.538 & 3.97(-3) & 1.60(-2) & 3.99(-3) & 1.62(-2) & 1.247 & 1.106\\
	32 	& 518.7 & 3.54 & 9.644 & 1.501 & 1.709 & 9.26(-4) & 5.81(-3) & 9.42(-4) & 5.91(-3) & 1.278 & 1.121\\
	64 	& 1638. & 2.68 & 7.60 & 1.607 & 2.181 & 1.71(-4) & 1.91(-3) & 1.75(-4)& 1.95(-3) & 1.298 & 1.123\\
\hline
	\multicolumn{12}{c}{Cores in the center of cylindrical clouds}\\
\hline
	1  	& 100.7 & 3.68 & 15.58 & 1.518 & 1.637 & 2.83(-2) & 7.83(-2) & 3.13(-2) & 8.66(-2) & 1.279 & 1.121\\
	2  	& 102.9 & 3.59 & 15.04 & 1.520 & 1.654 & 2.33(-2) & 6.51(-2) & 2.50(-2) & 6.98(-2) & 1.254 & 1.107\\
	4  	& 111.7 & 3.51 & 14.16 & 1.522 & 1.674 & 1.62(-2) & 4.72(-2) & 1.69(-2) & 4.92(-2) & 1.241 & 1.104\\
	8  	& 145.8 & 3.61 & 12.87 & 1.504 & 1.643 & 8.22(-3) & 2.73(-2) & 8.38(-3) & 2.79(-2) & 1.238 & 1.104\\
	16 	& 275.1 & 4.01 & 11.12 & 1.473 & 1.609 & 2.62(-3) & 1.20(-2) & 2.65(-3) & 1.20(-2) & 1.246 & 1.105\\
\hline
	\end{tabular}
	\tablefoottext{a}{Effective dust temperature of the modified black-body fit.}\\
	\tablefoottext{b}{Power of the effective emissivity $\kappa^{\rm em}_\lambda=\kappa_0(\lambda/\lambda_0)^{-\beta}$ where $\kappa_0$ is the effective emissivity at $\lambda_0=250\mu{\rm m}$}.\\
	\tablefoottext{c}{Effective emissivity per dust mass at $\lambda_0=250~\mu{\rm m}$.}\\
	\tablefoottext{d}{Luminosity of the modified black-body fit.}\\
	\tablefoottext{e}{Total luminosity of the dust re-emission.}\\
	\tablefoottext{f}{$FWHM$ of the Gaussian source approximation of the surface brightness at $250~\mu{\rm m}$.}\\
	\tablefoottext{g}{Flux of the Gaussian source approximation of the surface brightness at $250~\mu{\rm m}$.}
\end{table*}

\subsubsection{\label{sect_mbb_lum}Luminosity of passively heated cores and the $P-T$-relation}

\begin{figure*}[htbp]
	\includegraphics[width=0.49\hsize]{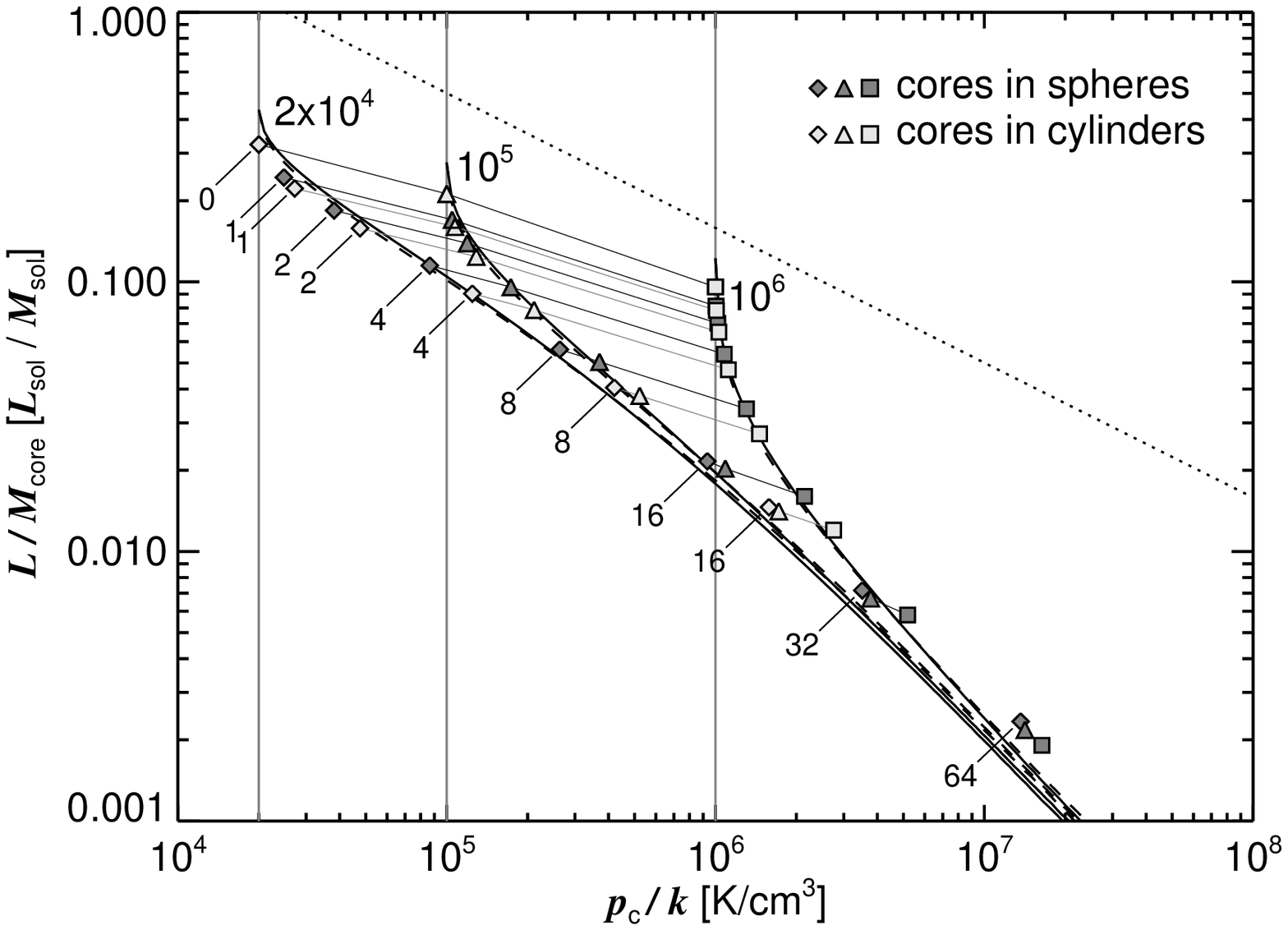}
	\includegraphics[width=0.49\hsize]{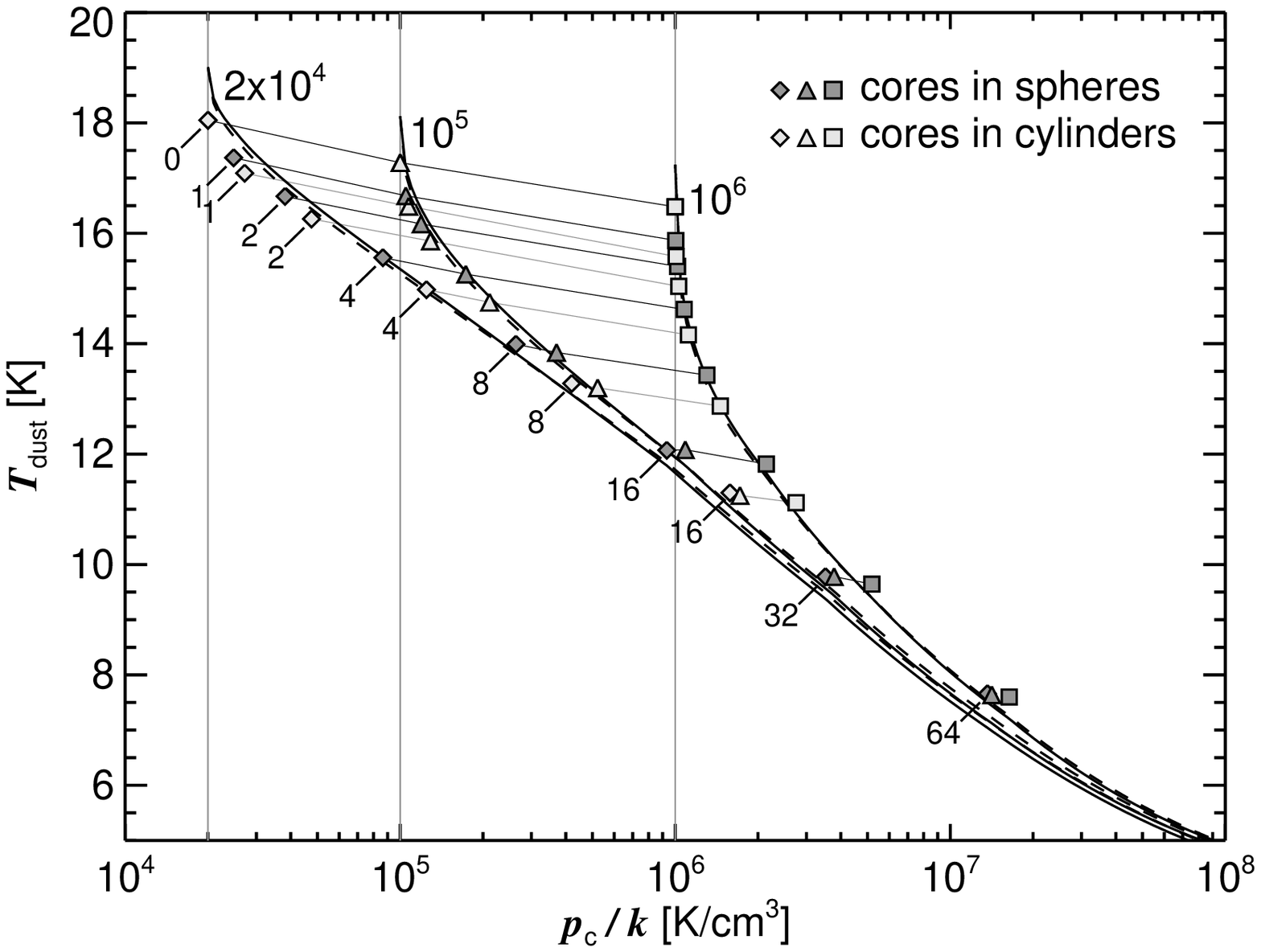}
	\caption{\label{fig_ptreldata}
	$L/M$-ratios (left-hand figure) and the effective dust temperatures (right-hand figure) 
	for embedded passively heated cores.
	The values obtained using ray-tracing (symbols) are compared with the approximate solutions for cores in
	spheres (solid curves) and cylinders (dashed curves).
	The different symbols refer to the external pressures (shown as vertical grey lines) assumed in the
	calculations: diamonds $2\times 10^4~{\rm K/cm^3}$, triangles $10^5~{\rm K/cm^3}$, 
	and squares $10^6~{\rm K/cm^3}$. 	They are labelled according to the extinction $A_V$ through the
	filament center. The dotted line in the left-hand figure shows the asymptote Eq.~\ref{eq_lmasymptote} for 
	non-embedded cores ($A_V=0~{\rm mag}$),
	where the external radiation heating the core is completely absorbed.
	}
\end{figure*}

The model relates the $L/M$-ratio and the effective dust temperature of the core to the central pressure 
in the filaments surrounding the core (Fig.~\ref{fig_ptreldata}). For a given shape, $ISRF$, external pressure, and extinction,
the ratio is independent of the true size of the core since both the luminosity and the mass, by mean of the mass-radius
relation, are proportional to $R_{\rm core}^2$.

If the flux heating a non-embedded core is completely absorbed, the luminosity becomes $L = 4\pi R_{\rm core}^2\,\pi J^{ISRF}$,
where $J^{ISRF}$ is the integrated flux of the $ISRF$. Ignoring the emission of the CMB, the $ISRF$ provided by \citet{Mathis1982} has
an energy density $u = \frac{4\pi}{c}J=0.46~{\rm eV/cm^3}$. 
Adopting the mass-radius relation (Eq.~\ref{eq_massradius}), the luminosity-mass ratio is given by
\begin{equation}
	\label{eq_lmasymptote}
	L/M_{\rm core}=1.6\left(\frac{p_{\rm ext}/k}{2\times 10^4~{\rm K/cm^3}}\right)^{-\frac{1}{2}}~L_{\odot}/M_{\odot}.
\end{equation}

As seen in Fig.~\ref{fig_ptreldata}, the curve provides an upper limit to the true $L/M$-ratios.
For non-embedded cores pressurized by the mean ISM pressure, the ratio is, despite the high extinction 
through the core center, lower by more then a factor of two, even if the total luminosity of the SED (as given in 
Table~\ref{table_fitparameters}) is considered.
One reason is the steep density profile of the cores and the consequently strong variation in the
optical thickness with the impact parameter of individual rays passing the core.
A core that appears opaque in the center can still be optically thin or marginally 
optically thick at the outskirts, so that only radiation towards the core center will be strongly attenuated. 
The amount of light absorbed in the core is further reduced by the clouds' albedo.

The overall increase in the extinction with pressure ($A_V\propto \sqrt{p_c}$) leads together with the decrease in the albedo to a 
stronger absorption in higher pressure regions. The
$L/M$-ratio asymptotically approaches, as shown in Fig.~\ref{fig_t_lightmass}, the curve given by Eq.~\ref{eq_lmasymptote}. 

As shown in Fig.~\ref{fig_ptreldata}, the simplified RT-model reproduces to first order not only the $L/M$-ratio but also the effective temperatures obtained using the full calculations.
The deviations are caused by the absence of the PAH emission in the simplified model, which is accurately accounted for in the full radiative transfer calculations. In the simplified model, all the absorbed UV and optical light is re-emitted in the FIR, which is otherwise re-emitted mostly by PAH molecules in the NIR and MIR. For extinction values below $A_V=4~{\rm mag}$, the heating of the dust grains in the cores is overestimated leading to a higher dust temperature. In the case of highly opaque filaments (above $A_V\sim 8~{\rm mag}$), the PAH emission of the filaments leads in the full RT calculation to an increasingly important heating source, which gives rise to the warmer dust temperatures (Sect.~\ref{sect_modelapprox}).


The curves for the lowest external pressure provide a lower limit to both the $L/M$-ratio
and the dust temperature of cores. Above the curves, passively heated cores are embedded in filaments, which are surrounded by a higher pressure gas. The cores
are also located in this region if they are heated by an additional source such as a progenitor star in the core or an additional external source. 


The resulting relation between the $L/M$-ratio and the effective dust temperature $T_{\rm dust}$ for a given external pressure 
is shown in Fig.~\ref{fig_t_lightmass}.
For the same $L/M$-ratio, the dust in embedded cores appears to be warmer in high pressure regions of the ISM because the SED is largely determined by the relative warm dust emission at the outskirts of the cores. 
The $L/M$-ratios for low external pressures lie on a curve that is slightly flatter than a simple power law as shown in the figure. The power-law fit provides a constant emissivity with $\beta=1.96$ and $\kappa_0=2.15~{\rm cm^2/g}$.

\begin{figure}[htbp]
	\includegraphics[width=\hsize]{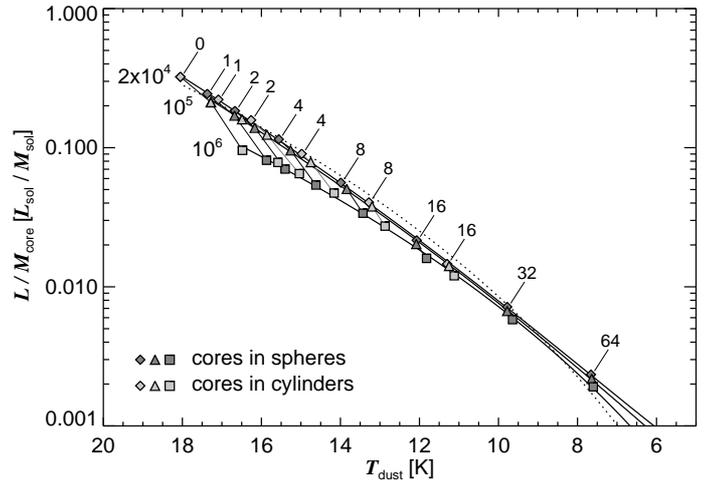}
	\caption{\label{fig_t_lightmass}
	Light-to-mass ratio ($L/M$) of the dust luminosity and the core mass as function of the dust temperature of embedded
	cores. The derived ratios are labelled with the assumed
	extinction values $A_V$ through the filaments center. 
	The ratios $L/M_{\rm core}(T_{\rm dust})$ obtained for cores that are embedded in cylindrical and spherical
	filaments for each given
	external pressure $p_{\rm ext}/k$ ($2\times 10^4$, $10^5$, $10^6~{\rm K/cm^3}$) are fitted by a polynomial
	function shown as solid black lines. The points for mean ISM pressure are fitted also by a simple power law 
	(dotted curve). }
\end{figure}

\begin{table*}[htbp]
	\caption{\label{table_coeff}Polynomial approximations}
	\begin{tabular}{cccccccc}
	$p_{\rm ext}/k$	& $a_0$ 	& $a_1$	& $a_2$	& $a_3$  & $a_4$ & $a_5$ \\
	${\rm [K/cm^3]}$ \\
	\hline
	\hline
		\multicolumn{7}{c}{$\beta = \Sigma_i a_i(\log_{10} T_{\rm dust}[{\rm K}])^i$} \\
	\hline
	$2\times 10^4$ &    -40.7509&      186.611&     -296.453&      202.709 &     -50.5764 & ---\\
	$1\times 10^5$ &     -41.0076&      190.600&     -307.465&      213.560&     -54.1691 & ---\\
	$1\times10^6$ &     -29.0751 &      144.608 &     -243.478 &      175.672 &     -46.2471 & --- \\
	\hline
	\hline
		\multicolumn{7}{c}{$\kappa_0 [{\rm cm^2/g}]= \Sigma_i a_i (\log_{10} T_{\rm dust}[{\rm K}])^i$} \\
	\hline
	$2\times 10^4 $ &  -133.777&      600.783&     -948.039&      639.586&     -156.646 & ---\\
	$1\times 10^5 $ &    -142.593 &      654.894&     -1060.36&      735.858&     -185.995 & ---\\
	$1\times 10^6$ &    -90.8491&      461.490&     -802.533&      592.289&     -158.761 & ---\\
	\hline
	\hline
		\multicolumn{7}{c}{$\log_{10} L/M_{\rm core} [L_\odot/M_{\odot}] = \Sigma_i a_i(\log_{10} T_{\rm dust}[{\rm K}])^i$} \\
	\hline
   	$2\times 10^4$		&     -2.63348&     -4.06995&      4.60187 & --- & ---  & ---\\
	$1\times 10^5$ 	&   -3.27281&     -2.87834&      4.03329  & --- & ---  & ---\\
	$1\times 10^6$		&	    -6.11354 &      2.87895 &       1.09179  & --- & --- & ---\\
	\hline
	\hline
		\multicolumn{7}{c}{$FWHM [R_{\rm cl}]= \Sigma_i a_i (T_{\rm dust}[{\rm K}])^i$}\\
	\hline
	 $2\times 10^4$ &    1.04042 &      0.193172 &    -0.0506557  &   0.00579236 &  -0.000312315 &  6.33987e-06 \\
    	$1\times 10^5 $& 	1.01551 &     0.198970 &   -0.0502195&   0.00571245& -0.000313287&  6.59132e-06 \\
	$1\times 10^6$&	1.29788 &   -0.0283919	&    0.0144707&  -0.00232888&  0.000140731& -2.79548e-06 \\
	\hline
	\end{tabular}
\end{table*}

\subsection{\label{sect_mbb_profile}Brightness profile}

Radiative transfer effects cause the brightness profiles of the dust emission spectrum to be broader than the intrinsic profile of the column density profile \citep{Fischera2008}. The profiles of the surface brightness at
three far infrared wavelengths, which correspond to the bolometers of the Herschel satellite's SPIRE-instrument 
(and of the balloon experiment BLAST), are shown in Fig.~\ref{fig_profiles}. The profile steepens towards longer wavelengths. This effect is caused
by the emission at longer wavelengths being dominated by larger grains, which have lower radial temperature 
variations than the smaller dust grains that dominate at shorter wavelengths. In addition, the emission at shorter wavelengths is more affected by temperature variations because these regions are blueward of the Rayleigh-Jeans limit. The steepening of the profile increases for highly embedded cores (as the cores are more optically thick) that are affected by stronger radial temperature variations.

The cores are often modeled observationally as Gaussian sources (see e.g., \citet{Netterfield2009} and \citet{Olmi2009}). 
To facilitate comparison of observations to my models, I approximate the surface brightness
at $250~\mu{\rm m}$ by a two-dimensional Gaussian function. A Gaussian function is 
broader than the profile of the column density, but 
as can be seen in Fig.~\ref{fig_profiles}, a Gaussian function can be a good representation of the true emission profile
as in case of cores embedded in filaments with $A_V<8~{\rm mag}$ that
are confined by the mean ISM pressure.
The FWHM of the Gaussian function is a good estimate of the physical radius of the core (Fig.~\ref{fig_fwhm}) albeit too large by $10\%$.
The Gaussian approximation becomes worse in high pressure regions or for highly embedded cores where the cores are more compact. The error in the size estimate based on the FWHM increases therefore towards  lower dust temperatures. The Gaussian approximation for highly embedded cores is too narrow, overestimating the flux passing through the core center and inferring a FWHM considerably larger than the intrinsic core radius (Fig.~\ref{fig_fwhm}). The estimated flux using a Gaussian approximation is slightly too high (6\% to 12\% overestimate) depending on the external and internal pressure of the filament. The flux estimate tends to be more accurate in the lower pressure regions of the ISM.

\begin{figure*}[htbp] 
%
	\includegraphics[width=0.49\hsize]{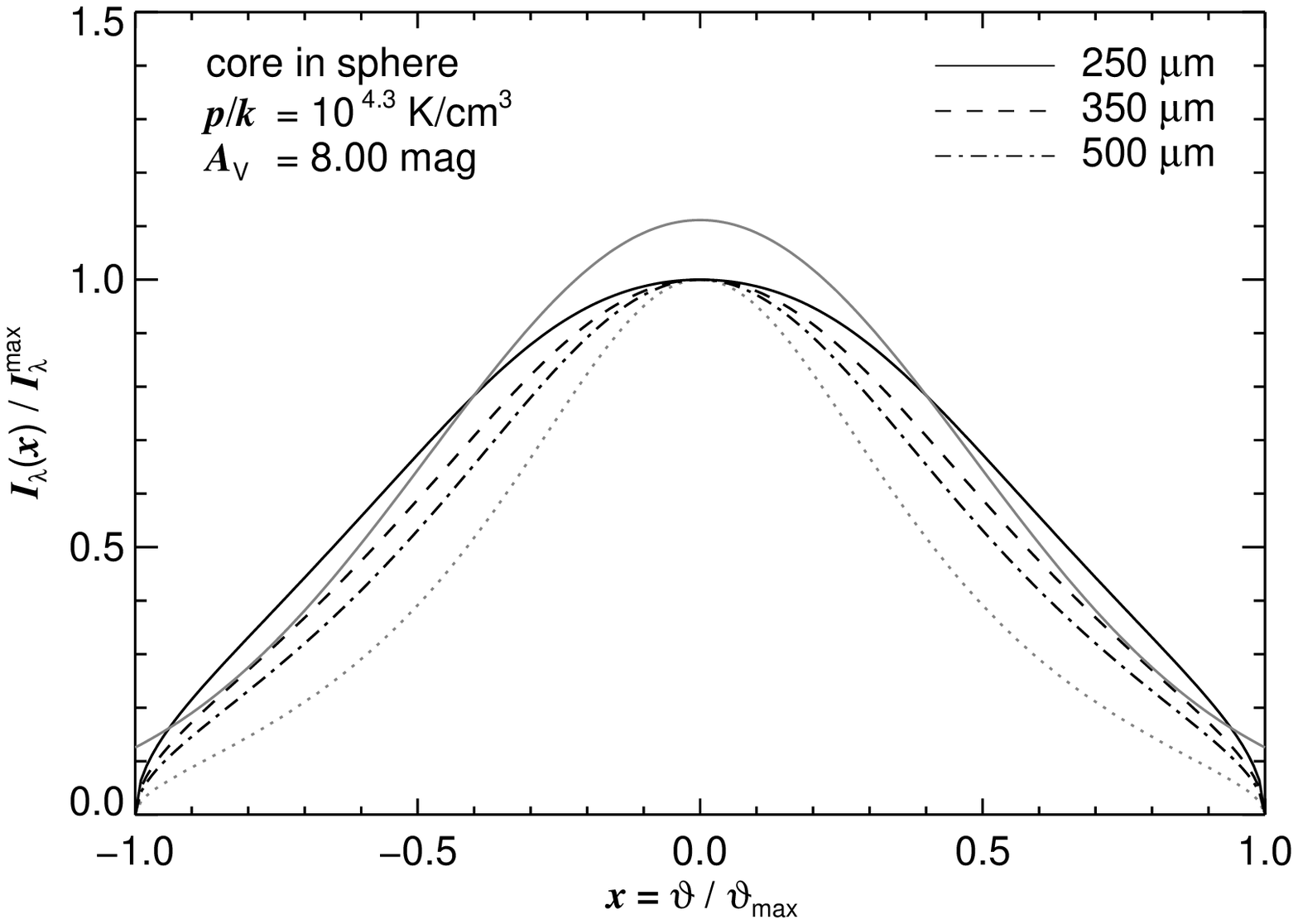}
	\hfill
	\includegraphics[width=0.49\hsize]{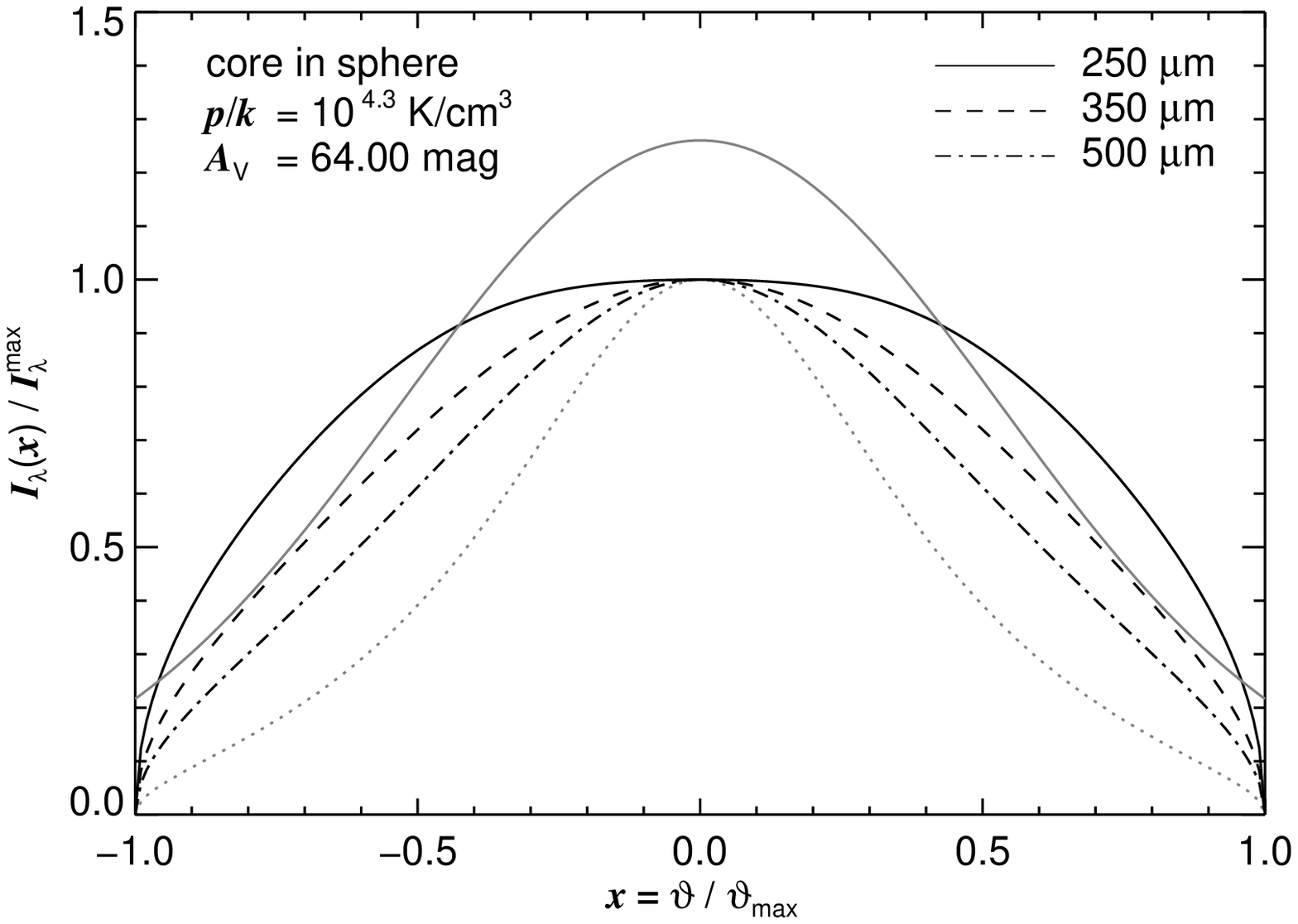}

	\includegraphics[width=0.49\hsize]{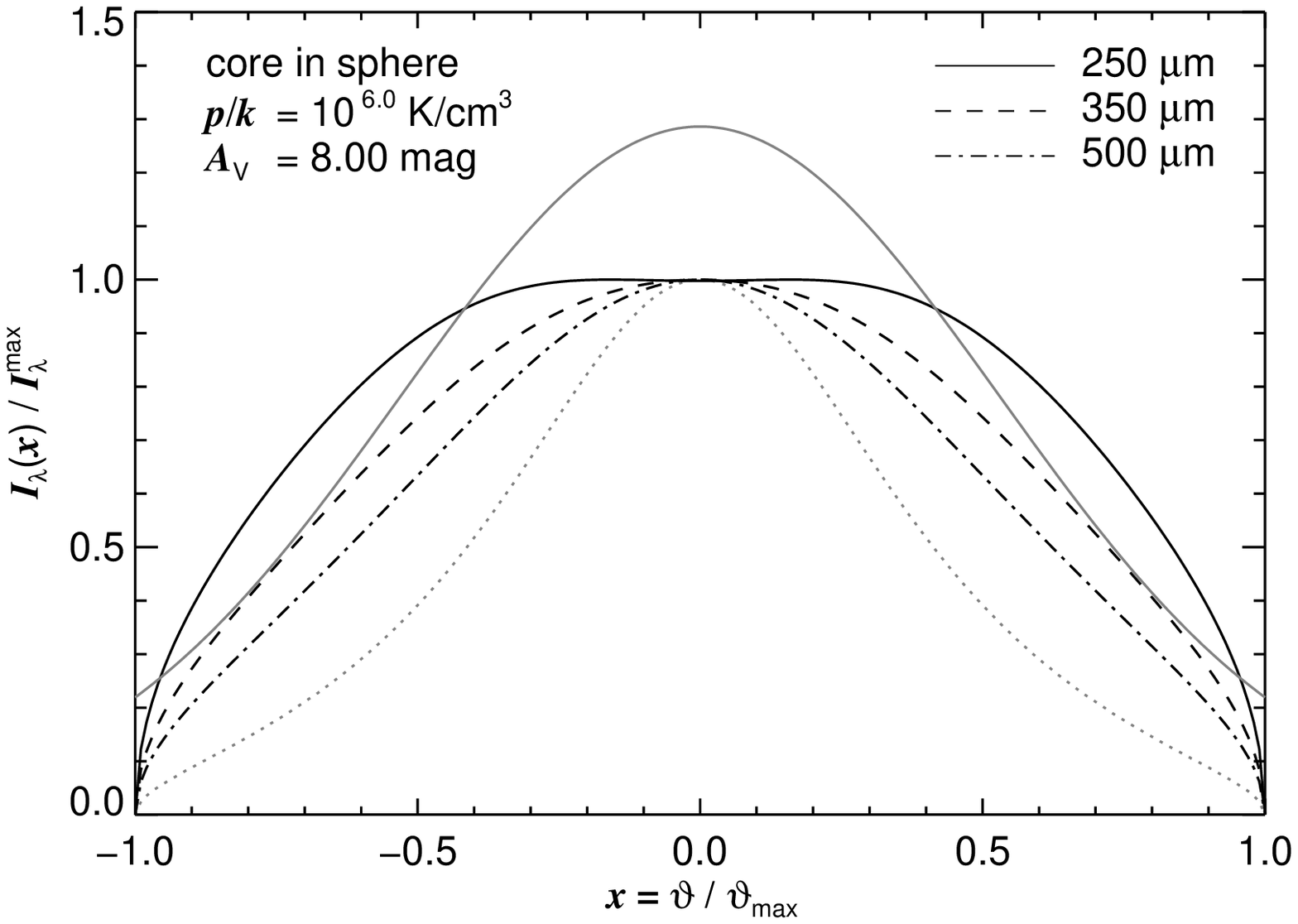}
	\hfill
	\includegraphics[width=0.49\hsize]{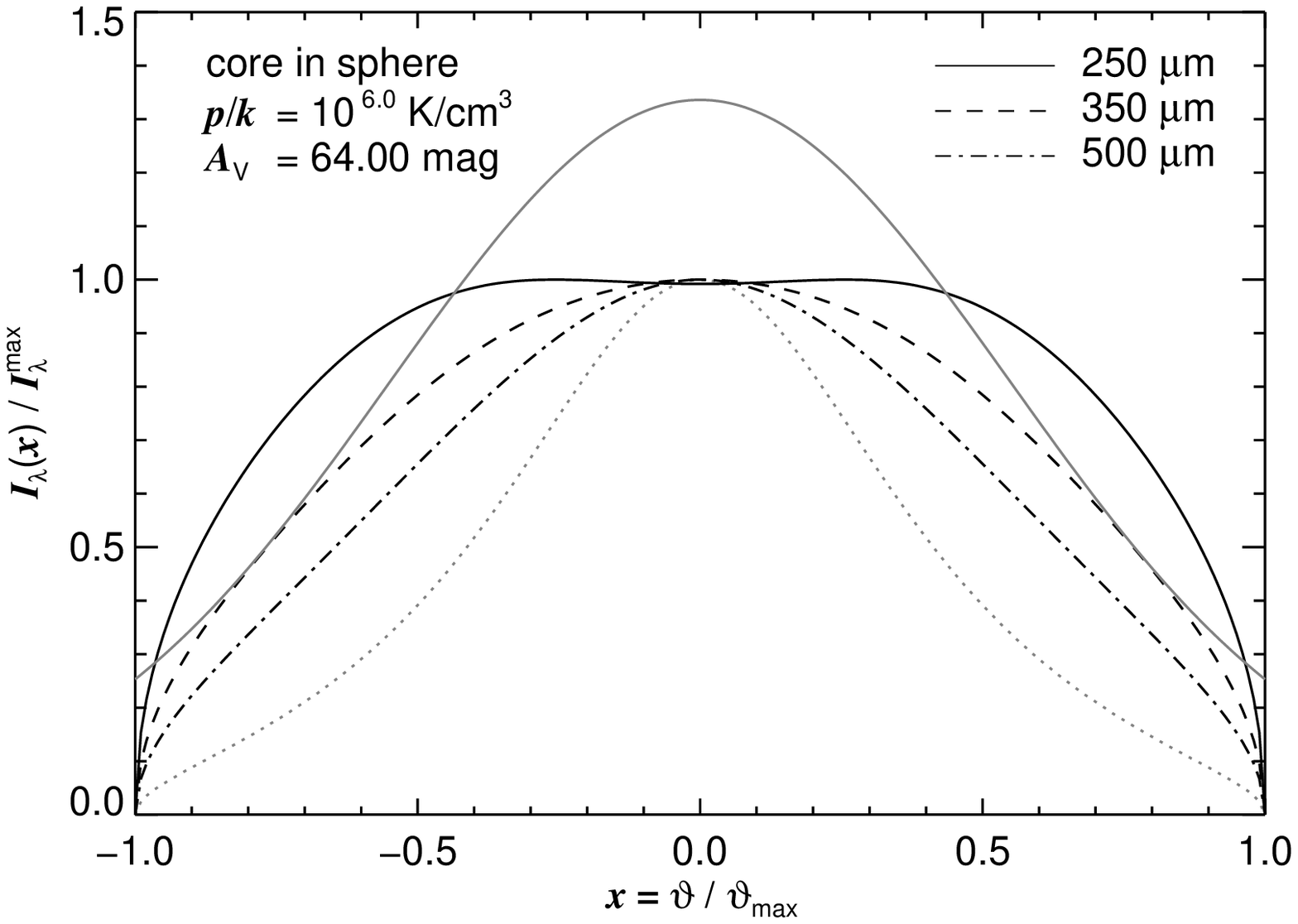}
	\caption{\label{fig_profiles}
	Brightness profiles at $250$, $350$, and $500~\mu{\rm m}$. They are compared with the profile of the column
	density (dotted line). The surface brightness at $250~\mu{\rm m}$ is fitted to a Gaussian source (grey line).
	The filaments are either confined by a medium with $p_{\rm ext}/k=2\times 10^4~{\rm K/cm^3}$ (top) or
	$10^6~{\rm K/cm^3}$ (bottom). The extinction through the filaments' center is either
	$A_V=8~{\rm mag}$ or $A_V=64~{\rm mag}$ (left-hand figure and right-hand figure, respectively).
	}
\end{figure*}

\begin{figure}[htbp]
	\includegraphics[width=\hsize]{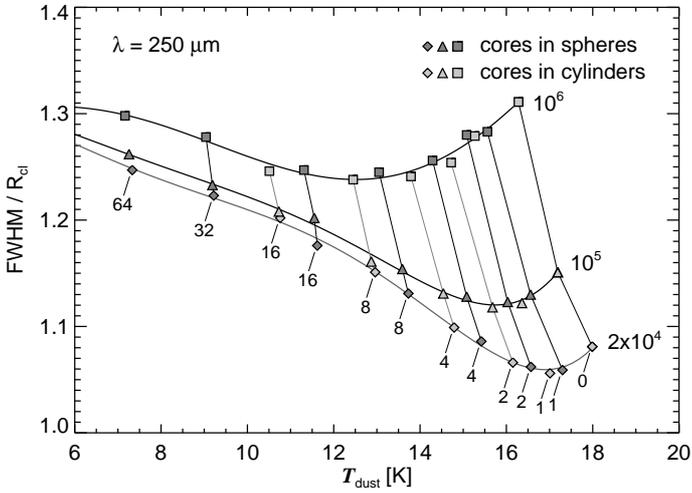}
	\caption{\label{fig_fwhm}
	FWHM at $\lambda=250~\mu{\rm m}$ of a Gaussian representation of the surface brightness profile
	of cores which are either embedded in spherical or cylindrical filaments. The data points for each assumed
	external pressure are fitted with a polynomial function shown as solid lines.
	}
\end{figure}

\section{\label{sect_discussion}Discussion}

My model connects the effective dust temperature to the physical properties of passively heated 
embedded cores. In particular, it establishes a relation between the dust temperature and the pressure of the medium surrounding the cores. The details of the relationship are 
determined by the ISRF, the dust properties in the cores, and the external pressure confining the filaments but is independent of the actual mass of the cores.  This model can be directly tested by
high resolution observations in the FIR/submm regime, which allow apart from measuring the core mass by modeling the dust emission spectrum also a determination of the size of the core and by applying the mass-radius relation (Eq.~\ref{eq_massradius}) the pressure around the core.

The model relates essential core parameters such as the $FWHM$, the emissivity ($\beta$ and $\kappa_0$), and the
luminosity-mass ratio to the pressure of the medium surrounding the filaments and the extinction through the filaments' center,
which corresponds (for a given filament shape) to a certain central pressure. The values of the  parameters are found to be (for a given external pressure) controlled mainly by the dust temperature, which allows in principle a simple correction for radiative
transfer effects.
Future work needs to determine how shapes of the ISRF, 
the dust properties, and the structure of the molecular clouds caused by the turbulent motion affect the core models, allowing a clearer understanding of the uncertainties associated with the measurement of core masses. The expected basic effects
are discussed in Sects.~\ref{sect_discussion1}, \ref{sect_discussion2}, and \ref{sect_discussion3}. A additional complication 
is caused by the coupling of gas and dust in high density environment, although the effect might be less important if the cores
are mainly supported by turbulent motion.


\subsection{\label{sect_discussionemissivity}The origin of the effective dust emissivity}

The emissivities of my models differ
considerably from the mean properties derived in Sect.~\ref{sect_dustapprox}. There are two major reasons why the dust emissivity seems to be less steep and
also lower. First, the variation in dust temperatures with size and composition at each radial distance in the core and 
second the smooth variation in the dust temperatures with position in the core.

Even without additional radiative transfer effects the parameters of the emissivity have been shown (Sect.~\ref{sect_mbb}) deviate from the intrinsic mean values. The reason is that the emission behavior depends on both grain size and grain composition. The mass of the dust grains is contained in the large grains, which are also the coldest with temperatures much below 19~K. The warm dust temperature is caused by the smallest dust grains, which do not contain much mass but efficiently absorb interstellar radiation. As a result their emission dominates the blue side of the SED and, to a large extend, its peak. At larger wavelengths, the 
overall dust emission is gradually enhanced by the emission of colder and larger grains, which flattens
the emission profile. To compensate 
for the warm dust and the low mass related to the dust emission, the emissivity is reduced to obtain the true dust mass.

A similar effect is caused by the gradually changing dust temperature with radius. The emission from warm dust grains
located at large cloud radii dominate the bulk of the SED and determine to a large part the position of the peak of the dust emission and therefore the effective dust temperature. This is most clearly demonstrated by the similar effective temperature shown in 
Fig.~\ref{fig_ptreldata} and the
dust temperature at the cores' edge as shown in Fig.~\ref{fig_heatingratio}. At longer wavelengths, the integrated dust emission
spectrum is gradually more affected by colder dust grains in the interior of the cloud. 
The shape of the dust re-emission overall 
appears to be flatter than expected for the underlying dust emissivity. As the dust emission is dominated by the warm dust grains, the 
absolute emissivity needs to be reduced to account also for the emission of cold dust grains. As seen, the effect can 
be larger than a factor of two depending on the radiation field heating the core and how opaque the core actually is. 

As seen, the effect of the temperature distribution inside the core on the apparent emissivity
increases for a given central extinction of the filaments with external pressure and (for given external pressure $p_{\rm ext}/k<10^{6}~{\rm K/cm^3}$) with extinction of the filaments until the filaments become highly opaque with $A_V>16~{\rm mag}$. The steepening of the apparent emissivity for embedded cores in filaments with $A_V>16$ is caused by an overall reduction in the temperature variation inside the core both in terms of the location as seen in Fig.~\ref{fig_heatingratio} and in terms of grain composition and grain size that occurs for of highly reddened radiation fields. The reason for the latter phenomenon 
is given in Sect.~\ref{sect_sed_grtemp}. The reduced radial temperature variation as seen in Fig.~\ref{fig_heatingratio} is caused by the additional heating of the grains in the core center by the dust emission from the filament. 
A similar effect is expected if the dust emission from the Galactic disc is considered.

A physical limit exists for embedded cores that become predominantly heated by the CMB background radiation. In this situation, the dust in the core approaches the temperature of the CMB independently of the position and composition or size and the apparent emissivity becomes equal to the intrinsic dust emissivity. 

The temperature variation is probably also the explanation of the flatter emissivity observed in other astrophysical objects such as
galaxies for example \citep{Dunne2001,Chakrabarti2008}. However, in the literature the possibility of a dependence 
of the emissivity on the dust properties is explored (see discussion by \citet{Dunne2001}).

\subsection{Basic model uncertainties}
\subsubsection{\label{sect_discussion1}Uncertainties in the ISRF}

\begin{figure}[htbp]
	\includegraphics[width=\hsize]{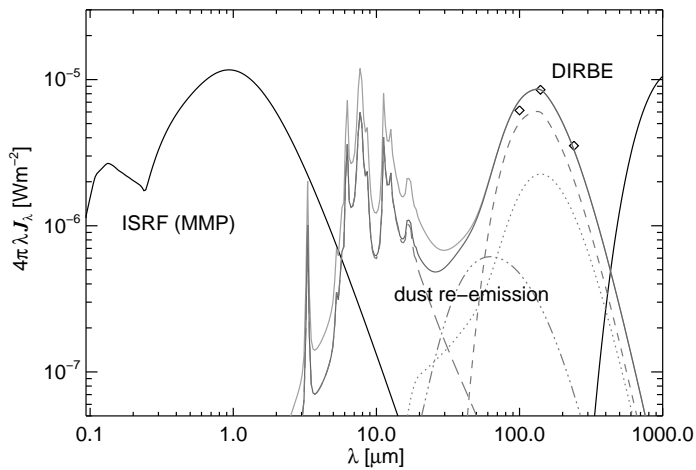}
	\caption{\label{fig_jmean_isrf}
	Mean interstellar radiation field assumed to model the SEDs of embedded cores. The solid black line is
	the ISRF caused by the stellar light as provided by \citet{Mathis1982}. The grey line shows the assumed angle-averaged 
	dust re-emission. The contribution of the PAH emission is reduced relative to the
	dust re-emission of the diffuse ISM  (Fischera, in prep.) shown as light grey curve. The strength of the radiation
	field is scaled to the DIRBE flux at $140~\mu{\rm m}$. The DIRBE fluxes at 100, 140, and $240~\mu{\rm m}$ 
	are shown as open diamonds. The dust emission is divided into its several components: PAH (long dashed line),
	graphite (short dashed line), silicates (dotted line), and iron (dashed-dotted line).
	}
\end{figure}

The $ISRF$ applied in the model is a canonical approach to calculate the dust re-emission from dust grains located in the ISM.
However, a large fraction of the radiation in the galactic disc is caused by dust grains and PAH molecules, which efficiently
absorb the light in the UV and optical light.
For comparison with the results presented in this paper, I considered an $ISRF$
where the dust re-emission from the Galaxy is taken into account. The most important contribution arises from the PAH
emission and possibly small stochastically heated grains. 
In general, this problem needs to be addressed more quantitatively based on all sky surveys such as the DIRBE experiment. 
The determination of the corresponding mean intensity inside the Galaxy is 
complicated by the spectral range being dominated by the zodiacal light. 
The approach presented here is qualitative, but demonstrates where the heating by dust grains and PAH molecules becomes important.

The mean intensity of the dust emission consists primarily of three components: emission from the diffuse gas, HII-regions, and dense clouds (which potentially contain a large percentage of the interstellar material considering that probably up to 30\% of hydrogen is molecular \citep{Ferriere2001}). One of the main differences between the three components is the strength of the PAH emission. In dense clouds, it is likely that the PAH molecules are condensed onto grains. If present, the PAH molecules would be heated by highly attenuated optical and UV radiation, which would reduce considerably the PAH emission relative to the dust emission peak. Additional PAH emission and warm dust emission (around 60 K) arises from HII-regions. I assume that the average PAH emission is lower than the PAH emission from the diffuse phase by a 
factor of 2. Otherwise the dust re-emission is assumed to be the same as in the diffuse ISM. I use the same dust properties as in the ray-tracing calculations. The SED of the dust emission is derived assuming a non-attenuated ISRF.
The strength of the dust emission is obtained by scaling the emission at $140~\mu{\rm m}$ to the corresponding mean intensity of the DIRBE experiment ($\nu J_{\nu} = 6.78 \times 10^{-7}~{\rm W/m^2}$).  The assumed total SED of the mean ISRF is displayed in Fig.~\ref{fig_jmean_isrf}. I derived the DIRBE fluxes shown in the figure using all sky DIRBE data from which the zodiacal light had been subtracted using a physical model. The colder 
dust emission from compact clouds relative to the emission from diffuse dust grains has only a small effect on the heating of the cores and is ignored. I also neglect the warm dust component from HII-regions. In this model (Fig.~\ref{fig_jmean_isrf}), 34\% of the Galactic mean intensity in the solar neighborhood is caused by dust grains and PAH molecules.

\begin{figure}[htbp]
	\includegraphics[width=\hsize]{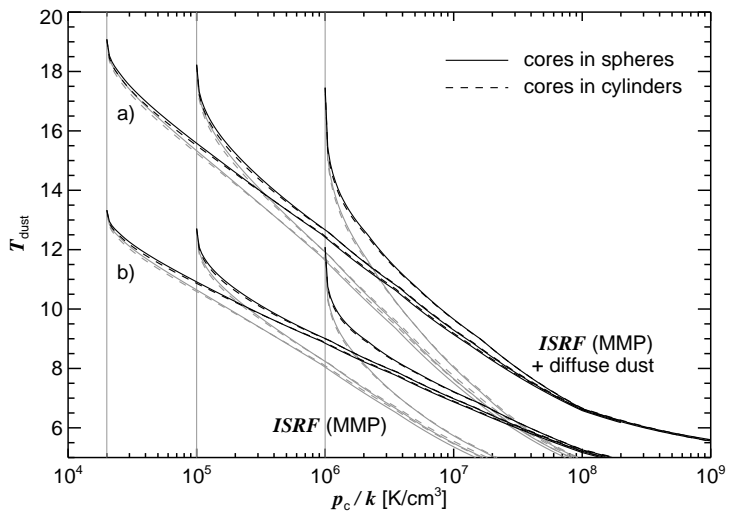}
	\caption{\label{fig_ptrel}
	Effective temperatures of embedded cores as a function of the central pressure in the filament, which is assumed to be either a self-gravitating sphere or a self-gravitating cylinder. In case $a)$, the emissivity is assumed to be the same as in the diffuse dust model, while in $b)$ the emissivity is increased by a factor 8. The external pressure outside of the filaments is assumed to be $2\times 10^4$, $10^5$, and $10^6~{\rm K/cm^3}$. The filaments are heated either by the 
	ISRF as provided by \citet{Mathis1982} (grey lines) or by an ISRF where the diffuse dust emission is added.	
	}
\end{figure}

In Fig.~\ref{fig_jmean_isrf}, the PAH emission of the assumed ISRF is clearly  stronger (despite it being fainter than the diffuse dust emission) than the PAH emission at the center of the filaments 
(Fig.~\ref{fig_jmean_ctr}). This implies that, even if PAH emission arises from the filaments, it is less important for heating the dust in the core than the Galactic IR emission. 
The result for the modified $ISRF$ is shown in Fig.~\ref{fig_ptrel}. The additional heating by IR photons becomes important only for highly embedded cores where the central extinction of the filaments is larger than $A_V\sim 8~{\rm mag}$. This additional heating is more important for the inner core regions, which lead to a steeper brightness profile.

\subsubsection{\label{sect_discussion2}Uncertainties in the emissivity}

The results presented in this paper depend to a certain degree on the dust model, which determines both the position of the peak of the dust re-emission spectrum and its shape. A different dust model will therefore lead to different effective dust parameters being
assigned to the derived dust emission spectra for embedded cores. This will also be true if the results are compared 
with calculations based on the dust model of \citet{Weingartner2001}, which has a different assumed dust composition and size distribution. However, since both models were constructed to reproduce the observed properties in the diffuse
ISM the differences should be small relative to the large uncertainties in the dust properties of the dense phase. 

The calculations are based on the dust properties of the diffuse medium merely as a reference point. 
Grains in the dense phase are likely to form icy layers that will, as a consequence, increase the sticking coefficient
and therefore accelerate the grain growth process by means of grain coagulation (see discussion in \citet{Fischera2008}). 
The process will naturally
alter the size distribution of the grains by removing the number of small grains while increasing the mass of the largest dust particles. This will result in a rather flat slope of the extinction curve in the UV, which is generally determined by the abundance of small grains. Its effect on the optical properties in the far infrared is uncertain but the emissivity may be considerably higher. 
According to \citet{Kruegel1994}, a factor of 10 higher emissivity is physically possible depending on the properties of the composites. In addition, the grain growth models presented by \citet{Ossenkopf1994} depend on the density inside the clouds. If correct, a higher emissivity is expected in denser clouds.
The mass estimate of cores is observationally based on an emissivity at $250~\mu{\rm m}$ of $\kappa_0=10~{\rm cm^2/g}$ \citep{Netterfield2009,Olmi2009} with $\beta=2$ and a gas-to-dust mass-ratio of $100$. Despite this relatively low gas-to-dust mass ratio, the cores appear to be far less massive than they would be if the masses 
were based on the emissivities derived in this paper. 
 
Figure \ref{fig_ptrel} shows a first-order effect of a higher emissivity on the core temperature. It is assumed that the spectral index $\beta$ has the same dependence on the pressure. In this case, the core temperature varies as $T_{\rm dust}^{(2)}=T_{\rm dust}^{(1)}(\kappa_{0,1}/\kappa_{0,2})^{1/(4+\beta)}$. 
In the special case where the 
emissivity of the dust grains has increased by a factor 8, the absorption coefficient above $60~\mu{\rm m}$ is scaled by the same factor.
Between $20~\mu{\rm m}$ and $60~\mu{\rm m}$, a smooth transition is made. The pressure $p_{\rm c}$ in the filaments obtained by measuring the mass by modeling the SED and the size of the core varies as $p_{\rm c}^{(2)} = p_{\rm c}^{(1)}(\kappa_{0,1}/\kappa_{0,2})^2$ and therefore depends strongly on the emissivity. Too low emissivities will produce 
too high pressure values. The 
uncertainty in the emissivity affects observationally only the mass estimate and the pressure. The $P-T$-relation 
might therefore be helpful to estimate the dust emissivity in cores.

A first-order correction to the dependence of $\beta$, $\kappa_0$, $L/M_{\rm core}$, and $FWHM$ on $T_{\rm dust}$ is
given by replacing the dust temperature with $T_{\rm dust}^{(2)}$. For the emissivities $\kappa_0$, the values have to be corrected by the corresponding factor. However, the dependence on dust temperature needs to be analyzed
quantitatively using more realistic assumptions about the dust properties inside the clouds.

\subsection{\label{sect_discussion3}Effect caused by the inhomogeneous medium}

As mentioned in the introduction, molecular clouds are turbulent, which produces a strong variation in the local density and as a consequence a variation in the column density seen through the molecular clouds (e.g. \citet{Kauffmann2010}). The effect of turbulence was studied in some detail by \citet{Fischera2004a, Fischera2005a} and discussed also by \citet{Fischera2008}, where
a method is described that considers how the turbulent density structure in radiative transfer studies through self-gravitating clouds can be taken into account. \citet{Fischera2005a} illustrated how the dust attenuation can be directly related 
to the physical condition of the turbulent medium. A statistical treatment of the radiative transfer through the clouds showing a smoothly varying density fluctuations as expected in turbulent clouds was presented by \citet{Hegmann2003}. 

Despite a number of studies about the effect of an inhomogeneous medium on the radiative transfer \citep{Boisse1990,Witt1996,Wolf1998,Varosi1999,Witt2000,Gordon2001,Hegmann2003,Doty2005},
the importance for the interpretation of infrared observations is still uncertain. Although similar in nature, not all studies can be directly applied to the problem of externally heated molecular clouds. The problem generally addressed in studies about the
radiative transfer through molecular clouds is the effect of the density structure on the dust scattering and dust absorption without considering the dust re-emission \citep{Boisse1990,Witt1996,Wolf1998,Witt2000,Hegmann2003}. While it is one of the main complications of radiative transfer calculations, \citet{Doty2005} stated (based on test calculations) that dust scattering plays only a very small role in the distributions of dust temperatures in the majority of their sources. They indeed ignored this problem altogether in their study of dust temperatures in molecular clouds. 

Because of the complexity of the problem, the model 
calculations have in most cases been based on rather simplified assumptions about the medium.  In general, a two-phase model is considered where randomly distributed dense `clumps' are embedded in an otherwise homogeneous thin medium \citep{Boisse1990,Witt1996,Wolf1998,Witt2000,Gordon2001,Doty2005}. The inhomogeneous medium is typically described
by the filling factor of the `clumps', their size relative to the extension of the whole cloud, the density ratio of
the `clumps' and the `inter-clump' medium, and in addition by the optical depth of the cloud. The application to
observations is complicated by neither the medium as a whole nor the individual clumps being described as physical entities with realistic density profiles. 

To some extent, the models of individual dense `clumps'  are comparable to the dense self-gravitating cores in my model of embedded cores. From  a physical standpoint, it seems, however, more realistic that they form in the inner center of the filaments where they experience the highest pressure rather than to place them randomly inside the filament.  
Because of the pressure profile, it is also expected that cores located closer to the cloud edges are less optically thick than 
at the center.
My model is less applicable to cores formed in a dense cluster where shadow effects by individual cluster members become important. Although this situation is often considered in two phase models by applying a high filling factor of the 'clumps', it becomes observationally difficult to separate the individual sources. 

The most important characteristic caused by in-homogeneities is that a non-homogenous dusty medium is more
transparent than an inhomogeneous medium
 \citep{Boisse1990,Hegmann2003,Doty2005,Fischera2003,Fischera2005a}. The effect disappears when the density enhancements become optically thin. 
For typical molecular clouds ($A_V\sim8~{\rm mag}$), the increase in transparency is highest in the UV and optical and is negligible in the infrared.
The effect leads therefore for a homogeneous extended source seen through a turbulent screen to a flattening of the
attenuation curve \citep{Fischera2005a}. This flattening is characterized by a large absolute-to-relative extinction, the $R_V$-value being defined as $A_V/(A_B-A_V)$, which is close to 3.1 in the diffuse ISM. The same effect is expected for the turbulent interstellar medium and \citet{Fischera2003} demonstrated that the turbulence can naturally explain the flatter curvature of the Calzetti extinction curve derived for local starburst galaxies.
Because of the inhomogeneities, the light most responsible for heating the gas and the dust can therefore penetrate
deeper into the filaments and consequently into the core. On the other hand, the total light absorbed inside the core 
is lower leading effectively to a lower dust temperature, solely by considering the energy balance in the core.

In the framework of the model, the SEDs of the cores are independent of the effective temperature, which determines for a given pressure  the mass and the size of the cores inside the filaments. If the cores are supported predominantly by turbulent pressure, more massive cores are also more turbulent, which creates a stronger variation in the local density around the mean. Their one-point statistic of the local density is described by a broader probability distribution approximately described by a log-normal function.
It is therefore expected that in a certain pressure region of a GMC, the effect of the turbulent density structure is stronger
for more massive cores than for less massive ones.

\section{\label{sect_summary}Summary and conclusions}

I have presented a physical model for the dust emission of condensed cores where the cores are described by critical stable self-gravitating clouds embedded at the center of self-gravitating filaments. The filaments are assumed to be either spherical or cylindrical in shape. For a set of dust properties and an external isotropic radiation field, the radiative transfer problem is determined by the external pressure confining the filaments and the extinction of the filaments. In the model, the extinction of either a sphere or a cylinder is directly related to the overpressure in the filament center and the external pressure. 

The problem has been solved using a ray-tracing technique where the effects of multiple scattering events and the anisotropy of the scattered light are accurately taken into account. The dust emission is derived using a physical dust model of stochastically heated grains of different sizes and compositions. The work focuses on the radiative transfer effects and is therefore based on the dust properties of diffuse dust grains.
The mean extinction at FIR wavelengths is approximately described by a simple power law with $\beta^{\rm ext}=2.024$ and an absolute extinction at $\lambda_0=250~\mu{\rm m}$ of $\kappa^{\rm ext}({\lambda_0})=5.232~{\rm cm^2/g}$. 

The variation in the dust temperatures inside embedded cores is found to depend on the external pressure and
the extinction of the filaments. The variation is not necessarily smaller than for non-embedded cores as stated by
\citet{Stamatellos2003}.

For comparison with observations, the bulk of the dust re-emission is approximated by a modified black-body spectrum using $\beta$ and $\kappa_0$ of the emissivity $\kappa^{\rm em}_\lambda=\kappa_0(\lambda/\lambda_0)^{-\beta}$ for
$\lambda_0=250~\mu{\rm m}$ and the dust temperature $T_{\rm dust}$ as free parameters.
As shown, the effective parameters of the emissivity can be applied using a single-zone model for the cores. This simplified model reproduces, to first order, the dust temperature of the cores.
On the basis of the model assumptions of self-gravitating clouds, the dust temperature of passively heated cores is independent of the mass of the cores.
It is found that:
\begin{itemize} 
	\item  The effective emissivity $\kappa_0$ is considerably lower than the intrinsic extinction. For filament 
	extinction $A_V<16~{\rm mag}$, the values of $\kappa_0$ are typically a factor 2 to 3 lower than expected
	using the mean	 dust properties in the cloud.
	\item The emissivity is flatter with $\beta<1.8$. For the mean value of ISM pressure and $A_V<16$, I found that 
		$1.52<\beta<1.74$.
	\item The effect on $\kappa_0$ and $\beta$ for a given dust temperature increases with external pressure.
	\item In highly embedded cores with $A_V>16$, the emissivity becomes equal to the mean extinction
	value.
\end{itemize}
	
The derived surface brightness is considerably broader than the underlying profile of column density, the effect being most prominent at shorter wavelengths. The brightness profiles become flatter for both more embedded cores and higher pressures of the external medium.
The surface brightness at $250~\mu{\rm m}$ has been compared with a Gaussian source approximation. The agreement is good for less embedded cores, but
is  poor for strongly embedded cores, where the Gaussian profile is steeper than the model. 
It is found that:
\begin{itemize}
	\item The Gaussian source approximation overestimates the total flux by $\sim 10\%$.
	\item In high pressure regions and highly embedded cores, the $FWHM$ of the Gaussian source
		approximation overestimates the size of the core by between 10\% and $30\%$.
\end{itemize}

For direct comparison with observations I have derived, for different ISM pressures, polynomial expressions of
the model relations for $\beta(T_{\rm dust})$, $\kappa_0(T_{\rm dust})$, $L/M_{\rm core}(T_{\rm dust})$, and 
$FWHM(T_{\rm dust})$ at $\lambda=250~\mu{\rm m}$. The relations are found to be only mildly affected by the 
geometry of the filaments.
For highly embedded cores, the work indicates that  the dust emission from the Galactic disc needs to be considered
to model accurately the SED and surface brightness profiles of the cores.

\begin{acknowledgements}
	The author likes to thank in particular Prof. Brian Schmidt for his overly great support and his encouragement. 
	I further like to thank namely Dr. R. Tuffs and Prof. Peter Martin for their support.
	 I like further
	thank CITA where I made most of the numerous radiative transfer calculations presented in this paper.
\end{acknowledgements}

\appendix

\section{\label{sect_modelapprox}Approximate solution for the dust temperature in cores}

In a single temperature model used by observers to interpret their observations and in this paper,
the individual dust grains
are, independently of their location inside the core, heated by the same mean intensity $J_\lambda^{\rm heat}$. 
In the framework of the model, this 
intensity is produced by the radiation field at the center of the filament and by the emission from the dust grains in the
core. That the dust temperature does not strongly depend on the heating flux (it varies as the sixth root of the heating rate) allows approximations that I outline in the following. The aim of the model is to provide an equivalent to the
time-consuming ray-tracing technique. To be widely applicable, the effects caused by scattering and re-heating by dust grains are taken into
consideration.
As in case of the more accurate approach using ray-tracing, the solution is derived in two independent consecutive steps by
estimating first the mean  intensity $J_\lambda(0)$ in the center of the filament and secondly the dust temperature inside the core.

To obtain a most accurate estimate of the attenuated flux caused by the self-gravitating cores and filaments
hence a more reliable estimate of the absorbed energy inside the clouds, I still consider clouds with physical density profiles rather than homogeneous clouds. The difference between  the important optical properties (their attenuation and escape probability) for the two assumptions about the density distribution inside the clouds is discussed in Sect.~\ref{sect_cloud_ext} and \ref{sect_cloud_pesc}, where it is shown that an assumption of a homogeneous cloud will lead to an over-estimate of the absorbed flux.

In both steps, the dust temperature in the corresponding cloud is derived.
To consider the radiative transfer effects caused by the emission from dust grains, it is assumed that the emissivity $\eta_\lambda^{\rm dust}=k^{\rm em}_\lambda\,B_\lambda(T_{\rm dust})$ in both the filament and the core is proportional to the hydrogen density $n_{\rm H}$.

There are in general three different contributions to the intensity $J_\lambda(0)$ at the center of the filament heating the core: (1) the attenuated radiation of the interstellar radiation field illuminating externally the filament, (2) the radiation of the scattered emission, and (3) the dust emission. To derive accurately the last two contributions, a proper radiative transfer treatment is needed. 

The light scattered away from the initial direction is partly compensated for by radiation from other initial directions scattered into it. This leads effectively to a lower extinction towards the cloud center. The approximation where the light is scattered in either a forward or backward direction and where the corresponding fraction is determined by the g-factor, the angle-averaged cosine of the scattered intensity, leads to a lower extinction
\begin{equation}
	\label{eq_redkappa}
	\bar k^{\rm ext}_\lambda = k^{\rm ext}_\lambda\left(1-\frac{1}{2}\omega_\lambda(1+g_\lambda)\right)
\end{equation}
where $\omega_\lambda=k^{\rm sca}_\lambda/k^{\rm ext}_\lambda$ is the dust albedo.
In the UV and optical, the light is strongly scattered in the forward direction with $g_\lambda>0.5$ and $\omega_\lambda>0.5$. This causes the extinction at these wavelengths to be considerably lower than nominal values. The assumption for scattered light
has overall less effect for highly embedded cores where the contribution of the scattered emission to the mean intensity 
inside the filament is weaker.

The dust grains  that provide most of the intensity at the center of an optical thick filament are located at the edge of the filament where the dust temperatures are the warmest. Further inside, the grains are heated by a strongly attenuated radiation field, which shifts the dust emission spectrum to longer wavelengths, where the absorption is also weaker. A conservative lower limit to the dust heating is provided by assuming that all the dust emission arises from a thin shell at the outer surface of the cloud. This provides
\begin{equation}
	J_\lambda^{\rm dust}(0) = 
		\frac{1}{3}\,(\eta_\lambda^{\rm dust}/n_{\rm H}) \frac{1}{2}\left<N_{\rm H}\right>(0)\,e^{-\tau_\lambda^{\rm ext}(R_{\rm cl})}
\end{equation}
for a sphere and
\begin{equation}
	J_\lambda^{\rm dust}(0) = 
		\frac{1}{2}\,(\eta_\lambda^{\rm dust}/n_{\rm H}) \frac{1}{2}\left<N_{\rm H}\right>(0)\,\left<e^{-\tau_\lambda^{\rm ext}}\right>
\end{equation}
for a cylinder, where $\left<N_{\rm H}\right>(0)$ is the corresponding column density of hydrogen at the center of the homogeneous cloud (Fig.~\ref{fig_colpresrelmean}). The mean attenuation is given by
\begin{equation}
	\left<e^{-\tau_\lambda^{\rm ext}}\right> = \int_0^{\pi/2}{\rm d}\vartheta\,\sin\vartheta\,e^{-\tau_\lambda^{\rm ext}(R_{\rm cl})/\sin\vartheta},
\end{equation}
where the optical depth given by
\begin{equation}
	\tau^{\rm ext}_\lambda = \int_0^{R_{\rm cl}}(\bar k_\lambda^{\rm ext}/n_{\rm H})\,n_{\rm H}(r)\,{\rm d}r.
\end{equation}

The dust emission of the filament and the core are calculated using energy conservation between the absorbed and the 
emitted flux.
%
The absorbed luminosity in the cloud caused by the external radiation is given by
\begin{equation}
	L_\lambda^{\rm abs,a} = S_{\rm cl}\pi J_\lambda (1-\exp\{-\tau^{\rm eff}_\lambda(\hat p,\tau_\lambda)\})(1-\omega_\lambda^{\rm cl}(\hat p,\tau_\lambda)),
\end{equation}
where $S_{\rm cl}$ is the cloud surface, $\tau^{\rm eff}_\lambda$ the effective optical depth (see \ref{sect_cloud_ext}), and $\omega^{\rm cl}_\lambda$ the albedo of the cloud (see \ref{sect_cloud_albedo}). The attenuation and the cloud albedo depend (apart from the shape) on the overpressure $\hat p=p_{\rm c}/p_{\rm ext}$
and the radial extinction $\tau_\lambda$. The additional dependence on the non-isotropy of the scattered light is ignored. The luminosity of the dust emission absorbed inside the cloud is obtained using the escape probability, $p_{\rm esc}$ (see~\ref{sect_cloud_pesc}),
\begin{equation}
	L_\lambda^{\rm abs,b} = L_\lambda^{\rm dust}(1-p_{\rm esc}(\hat p,\tau_\lambda))(1-\omega_\lambda^{\rm cl}(\hat p,\tau_\lambda)),
\end{equation}
where $L_\lambda^{\rm dust}=V_{\rm cl}\left<n_{\rm H}\right>\,4\pi \,\eta_\lambda^{\rm dust}/n_{\rm H}$, $V_{\rm cl}$ is the  cloud volume, and $\left<n_{\rm H}\right>$ the mean hydrogen density.
The absorbed luminosity is, on the other hand, equal to 
\begin{equation}
	L_\lambda^{\rm abs} = V_{\rm cl} \left<n_{\rm H}\right>(k_{\lambda} ^{\rm abs}/n_{\rm H})\,4\pi\,J_\lambda^{\rm heat}.
\end{equation}
The external radiation field is therefore reduced to
\begin{equation}
	J_\lambda^{\rm heat,a} = f_\lambda(\hat p,\tau_\lambda)\,J_{\lambda},
\end{equation}
where
\begin{equation}
	f_\lambda(\hat p,\tau_\lambda) = \frac{\left(1-\exp\{\tau^{\rm eff}_{\lambda}(\hat p,\tau_\lambda)\}\right)}{\left<\tau_\lambda\right>}\frac{1-\omega_\lambda^{\rm cl}(\hat p,\tau_\lambda)}{1-\omega_\lambda}
\end{equation}
with the mean extinction of the cloud defined as $\left<\tau_\lambda\right>=4V_{\rm cl}/S_{\rm cl}\,\left<n_{\rm H}\right>k_\lambda^{\rm abs}/n_{\rm H}$.
The intensity produced by the dust grains emission is given by
\begin{equation}
	J_{\lambda}^{\rm heat,b} = \eta_\lambda^{\rm dust}/k_{\lambda}^{\rm abs}\,(1-p_{\rm esc}(\hat p,\tau_\lambda))(1-\omega_\lambda^{\rm cl}(\hat p,\tau_\lambda)).
\end{equation}
The contribution of the thermal dust emission to the total heating rate from dust grains located in the cloud is derived in an iterative manner.

\begin{figure}[htbp]
	\includegraphics[width=\hsize]{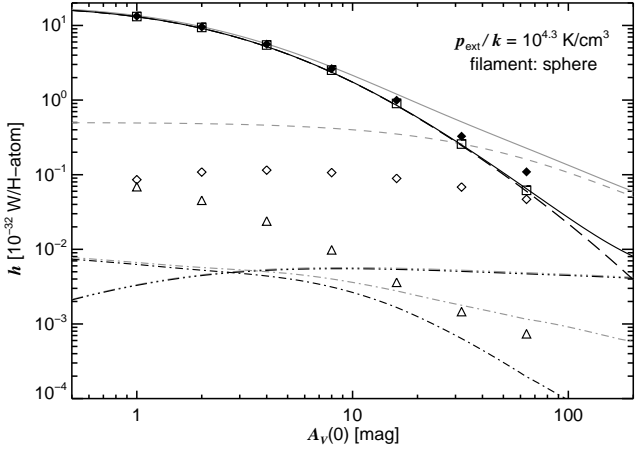}
	\caption{\label{fig_heatapprox}
	Heating rates of the approximation for cores embedded in 
	spherical filaments, given as function of the extinction $A_V(0)$
	through the filament center. The assumed external pressure is $p_{\rm ext}/k=2\times10^4~{\rm K/cm^3}$. 
	The heating rates correspond to the considered $ISRFs$ without (black curves) and with dust emission 
	from the ISM (grey lines).  The heating rates obtained for an $ISRF$ without dust emission from the Galactic disc are
	compared with the accurate ray-tracing calculation (symbols).
	The total heating rates (solid lines, filled diamonds) are separated into the several components of the radiation: 
	attenuated external radiation field, which includes the heating
	by scattered radiation (long dashed lines, open squares); 
	heating by dust emission from the filaments (dashed-three dotted curves, open diamonds);
	heating by dust emission from the cores (dashed-single dotted curves, open triangles); and dust emission 
	from the Galactic disc (short dashed curve). } 
\end{figure}

The contribution of the different components of the radiation considered in the approximation to the total heating rate 
are shown in Fig.~\ref{fig_heatapprox} for cores embedded in spherical filaments, which are pressurized by a medium with $p_{\rm ext}/k=2\times 10^{4}~{\rm K/cm^3}$.
As shown, the simplification for the scattered radiation (Eq.~\ref{eq_redkappa}) leads to highly accurate heating rates 
of the scattered light inside the core.
The heating by the dust emission from the filament and the core is considerably
lower in this approximation because of the absence of PAH emission. The difference is more prominent for the dust heating from the filaments, also because the approximation provides only a lower limit. Overall the heating by self-absorption inside the cores plays only a minor role in the total heating rate. 
The heating caused by the dust emission from the filaments provides a more important contribution, which gradually increases with extinction $A_V$.
On the basis of the ray-tracing calculation, this contribution at $A_V=64~{\rm mag}$  is approximately half of the total heating
rate. However, as the figure suggests, for highly embedded cores the dust emission of the
galactic disc, in particular the emission from PAH molecules, possibly becomes the predominant heating source. 

The effective temperature of the dust emission spectrum is obtained in a procedure often used in radiative transfer problems where it is assumed that grains at a certain location inside the cloud all have the same temperature. The basic difference
here is the replacement of the dust emissivity by its effective value to obtain the correct dust temperature in cores. For the
filaments, the intrinsic dust emissivity is used. The equations are given here for completeness.
The effective temperature is determined by the equilibrium of the heating and cooling rates
\begin{eqnarray}
	\label{eq_equilibrium}
	L^{\rm heat} & = & M_{\rm dust}\int{\rm d}\lambda\, \kappa_\lambda^{\rm abs} 4\pi J_\lambda^{\rm heat}\\\nonumber
	& =& M_{\rm dust}\int{\rm d}\lambda \,\kappa_\lambda^{\rm em} 4\pi B_\lambda(T_{\rm dust})=L^{\rm dust}.
\end{eqnarray}
The cooling rate can be written as
\begin{equation}
	L^{\rm dust} = M_{\rm dust}\,4\sigma_{\rm SB}\left<\kappa^{\rm em}(T_{\rm dust})\right>\,T_{\rm dust}^4,
\end{equation}
where $\sigma_{\rm SB}$ is the Stefan-Boltzmann constant. 
The Planck average emission coefficient is given, for a power law ($\kappa^{\rm em}_\lambda=\kappa_0(\lambda/\lambda_0)^{-\beta}$), by
\begin{equation}
	\label{eq_qpl}
	\left<\kappa^{\rm em}(T_{\rm dust})\right>=\frac{\kappa_0}{\lambda_0^{-\beta}}
		\frac{\Gamma(4+\beta)\zeta(4+\beta)}{\Gamma(4)\zeta(4)}\,\left(\frac{k\,T_{\rm dust}}{h\,c}\right)^{\beta}
\end{equation}
where $k$, $h$, and $c$ are the Boltzmann constant, the Planck constant, and the velocity of light in the vacuum and where $\Gamma(x)$  and $\zeta(x)$ are the $\Gamma$-function and the $\zeta$-function.

\section{Effective dust properties of the clouds}

In the approximation of the radiative transfer, the absorbed flux inside the core or the filament is derived using an effective
value of the dust extinction and an albedo that is based on escape probabilities. To obtain a more accurate estimate of the dust temperature, the values are consistently based on the physical density profile of the clouds rather than assuming that the dusty medium is homogeneously distributed within the given cloud volume. The column density related to the corresponding homogeneous sphere and cylinder is given in Fig.~\ref{fig_colpresrelmean}. In the limit of high overpressure ${\hat p=p_{\rm c}/p_{\rm ext}}$, the mean column densities through spheres and cylinders are given approximately by
\begin{eqnarray}
	\left<N_{\rm H}\right>(0) &\sim & \sqrt{\frac{p_{\rm ext}/k}{2\times 10^4~{\rm K/cm^3}}}\nonumber\\
		&&\times \left\{\begin{array}{lc}
		6.50\times 10^{21} \, \hat p^0\,{\rm cm^{-2}} & {\rm (sphere)},\\
		4.33\times 10^{21} \, \hat p^{1/4}\,{\rm cm^{-2}} & {\rm (cylinder)}.
	\end{array}\right.
\end{eqnarray}

\begin{figure}[htbp]
	\includegraphics[width=\hsize]{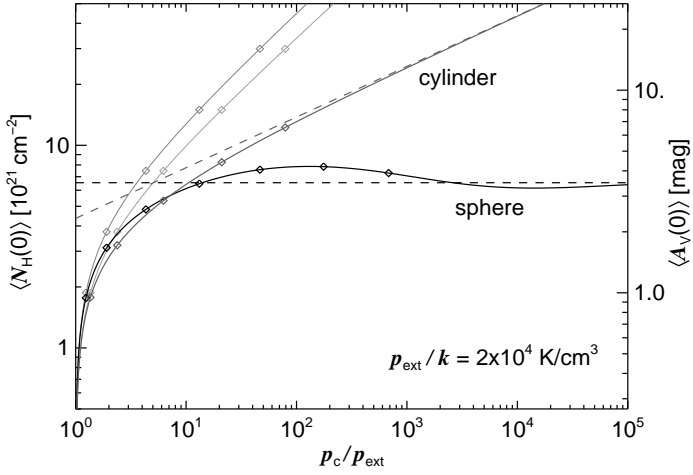}
	\caption{\label{fig_colpresrelmean}
	Column density at the center of a self-gravitating sphere or a cylinder as a function of the over pressure 
	$\hat p=p_{\rm c}/p_{\rm ext}$, where the mass is homogeneously distributed within the cloud radius. 
	The column density and the extinction are given for an external pressure
	 $p_{\rm ext}=2\times 10^4{\rm K/cm^3}$. The asymptotic behavior is shown as dashed lines. 
	 The symbols mark the positions
	of the accurate radiative transfer calculations. For comparison also the true column densities at the 
	cloud centers  is shown (light grey curves).
	}
\end{figure}

\subsection{\label{sect_cloud_ext}Attenuation}

The total attenuation described in terms of an effective optical depth caused by the cloud is given by the ratio of the flux density entering the cloud from outside by an isotropic radiation field to the flux density leaving the cloud without being scattered and absorbed. The effective optical depth is therefore given by
\begin{equation}
	\tau_\lambda^{\rm eff} = -\ln\left\{\frac{F_\lambda^{-}}{F_{\lambda}^+}\right\} = -\ln\left\{\frac{1}{\pi} \int{\rm d}\Omega\,\vec n_F \cdot \vec n_{J}
	(\Omega)\,
	e^{-{\tau_\lambda(\Omega)}}\right\},
\end{equation}
where $\vec n_F$ is the normal vector to the surface element the flux density is derived for, $\vec n_J$ is the vector of the 
photons illuminating the surface element, and $\tau_\lambda(\Omega)$ the total optical depth in direction $\Omega$.
For a self-gravitating cloud, the attenuation depends (apart from the actual shape) on the density profile determined by the overpressure $\hat p=p_{\rm c}/p_{\rm ext}$ and the radial optical depth $\tau_\lambda$.
For the sphere, the effective optical depth is
\begin{equation}
	\tau^{\rm eff}_\lambda(\hat p,\tau_\lambda) = -\ln\left\{2\int\limits_0^{\pi/2}{\rm d}\vartheta\,\sin\vartheta \cos{\vartheta}\,
	e^{-\tau_\lambda(\sin\vartheta)}\right\}
\end{equation}
and for the cylinder
\begin{eqnarray}
	\tau^{\rm eff}_\lambda(\hat p,\tau_\lambda) &=& -\ln\Bigg\{2\int\limits_0^{\pi/2}{\rm d}\vartheta\,\sin\vartheta\nonumber\\
	&&\times \int\limits_0^{\pi/2}{\rm d}\varphi\,\sin{\vartheta}\,\cos{\varphi}\,
	e^{-\tau_\lambda(\sin\varphi)/\sin(\vartheta)}\Bigg\},
\end{eqnarray}
where the optical depth in the exponent is given by
\begin{equation}
	\tau_\lambda(\sin\varphi) = R_{\rm cl}\,k_\lambda^{\rm ext}/n_{\rm H} \int\limits_{-\cos\varphi}^{\cos\varphi}{\rm d}l\,n_{\rm H}\left(R_{\rm cl}\sqrt{l^2+\sin^2\varphi}\right).
\end{equation}
Since for self-gravitating clouds, $R_{\rm cl}\propto T/\sqrt{p_{\rm ext}}$ and $n_{\rm H}\propto p_{\rm ext}/T$, where $T$ is the effective temperature of the gas in the cloud that combines the different velocity components the extinction depends for a given overpressure only on the square root of the external pressure ($\tau_\lambda\propto \sqrt{p_{\rm ext}}$).

At low optical depths, the effective optical depth becomes equal to the mean optical depth
\begin{equation}
	\tau_\lambda^{\rm eff}(\hat p, \tau_\lambda\ll 1) = \left<\tau_\lambda\right>=4(V_{\rm cl}/S_{\rm cl})\left<n_{\rm H}\right>(k_\lambda^{\rm ext}/n_{\rm H}).
\end{equation}
In case of highly opaque clouds, most of the light contributing to the flux density $F_\lambda^-$ arises from the rim where the extinction is lowest. The optical depth close to the edge is $\tau_\lambda\approx R_{\rm cl}\,(k_{\lambda}^{\rm ext}/n_{\rm H})\,n_{\rm H}(R_{\rm cl})\,2(\pi/2-\varphi)$. By considering only the contributions close to the rim, the effective optical depth of a sphere becomes
\begin{eqnarray}
	\tau_\lambda^{\rm eff}(\hat p,\tau_\lambda\gg 1) &\approx& 2\ln
	\left\{2R_{\rm cl} k_\lambda^{\rm ext}/n_{\rm H}\,n_{\rm H}(R_{\rm cl})\right\} \nonumber\\
	&&- \left\{
	\begin{array}{ll}
		\ln 2,\quad \mathrm{sphere}\\
		\ln \frac{3}{4}, \quad \mathrm{cylinder}.
	\end{array}
	\right.
\end{eqnarray}
The effective extinction of a sphere and a cylinder for a wide range of overpressures are given in Fig.~\ref{fig_optprop1}. 
For high overpressure $\hat p$, the effective optical depth can be considerably lower than the corresponding
mean optical depth through a comparable homogeneous cloud of the same mass.

\subsection{\label{sect_cloud_pesc}Escape probability}

The escape probability is based on the assumption that the emissivity is proportional to the density. It is given by the
ratio of the luminosity escaping the cloud without being scattered or absorbed to the total luminosity produced in the cloud
\begin{equation}
	p_{\rm esc}(\hat p, \tau_\lambda) = \frac{F_\lambda^{-}(\hat p,\tau_\lambda)}{F_\lambda^{-}(\hat p, 0)},
\end{equation}
where
\begin{equation}
	F_\lambda^{-}(\hat p, 0)=V_{\rm cl}4\pi\,(\eta_\lambda/n_{\rm H}) \left<n_{\rm H}\right>/S_{\rm cl}
\end{equation}
and
\begin{eqnarray}
	F_\lambda^{-}(\hat p, \tau_\lambda) &=& \int{\rm d}\Omega\,\vec n_F\cdot \vec n_J\,
		\int_0^{\infty}{\rm d}l\,\eta_{\lambda}(l,\Omega)\,\nonumber\\&&\times 
		\exp\left\{-\int_0^{l}{\rm d}l'\,k_\lambda^{\rm ext}(l',\Omega)\right\}.
\end{eqnarray}
For the sphere, the escape probability becomes
\begin{eqnarray}
	p_{\rm esc}^{\rm sph}(\hat p,\tau_\lambda) &=& \frac{3}{2\left<n_{\rm H}\right>}
		\int_0^{\pi/2}{\rm d}\vartheta\,\sin\vartheta\,\cos\vartheta\nonumber\\&&\times
		\int\limits_{-\cos\vartheta}^{\cos\vartheta}{\rm d}l\,n_{\rm H}(R_{\rm cl}\sqrt{l^2+\sin^2\vartheta})
		\,e^{-\tau_\lambda(l,\sin\vartheta)}
\end{eqnarray}
and for the cylinder
\begin{eqnarray}
	p_{\rm esc}^{\rm cyl}(\hat p,\tau_\lambda) &=& \frac{2}{\pi\left<n_{\rm H}\right>}
		\int\limits_{0}^{\pi/2}{\rm d}\vartheta\,\sin\vartheta\int\limits_0^{\pi/2}{\rm d}\varphi\,\cos\varphi\nonumber\\&&\times
		\int\limits_{-\cos\vartheta}^{\cos\vartheta}{\rm d}l\,n_{\rm H}(R_{\rm cl}\sqrt{l^2+\sin^2\varphi})
		\,e^{-\tau_\lambda(l,\sin\varphi)/\sin\vartheta},
\end{eqnarray}
where the optical depth in the exponent is given by
\begin{equation}
	\tau_\lambda(l,\sin\vartheta) =R_{\rm cl} (\kappa_\lambda^{\rm ext}/n_{\rm H})\int\limits_{-\cos\vartheta}^{l}{\rm d}l'\,
		n_{\rm H}(R_{\rm cl}\sqrt{l'^2+\sin^2\vartheta}).
\end{equation}

\begin{figure*}[htpb]
	\includegraphics[width=0.49\hsize]{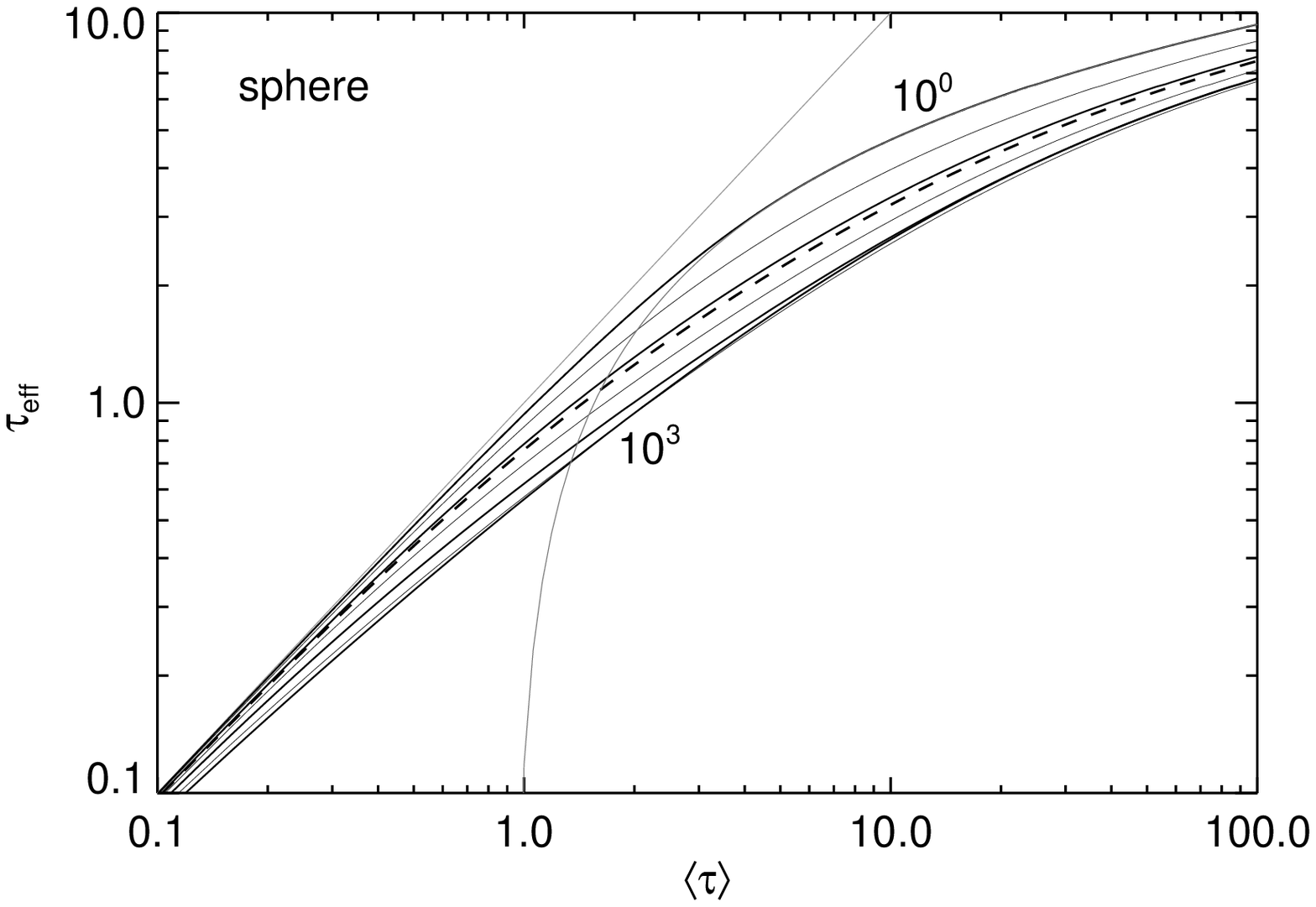}
	\hfill
	\includegraphics[width=0.49\hsize]{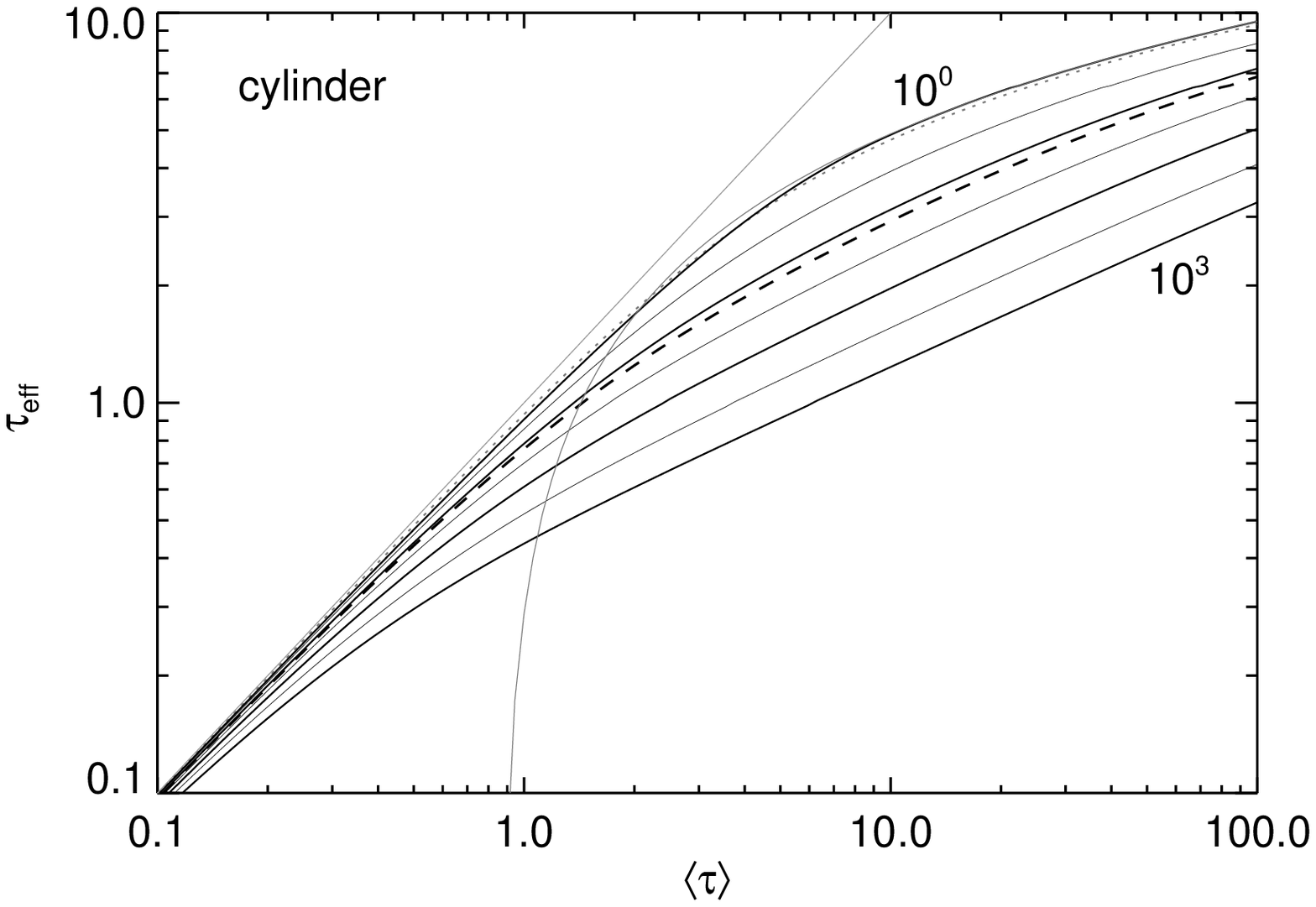}
	\caption{\label{fig_optprop1}
	Effective optical depth of self-gravitating spheres and cylinders as function of the mean optical
	depth for several assumptions for the over pressure $\hat p=p_{\rm c}/p_{\rm ext}$. The overpressure
	is varied from $10^0$ to $10^3$ in steps of $\Delta \log \hat p=0.5$. The dashed line shows the 
	effective optical depth for clouds with over pressure $\hat p= 14.04$, the over pressure for critical
	stable spherical clouds. The thin grey lines show the approximations for low and high mean optical depths.
	The effective extinction of the cylinders is also compared to the corresponding values of a homogeneous
	sphere of the same mean optical depth (dotted line).
	}
	\includegraphics[width=0.49\hsize]{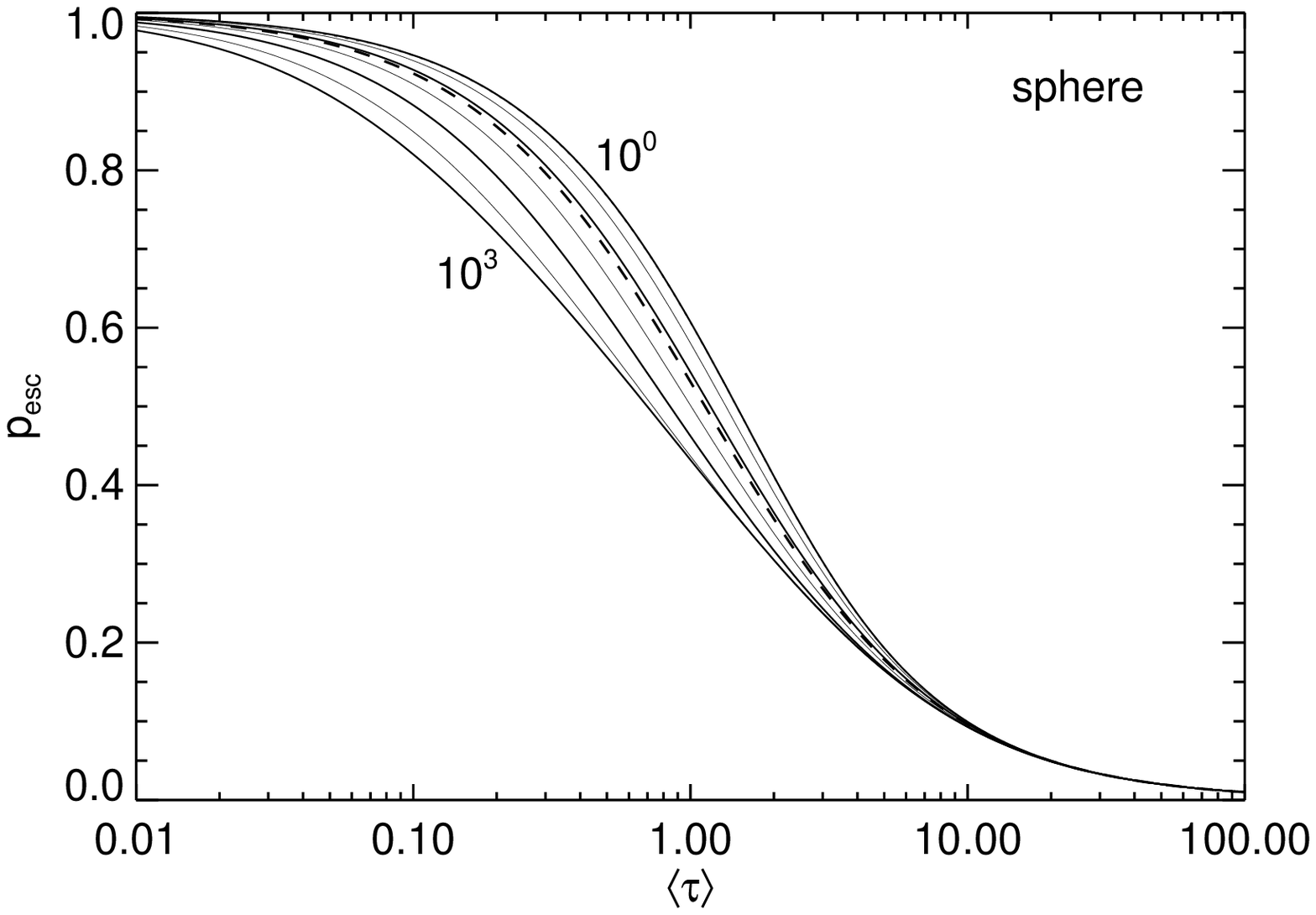}
	\hfill
	\includegraphics[width=0.49\hsize]{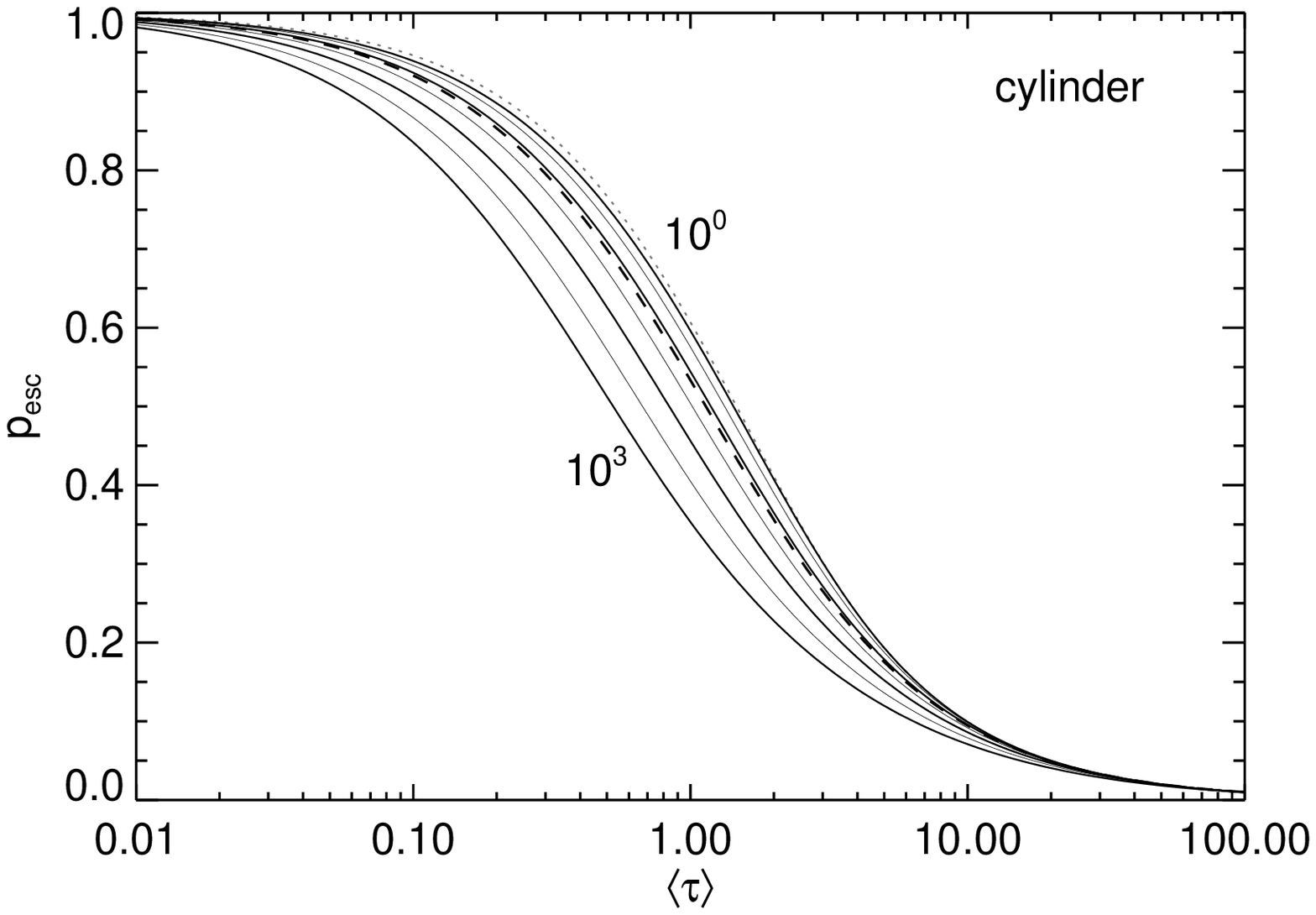}
	\caption{\label{fig_optprop2}
	Escape probability of self-gravitating spheres and cylinders as a function of the mean optical
	depth. The overpressure is varied in the same range as in Fig.~\ref{fig_optprop1}. The dashed
	line shows the escape probability of a critical stable cloud with $\hat p=14.04$ and
	dotted line in the right-hand figure the escape probability of a homogeneous sphere.
	}
\end{figure*}

The escape probability for self-gravitating spheres and cylinders is shown in Fig.~\ref{fig_optprop2}.
For intermediate values, the escape probability is lower than for the homogenous clouds.

\subsection{\label{sect_cloud_albedo}Cloud albedo}

The cloud albedo is based on the escape probability described in the previous section. It is assumed that the scattered 
light is produced in proportion to the density. For multiple scattered events, this gives the cloud albedo
\begin{equation}
	\omega_\lambda^{\rm cl}(\hat p, \tau_\lambda) = 
	\frac{\omega_\lambda p_{\rm esc}(\hat p, \tau_\lambda)}{1-\omega_\lambda(1-p_{\rm esc}(\hat p, \tau_\lambda))}.
\end{equation}
These values are certainly very approximately. However, the formula indicates to first order the general way in
which the cloud albedo decreases for higher extinction values. As a consequence, a larger fraction of light is absorbed inside
the cloud.

\bibliographystyle{aa} 
\bibliography{14785.bib}

%

\end{document}